\documentclass[twocolumn]{aastex631}

\usepackage{graphicx}
\usepackage{txfonts}
\usepackage{multirow}
\usepackage{array}
\usepackage[toc,page]{appendix}

\definecolor{db}{rgb}{0.0, 0.3, 0.8}
\definecolor{dg}{rgb}{0.0, 0.5, 0.0}
\definecolor{dr}{rgb}{0.7, 0.0, 0.0}

\begin{document}

\title{Blueshifted \ion{Si}{4} 1402.77\,\AA~line profiles in a moving flare kernel observed by IRIS}

\correspondingauthor{J. L\"{o}rin\v{c}\'{i}k}
\email{lorincik@baeri.org}

\author[0000-0002-9690-8456]{Juraj L\"{o}rin\v{c}\'{i}k}
\affil{Bay Area Environmental Research Institute, NASA Research Park, Moffett Field, CA 94035, USA}
\affil{Lockheed Martin Solar \& Astrophysics Laboratory, Org. A021S, Bldg. 252, 3251 Hanover St., Palo Alto, CA 94304, USA}
\author[0000-0003-1308-7427]{Jaroslav Dud\'{i}k}
\affil{Astronomical Institute of the Czech Academy of Sciences, Fri\v{c}ova 298, 251 65 Ond\v{r}ejov, Czech Republic}
\author[0000-0002-4980-7126]{Vanessa Polito}
\affil{Bay Area Environmental Research Institute, NASA Research Park, Moffett Field, CA 94035, USA}
\affil{Lockheed Martin Solar \& Astrophysics Laboratory, Org. A021S, Bldg. 252, 3251 Hanover St., Palo Alto, CA 94304, USA}

\begin{abstract}

We analyze spectra of a slipping flare kernel observed during the 2015 June 22 M6.5-class flare by the Interface Region Imaging Spectrograph (IRIS). During the impulsive and peak phases of the flare, loops exhibiting an apparent slipping motion along the ribbons were observed in the 131\,\AA~channel of SDO/AIA. The IRIS spectrograph slit observed a portion of the ribbons, including a {moving kernel corresponding to a flare loop footpoint} in \ion{Si}{4}, \ion{C}{2}, and \ion{Mg}{2} at a very-high 1\,s cadence. The spectra observed in the kernel were mostly redshifted and exhibited pronounced red wings, as typically observed in large flares. However, in a small region in one of the ribbons, the \ion{Si}{4} 1402.77\,\AA~line was partially blueshifted, with the corresponding Doppler velocity $|v_{\text{D}}|$ exceeding $50$\,km\,s$^{-1}$. In the same region, the \ion{C}{2} 1334.53\,\AA, 1335.66\,\AA~and 1335.71\,\AA~lines were weakly blueshifted ($|v_{\text{D}}| < 20$\,km\,s$^{-1}$) and showed pronounced blue wings, which were observed also in the \ion{Mg}{2} k 2796.35\,\AA~as well as the \ion{Mg}{2} triplet 2798.75\,\AA~and 2798.82\,\AA~lines. Using high-cadence AIA observations we found that the region where the blueshifts occurred corresponds to the accelerating kernel front as it {moved} through a weak-field region. The IRIS observations with high resolution allowed us to capture the acceleration of {the kernel} under the slit for the first time. The unique observations of blueshifted chromospheric and TR lines provide new constrains for current models of flares.

\end{abstract}


\section{Introduction} \label{sec_introduction}

Solar flares are violent manifestations of energy release in the atmosphere of the Sun \citep[e.g.,][]{fletcher11}. They occur due to evolution of usually complex magnetic fields in the solar atmosphere and are governed by magnetic reconnection, a mechanism guiding the release of the magnetic energy accumulated along magnetic field lines and their subsequent reconfiguration into a lower energetic state. 

According to the standard two-dimensional CSHKP model of flares \citep[][]{carmichael64, sturrock66, hirayama74, kopp76}, energetic particles, that are accelerated due to the release of magnetic energy following reconnection, precipitate into the chromosphere along the newly-reconnected field lines and heat the dense chromospheric plasma at flare ribbons. The ribbons are the footpoints of the magnetic field lines associated with the flare loop arcade \citep[e.g.,][]{heyvaerts77, qiu02,dudik14}, and are the locations where the bulk of the energy is deposited. Being unable to radiate away all the energy, the hot plasma in the chromosphere and transition region is subject to overpressure that causes it to expand along the field lines, which are being filled, via the chromospheric evaporation process \citep[e.g.,][]{acton82, antonucci83, zarro88}. Key signatures of plasma upflowing from flare loop footpoints are blueshifts of lines formed at high temperatures, in some cases exceeding log \mbox{($T$ [K]) = 7} \citep[e.g.,][]{mason86, delzanna06, polito16a}. Upflows at flare loop footpoints have also been observed in lines formed in the solar corona above temperatures of log($T$ [K]) $\approx6$ \citep[e.g.,][]{delzanna02, young13} and sometimes at the solar transition region temperatures of log($T$ [K]) $\approx5$ \citep[][]{brosius03}. Such studies have usually shown a dependence between the Doppler velocities of the upflows and the formation temperature of the line exhibiting the blueshifts, with the strongest upflows observed for the hottest flare lines \citep[see e.g., the review of][]{milligan15}, which can be blueshifted by hundreds of km\,s\,$^{-1}$ \citep[][]{milligan06a, watanabe10, young13}. As flares progress, the velocity of the evaporation decreases until it completely diminishes, usually over the course of a few minutes \citep[e.g.,][]{polito15, lining15, brosius15, grahamcauzzi15}. It ought to be noted that the evaporation has also sometimes been observed at lower velocities of few tens km\,s\,$^{-1}$ \citep[e.g.,][]{czaykowska99, milligan06b, brosius09}.

The expansion of the heated chromospheric plasma further drives flows of plasma below the heated region in what is known as chromospheric condensation \citep[e.g.,][]{fischer85a}. The chromospheric condensation is manifested as redshifts of lines formed in the chromosphere and the transition region log($T$ [K]) $\approx 4-5$ at speeds of several tens km\,s$^{-1}$ \citep{wuelser94, tian15, graham20}. The redshifts of these lines trace early signatures of the energy deposition in the chromosphere and their emission slightly precedes that of the hot flare lines \citep{grahamcauzzi15}. 

Rarely, modest upflows of lines formed in the chromosphere and the transition region \citep[e.g.,][]{milligan06b, brosius09, brosius13,li15, li19} are reported in GOES-class flares. In addition, blueshifts of transition region lines have also been observed in micro- or nano-sized flare events \citep[][]{testa14, polito18, testa20}, reproduced assuming heating by accelerated electrons with weaker energy fluxes ($\approx$~10$^9$ ergs/cm$^2$/s). Peculiar chromospheric \ion{Mg}{2} line profiles with an enhanced blue wing have been reported by \citet{tei18}. Their interpretation is based on a cloud model in which the upflows visible in the \ion{Mg}{2} line arise as a consequence of expansion of a portion of the chromosphere heated by accelerated electrons.  

Some of the results mentioned above have only been possible with the advent of the Interface Region Imaging Spectrometer \citep[IRIS;][]{depontieu14}, which has allowed us to study flare ribbon spectra at unprecedented detail. Its broad temperature coverage (log \mbox{($T$ [K]) = 3.7 -- 7)} allows for simultaneous observations of lines formed across different atmospheric layers, significantly contributing to our understanding of the energy balance in the lower solar atmosphere \citep[see the review of][]{depontieu21}. Thanks to their high spatial, temporal, and spectral resolutions, IRIS data revealed sometimes periodic fluctuations of characteristics of lines observed in ribbons, such as their intensities, widths, Doppler shifts, and line wing enhancements \citep[e.g.,][to name a few]{lizhang15, brosius15, warren16, tian18, jeffrey18}. In the observations presented by \citet{lizhang15}, the quasi-periodic pattern observed in the properties of the \ion{Si}{4} 1402.77\,\AA~line can be attributed to the magnetic slipping reconnection. 

Magnetic slipping reconnection \citep[][]{aulanier06, aulanier12, janvier13} is the generalisation of magnetic reconnection in three dimensions. It is characterized by a sequential change of connectivity between neighboring field lines, translated in the apparent slipping motion of the reconnecting field lines \citep[see e.g., the review of][]{janvier17}. In observations, the slipping reconnection is manifested in the apparent slipping motion of flare loops and flare kernels \citep[e.g.,][]{dudik14, lizhang14, lorincik19a} corresponding to the flare loop footpoints \citep[e.g.,][]{berlicki04, fletcher04} along flare ribbons.

Even though the resolution of IRIS makes it an ideal instrument to study the spectra of ribbons during the slipping reconnection, only a few spectroscopic studies focused on the emission of {slipping loops and kernels at their footpoints} have been published so far. \citet{dudik16} found that the strongest sources of the evaporation observed in the 1354.10\,\AA~line of \ion{Fe}{21} were located at the footpoints of flare loops slipping along the analyzed ribbon. The emission of the \ion{Si}{4} 1402.77\,\AA~line was studied by \citet{li16} who found that this line exhibits redshifts and broadening at the footpoints of slipping loops during a build-up of an eruptive flux rope in another event. In the analysis of a confined flare of \citet{li18b}, the period of the motion of the {observed kernels} was found to correspond to the period of strong upflows observed in \ion{Si}{4} along a filament located between a pair of ribbons. 

These studies show that various properties of spectra observed in flare ribbons, related to different aspects of flare evolution, can be attributed to the slipping reconnection. However, these studies do not focus on both the spatial and temporal evolution of spectra in {individual kernels} and/or portions of ribbons along which they slip. To our knowledge, analysis of this kind is still missing in the literature.

In this work, we present rare observations of blueshifted chromospheric and transition region lines observed simultaneously in a slipping flare kernel at a high 1\,s cadence. This manuscript is structured as follows. Data used here are described in Section \ref{sec_observations}. In Section \ref{sec_event}, we introduce the event analyzed in this manuscript. Section \ref{sec_kernels_spec} contains the spectroscopic analysis of a flare kernel. Discussion of our observations and results is presented in Section \ref{sec_discussion}. Finally, Section \ref{sec_summary} summarizes our results. %

\section{Data} \label{sec_observations} 

\subsection{Spectroscopic observations}

In this study we analyze an eruptive flare that occurred on 2015 June 22. The spectroscopic data used here were acquired by the IRIS satellite. IRIS observes in two far-ultraviolet (FUV) bands at 1331.6\,\AA~-- 1358.4\,\AA~and 1380.6\,\AA~-- 1406.8\,\AA~and in a near-ultraviolet (NUV) band covering the range of 2782.6\,\AA~-- 2833.9\,\AA. These passbands contain numerous emission lines formed range of temperatures corresponding to \mbox{(log $T$ [K]) = 3.7 -- 7}. The maximal field-of-view (FOV) of IRIS is \mbox{175\arcsec $\times$ 175\arcsec}, with pixel size of 0.167\arcsec, and the effective spatial resolution of 0.33\arcsec and 0.4\arcsec in FUV and NUV bands, respectively. The spectral resolutions of IRIS are 0.026\,\AA~and 0.053\,\AA~in the FUV and NUV bands, respectively. The temporal resolution of the instrument depends on the observational mode and can be lower than 1\,s. IRIS simultaneously produces high-resolution imaging observations with its slit-jaw imager (SJI).

IRIS data were downloaded from the LMSAL data archive\footnote{\url{https://iris.lmsal.com/search/}}. The dataset contains nine spectral windows centred at 1336\,\AA, 1343\,\AA, 1349\,\AA, 1356\,\AA, 1403\,\AA, 2796\,\AA, 2814\,\AA, 2826\,\AA, and 2832\,\AA. The rasters are supplemented by four series of SJI images observed at 1336\,\AA, 1403\,\AA, 2796\,\AA, and 2862\,\AA~with a cadence of 17\,s. IRIS observations span a period between $\approx$17:00 UT and 21:15 UT and were acquired in the sparse raster mode consisting of 16 slit positions separated by $\approx$1\arcsec. The exposure time of this observation was 1\,s.

The level-2 science-ready data were processed using the standard routines contained within the SolarSoft package in IDL and the IRISpy package \citep{sunpy20} in Python. In order to maintain the information about unsaturated pixels, the intensities were kept in observed data number (DN) units and we did not calibrate them to physical intensity units. Missing pixels were excluded from the analysis. They corresponded to roughly 11\% of the total number of pixels in the rasters. Finally, despite utilising level-2 data which already account for this issue, we investigated the wavelength drift induced by the orbital variations of the instrument. The analysis of the variations of the centroid of the \ion{Ni}{1} 2799.47\,\AA~line, typically used for this purpose, was performed with the \texttt{iris\_orbitvar\_corr\_l2.pro} routine. It revealed only small indications of the wavelength drift below $\pm 1$\,km\,s$^{-1}$ which is negligible for the purposes of our study as the Doppler velocities in flare ribbons reach much higher values.
 
\subsection{Imaging observations \label{sec_data_aia}}

Context imaging observations used in this study were provided mainly by the Atmospheric Imaging Assembly \citep[AIA;][]{lemen12} onboard the Solar Dynamics Observatory \citep[SDO;][]{pesnell12}. AIA consists of four telescopes observing the Sun in 10 filter channels, 7 operating in EUV, 2 in UV, and one in the visible-light part of the spectrum. The channels of AIA were designed to observe plasma originating over a broad range temperatures. AIA observes the solar photosphere in the 1700\,\AA~and 4500\,\AA~channels, the chromosphere in the 304\,\AA~channel, and the transition region in the 1600\,\AA~filter channel. The solar corona is imaged in the 171\,\AA, 193\,\AA, 211\,\AA, and 335\,\AA~channels. Finally, the 94\,\AA~and 131\,\AA~channels are typically used to study plasma radiating at flare temperatures.
 
AIA provides full-disk images consisting of 4096 $\times$ 4096 pixels with a projected pixel size of 0.6\arcsec. The instrument has a spatial resolution of 1.4--2.4\arcsec and a cadence of 12\, or 24\,s, where both quantities depend on the filter channel \citep{boerner12}. Under normal circumstances, the exposure times of these channels range between 1 and 3\,s but drop substantially in the flare mode in order to avoid image saturation.

AIA data were converted to level 1.5 using the standard \texttt{aia\_prep.pro} routine provided in the SolarSoft package, normalized to exposure times, and corrected for the stray light using the PSF deconvolution method of \citet{poduval13}. The AIA data were then co-aligned with data observed by IRIS, with IRIS 1400\,\AA~SJI images being taken as reference ones. After applying a rotation for a roll-angle of 44.22$^{\circ}$ found using the \texttt{auto\_align\_images.pro} routine, data were shifted by $\delta X = -11.55"$ and $\delta Y = -13.23"$. These quantities were measured manually by comparing the positions of bright flare kernels in different parts of flare ribbons (Section \ref{sec_ribbons}) observed in the 304\,\AA~channel of AIA and the 1400\,\AA~SJI of IRIS. In the remainder of this paper, we use the coordinate system of IRIS, with axes termed `IRIS X' and `IRIS Y'. We indicate the solar north using arrows.

We finally note that in this manuscript we also use images of line-of-sight magnetic field $B_{\text{LOS}}$ provided by the Helioseismic and Magnetic Imager \citep[HMI;][]{scherrer12} onboard SDO.

\begin{figure*}[!t]
\includegraphics[width=17.24cm, clip,   viewport=40 05 460 115]{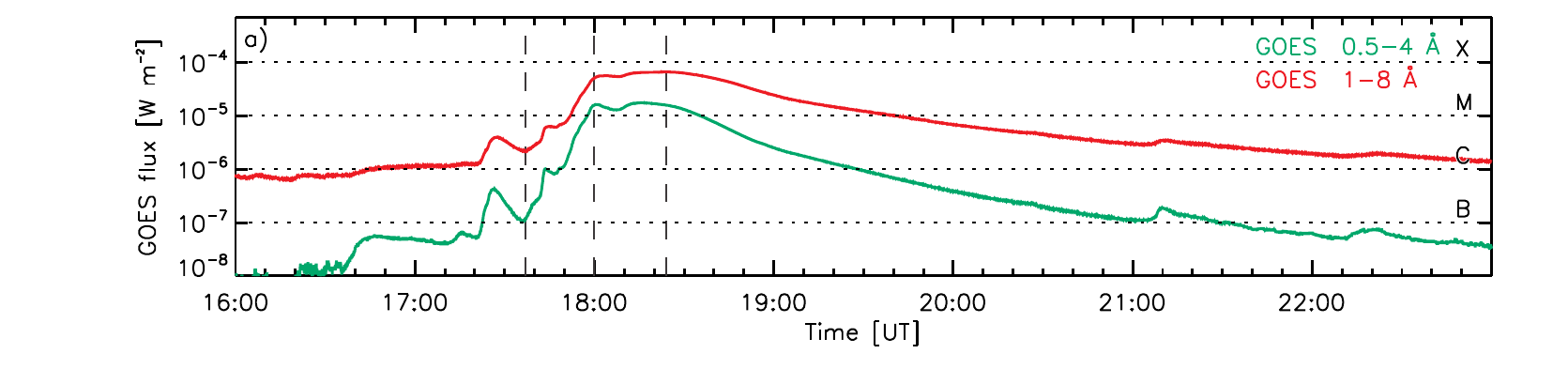}
\\
\includegraphics[width=6.50cm, clip,   viewport=05 36 235 270]{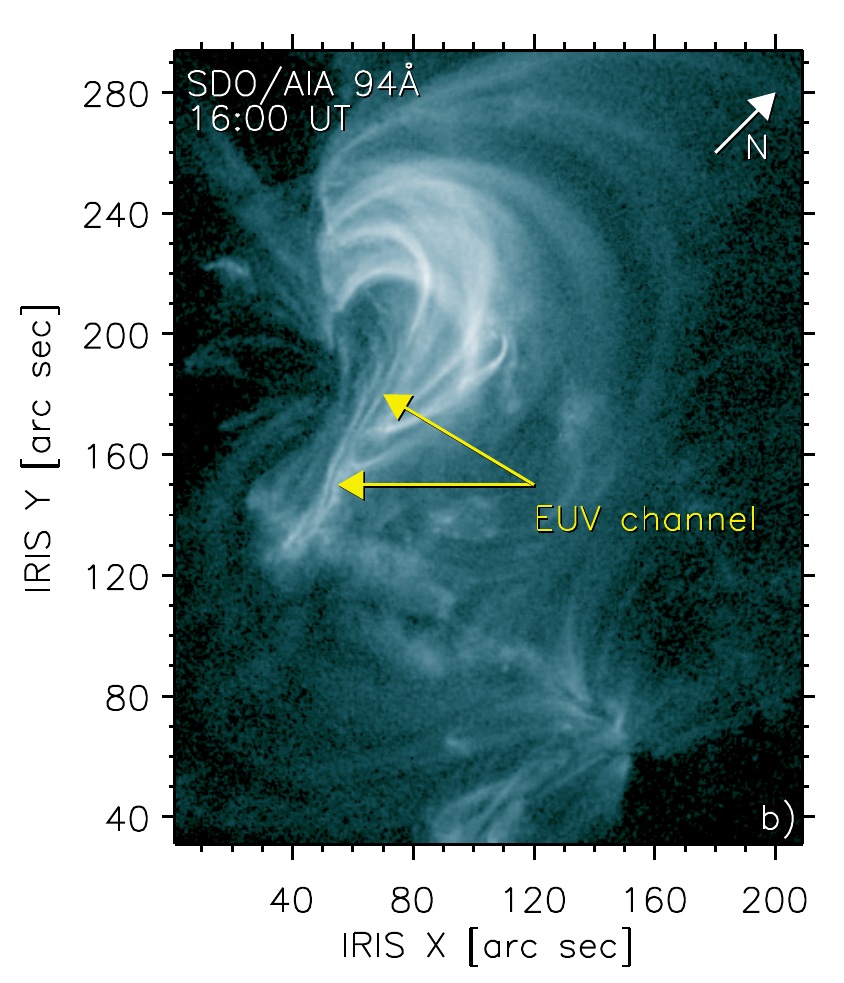} 
\includegraphics[width=5.37cm, clip,   viewport=45 36 235 270]{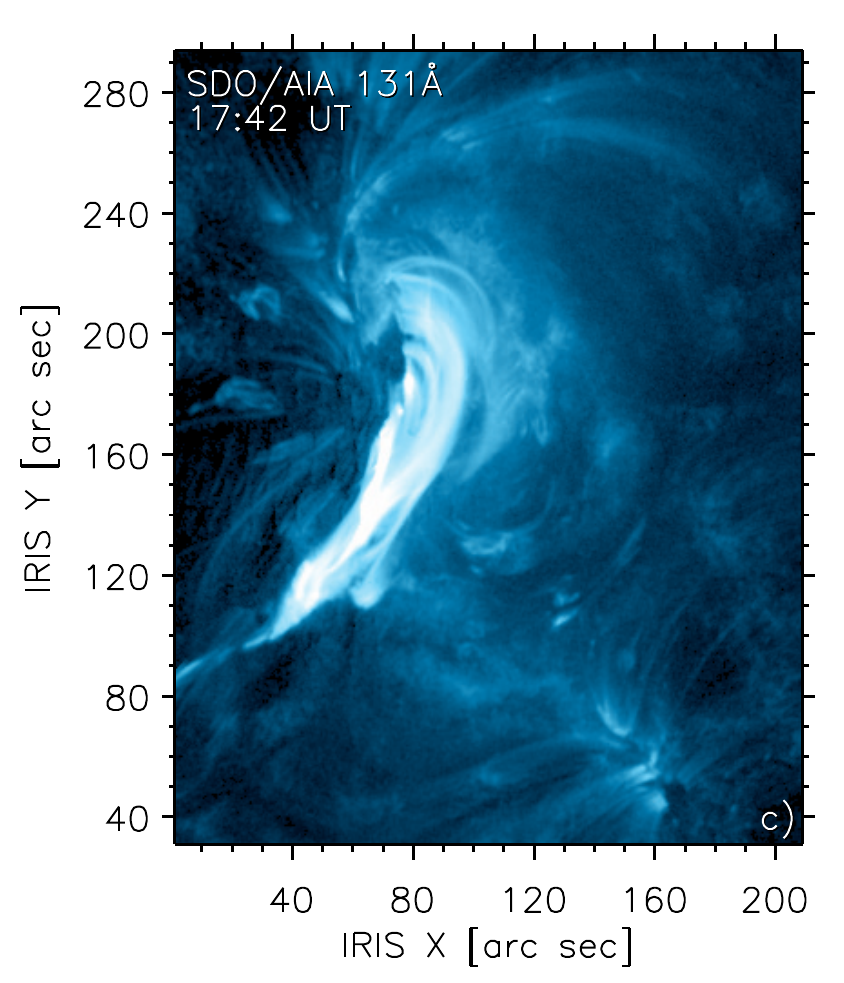} 
\includegraphics[width=5.37cm, clip,   viewport=45 36 235 270]{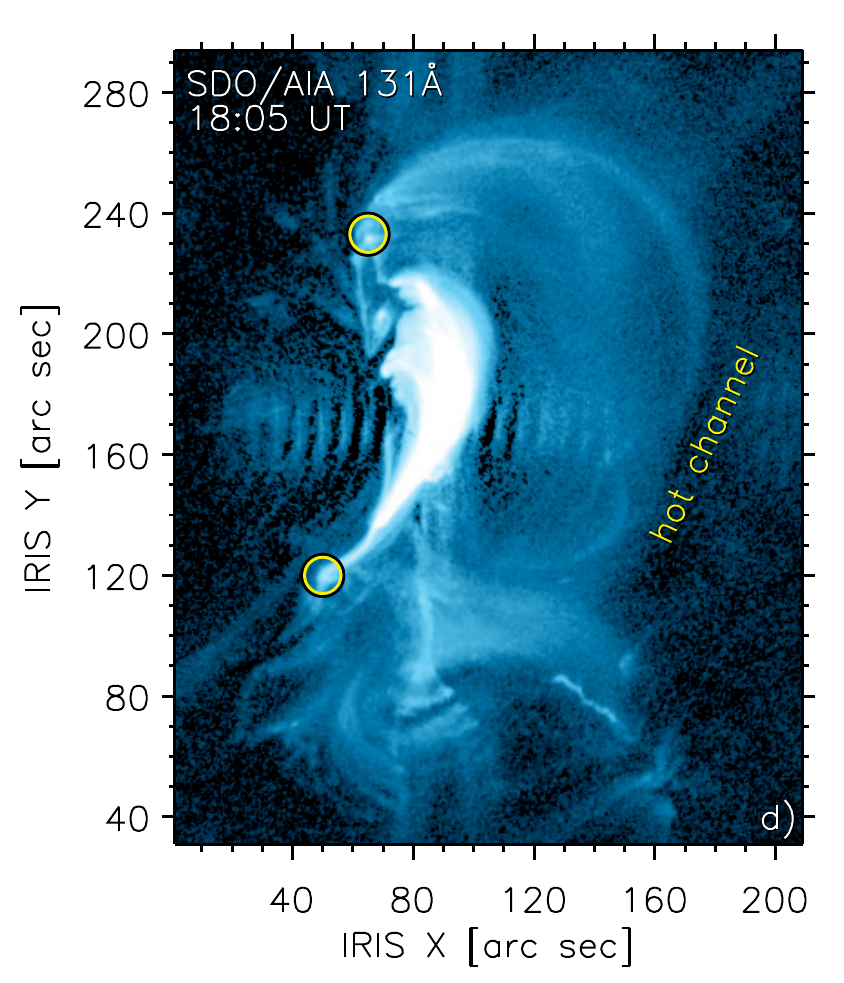}
\\
\includegraphics[width=6.50cm, clip,   viewport=05 04 235 275]{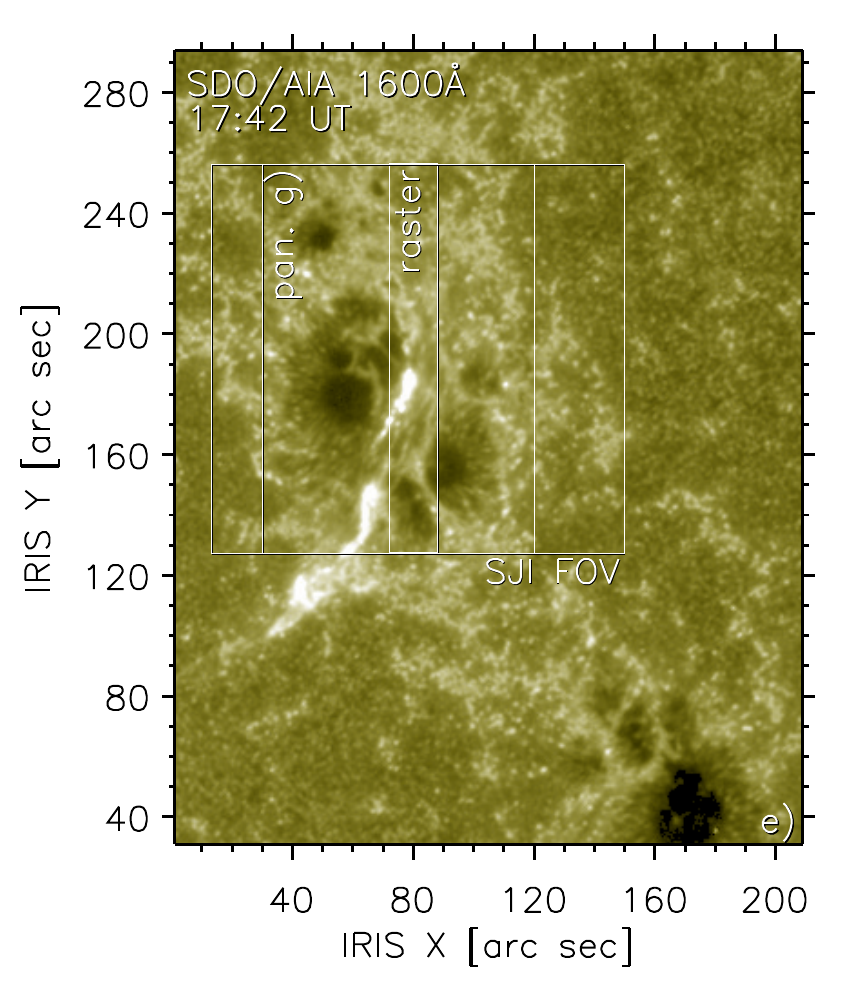} 
\includegraphics[width=5.37cm, clip,   viewport=45 04 235 275]{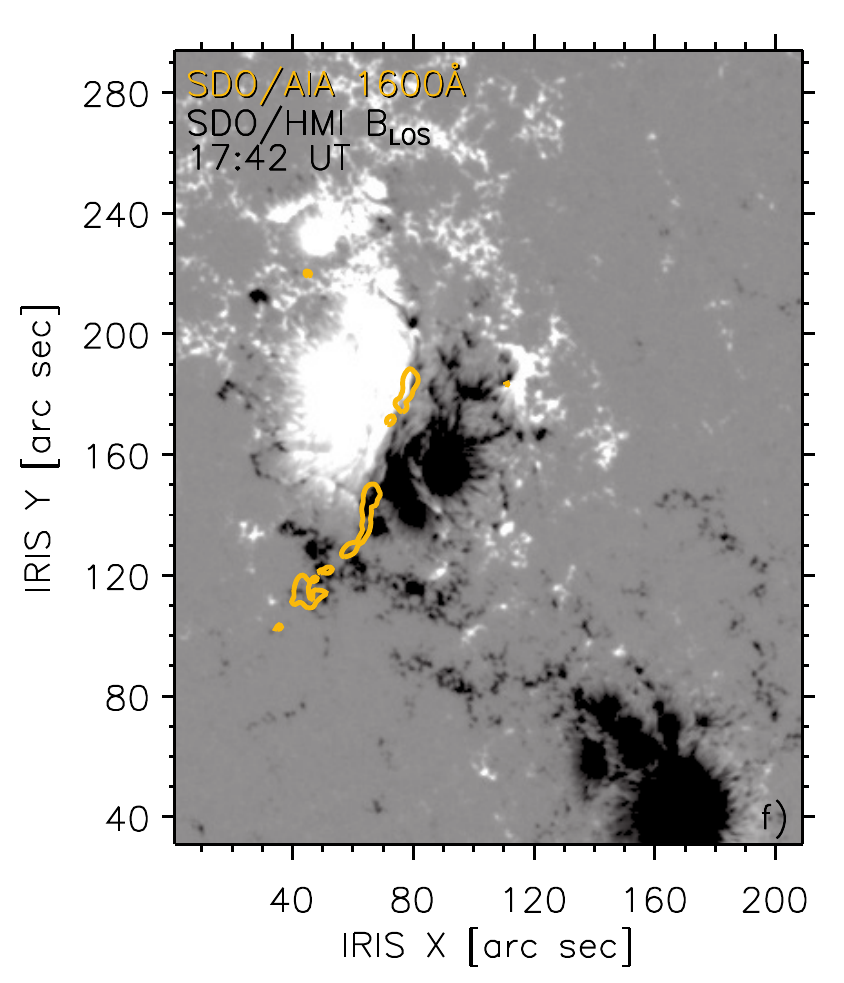} 
\includegraphics[width=5.30cm, clip,   viewport=25 04 215 280]{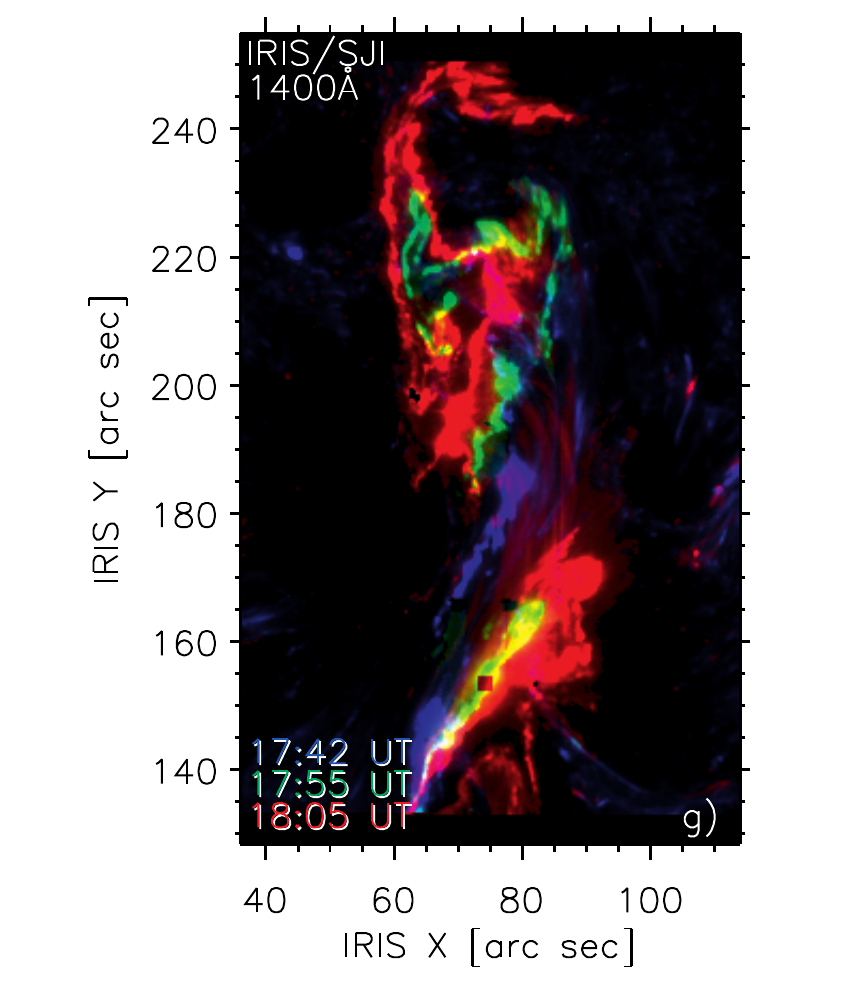}
\\
\caption{Context observations of the 2015 Jun 22 flare. Panel (a) shows the X-ray flux curve measured in the \mbox{1 -- 8\,\AA}~(red) and \mbox{0.5 -- 4\,\AA}~(green) channels of the GOES satellite. The vertical dashed lines mark different phases of the flare. In panel (b), the flaring active region and the pre-eruptive EUV channel are viewed in the AIA 94\,\AA~channel. Panels (c) and (d) display the flare during its impulsive and peak phases, respectively, as observed in the AIA 131\,\AA~channel. The erupting hot channel and its footpoints are indicated in panel (d). Flare ribbons, observed in the AIA 1600\,\AA~channel, are shown in panel (e). Orange contours corresponding to 300\,DN, higlighting the shape of the ribbons, are plotted over the $B_{\text{LOS}}$ map from SDO/HMI in panel (f). Panel (g) presents temporal evolution of the north-eastern ribbon, as observed in the 1400\,\AA~SJI images of IRIS at three different stages of its evolution. Note the direction of the solar north shown by the white arrow on panel (a).
\\Animated version of the 131\,\AA~channel observations is available in the online journal. The observations cover the period between 16:00 UT and 19:00 UT and the real-time video duration is 30\,s.  \label{fig_overview}}
\end{figure*}
\section{2015 June 22 Eruptive flare} \label{sec_event}

The 2015 Jun 22 eruptive flare occurred in the active region NOAA 12371. The eruption was accompanied by a M6.5-class flare identified as SOL2015-06-22T17:39:00. This event is well-known and has been analyzed in numerous studies, often using the IRIS satellite. For example, \citet{cheng16} analyzed hot emission of the erupting flux rope, \citet{polito16b} performed density diagnostics in one of the flare ribbons which developed during the flare, \citet{mikula17} analyzed the structure of cooling flare loops, \citet{panos18} used machine learning to cluster profiles of the chromospheric \ion{Mg}{2} h \& k lines forming in the ribbons. Characteristics of the \ion{Mg}{2} lines were also analyzed by \citet{huang19}. The event has also been extensively studied using other instruments. For example, \citet{kuroda18} detail the high-X ray (HXR) and microwave emission of the flare, \citet{liu18} and \citet{wang18} analyzed the evolution of the magnetic fields of the flaring active region, and \citet{kang19} as well as \citet{sahu20} focus on the precursor activity of this flare as well as build-up of the erupting flux rope.

\subsection{Eruption of the hot channel} \label{sec_overview}

Figure \ref{fig_overview} provides context observations of the flare we analyze here. Panel (a) presents the X-ray flux measured in the 0.5 -- 4\,\AA~(green) and 1 -- 8\,\AA~(red) channels of the GOES satellite. The second row of the figure contains observations of the flaring active region in AIA 94\,\AA~and 131\,\AA~channels before the flare (panel (b)), and during its impulsive (panel (c)) and peak (panel (d)) phases. Panels (e)--(g) in the bottom row of the figure show flare ribbons that developed during this flare.

The active region NOAA 12371 exhibited high activity prior to the flare we analyze here, related to a transformation of the `EUV channel' (Figure \ref{fig_overview}(b)) in its core to an erupting flux rope \citep{sahu20}.

The flare started roughly at 17:37 UT, when slipping flare loops (Figure \ref{fig_overview}(c), Section \ref{sec_sliprecco}) started to appear among the flare loops that formed during one of the precursors described by \citet{sahu20}. The onset of the impulsive phase of the flare is indicated using the left dark-grey dashed line in panel (a). The X-ray flux was increasing roughly until 18:00 UT (middle grey line in panel (a)) when the X-ray flux reached a broad plateau at which the flare peaked (panel (d)). The impulsive and peak phases were characterized by the development of $J$-shaped (hooked) flare ribbons (Section \ref{sec_ribbons}) and an arcade of flare loops best visible in the animated version of this figure. In panel (d) we marked the erupting flux rope manifested as the hot channel \citep[see e.g.,][]{zhang12}. This bundle of hot and twisted loops appeared from above the developing arcade of flare loops. Two patches of footpoints of the hot channel were identified, one rooted toward the north-east and the other toward the south-west (yellow circles in panel (d)). These footpoints were observed to drift \citep[c.f.,][]{aulanier19} for several tens of arc-seconds along the ribbon hooks evolving in the meantime (Section \ref{sec_ribbons}).

After $\approx$18:20 UT (right grey line in panel (a)), the X-ray flux started to progressively drop during a prolonged gradual phase lasting for several hours. Note that the small peaks visible at 21:10 UT and 22:20 UT originated in sources located elsewhere on the Sun.

\subsection{Flare ribbons \label{sec_ribbons}}

Observations in AIA filters which are sensitive to chromospheric and transition region plasma reveal a pair of flare ribbons which formed and elongated during the impulsive phase of the flare. Figure \ref{fig_overview}(e) shows the ribbons as viewed in the AIA 1600\,\AA~channel as well as the IRIS raster position and the field-of-view (FOV) of the IRIS SJI at the corresponding time (white frames). 

One ribbon is observed in the north-east and the other one in the south-west. In order to find in which magnetic polarities the ribbons resided, we overlaid the contours of the ribbons at 17:42 UT on the HMI $B_{\text{LOS}}$ image at the closest time (panel (f)). The AIA 1600\,\AA~contours (orange) correspond to 300\,DN\,s$^{-1}$\,pix$^{-1}$ and the $B_{\text{LOS}}$ map is saturated to $\pm$ 500\,G. According to this figure, the north-eastern ribbon was located in a region dominated by the positive-polarity flux concentration, while the south-western ribbon was located in the negative-polarity flux region.

The ribbons started to form after the onset of the flare, when a series of brightenings appeared at $\approx$[80\arcsec, 180\arcsec]. Some of these brightenings expanded in the north-east direction to form the north-eastern ribbon, the conjugate ones expanded towards the south-west to form its south-western counterpart. Since IRIS did not image the south-western ribbon in its entirety and the visible portions were partially saturated during the impulsive and peak phases of the flare, we focused our analysis on the north-eastern ribbon.

We study the temporal evolution of the north-western ribbon using the IRIS SJI 1400\,\AA~images. As can be seen in Figure \ref{fig_overview}(g), the ribbon was highly structured and dynamic. After the ribbon elongated during the impulsive phase (blue patches), it formed a hooked extension consisting of several smaller hooks (green). A few minutes later (red), this hook disappeared, the ribbon propagated towards the south-east direction and further elongated towards the north-east direction, and finally developed another hook at the north-west. This hook disappeared after $\approx$18:20 UT after the flare had peaked.
 
The overall evolution of this ribbon was complicated due to the appearance, diminishing, and consequent re-appearance of its hooked extension. It corresponds to previously-reported signatures of the drift of flux rope footpoints across the surface of the Sun \citep[e.g.,][]{aulanier19, zemanova19, lorincik19b}.

\begin{figure*}[!t]
\centering     
\includegraphics[width=4.407cm, clip,   viewport=00 30 169 220]{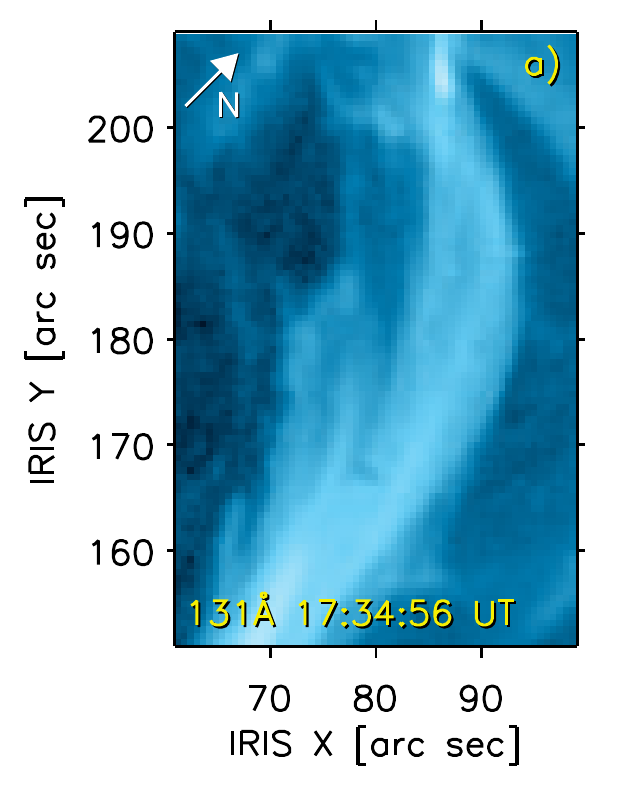} 
\includegraphics[width=3.155cm, clip,   viewport=48 30 169 220]{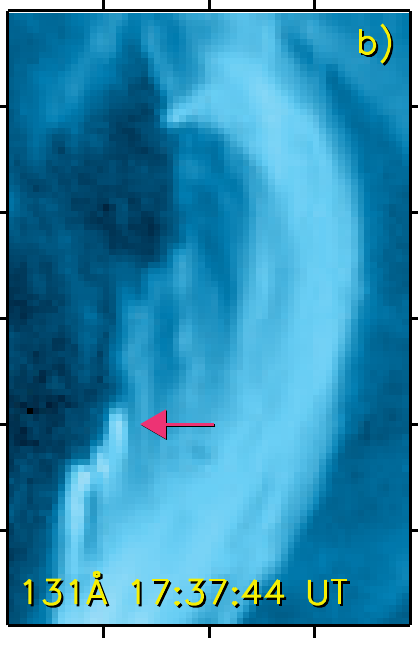}
\includegraphics[width=3.155cm, clip,   viewport=48 30 169 220]{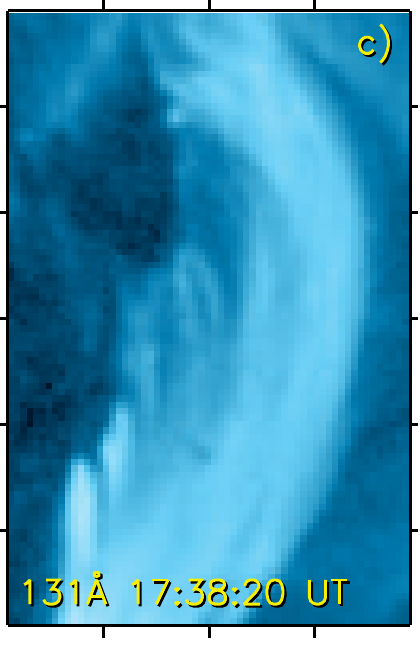}    
\includegraphics[width=3.155cm, clip,   viewport=48 30 169 220]{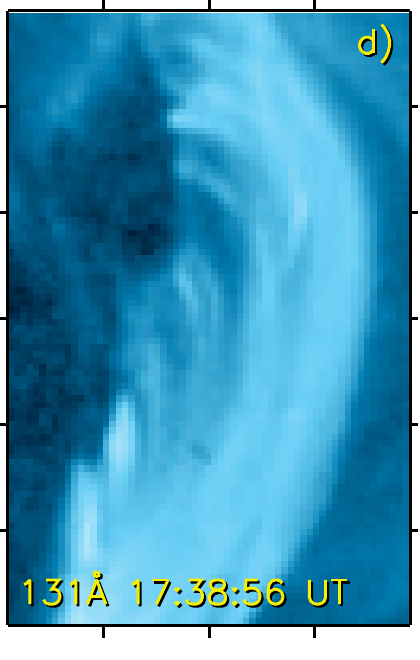}
\includegraphics[width=3.155cm, clip,   viewport=48 30 169 220]{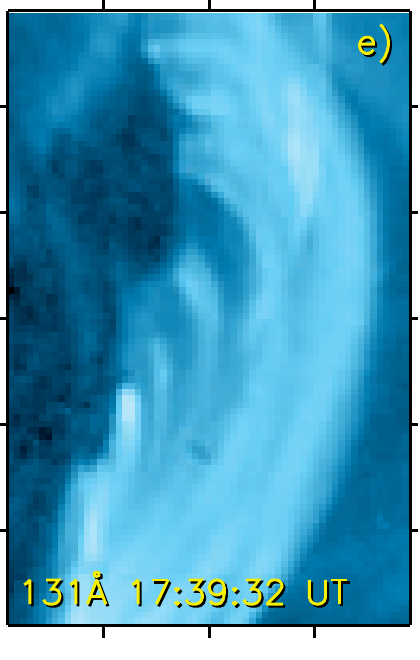}       
\\
\includegraphics[width=4.407cm, clip,   viewport=00 00 169 220]{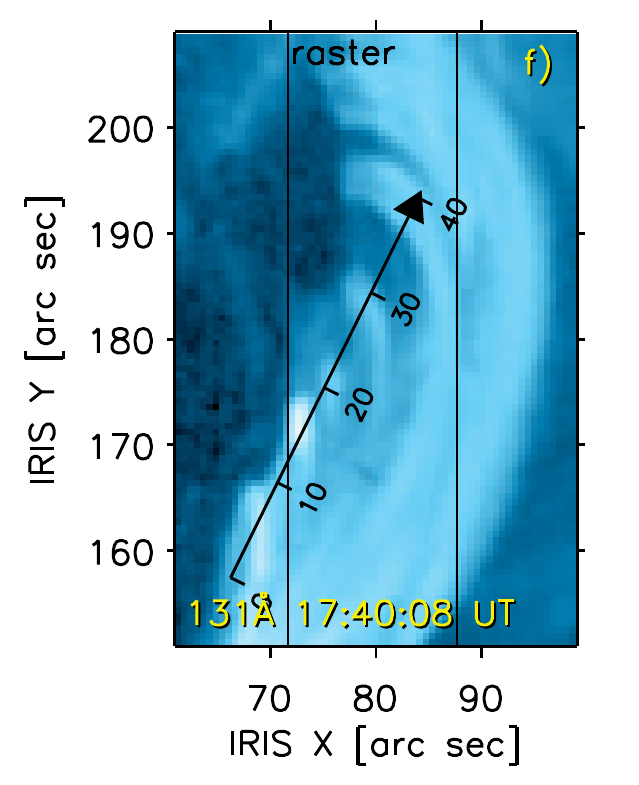} 
\includegraphics[width=3.155cm, clip,   viewport=48 00 169 220]{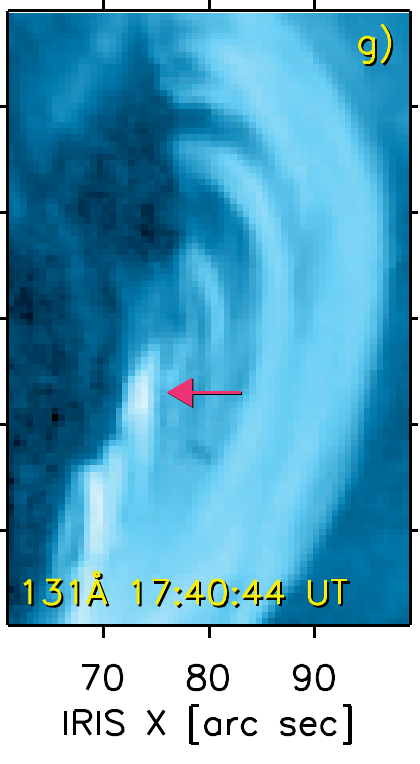}
\includegraphics[width=3.155cm, clip,   viewport=48 00 169 220]{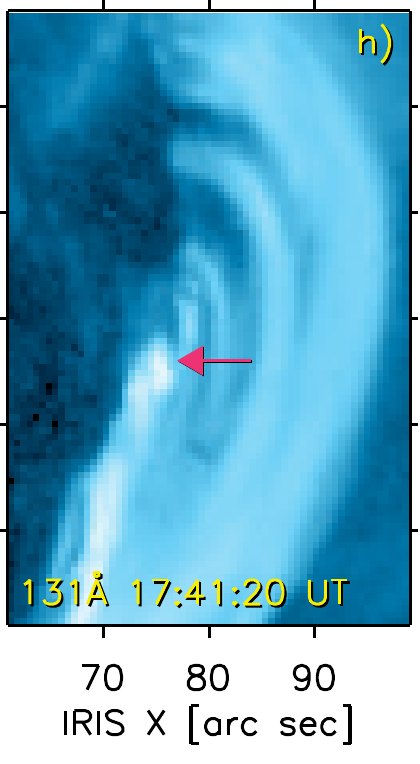}    
\includegraphics[width=3.155cm, clip,   viewport=48 00 169 220]{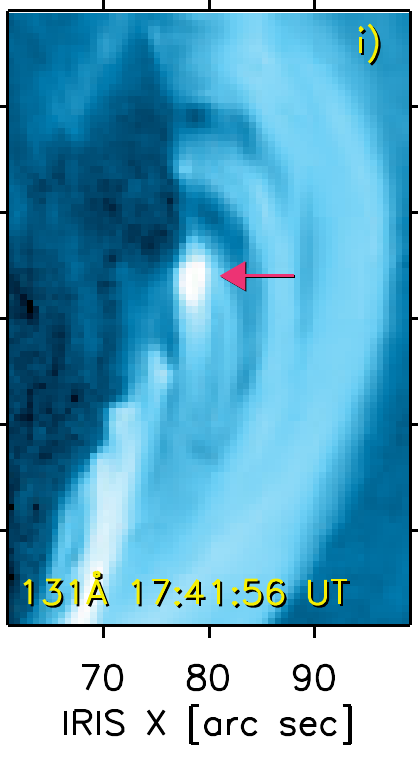}
\includegraphics[width=3.155cm, clip,   viewport=48 00 169 220]{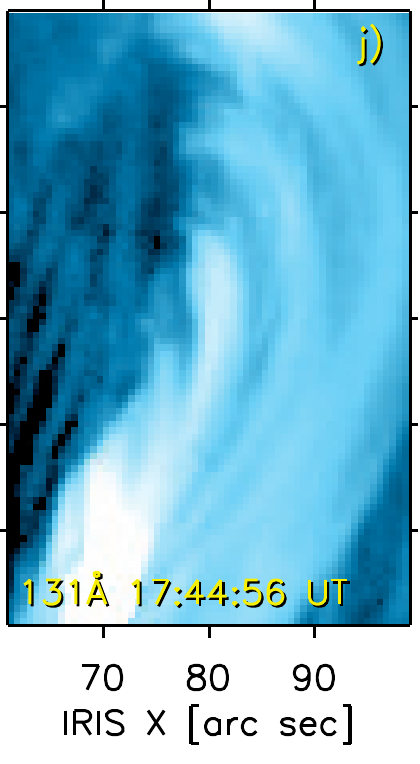}       
\caption{SDO/AIA 131\,\AA~observations of the apparently-slipping flare loops. The frame in panel (f) marks the position of the raster of IRIS at the corresponding time as well as the cut used for measuring velocities of the slipping flare loops. The pink arrows mark the footpoints of loops which correspond to the kernels visible in the SJI images of IRIS (Figure \ref{fig_ribbon_sji}). Note the direction of the solar north shown by the white arrow on panel (a).\\ Animated version of this figure is available online. The video contains observations from the period between 17:30 UT and 17:50 UT and the real-time video duration is 10\,s. \label{fig_ribbon_sliploops}}
\end{figure*}

\subsection{Slipping reconnection observed by AIA and IRIS} \label{sec_sliprecco}

Along the developing ribbon and its hooks, flare loops exhibiting apparent slipping motion were observed in the AIA 94\,\AA, and 131\,\AA~channels. The slipping flare loops can be distinguished in numerous locations and instants during the impulsive and peak phases of the flare until $\approx$18:20 UT, when the arcade of flare loops developed in its entirety. Here we focus on the flare loops and kernels which are slipping along the developing ribbon during the impulsive phase of the flare. 

\begin{figure}[h]
\centering   
\includegraphics[width=8.5cm, clip,   viewport= 17 0 275 136]{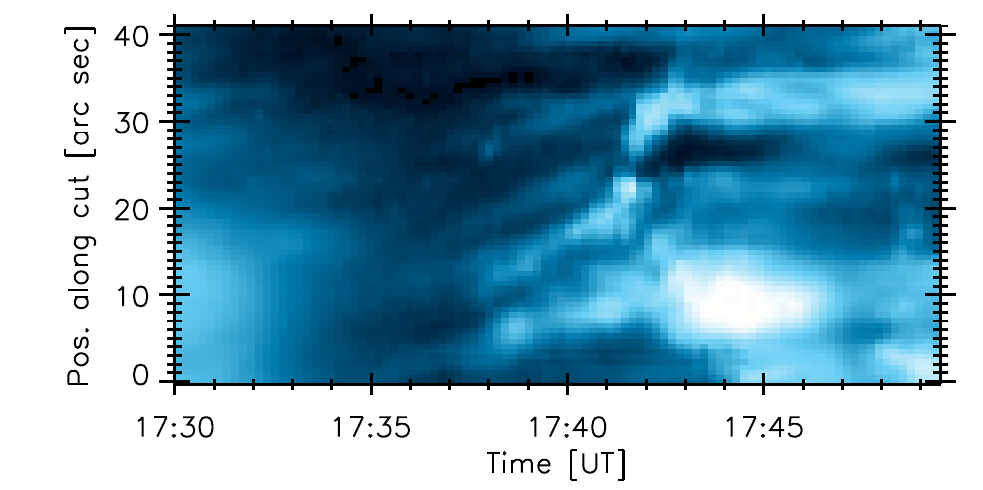} 
\includegraphics[width=8.5cm, clip,   viewport= 17 0 275 136]{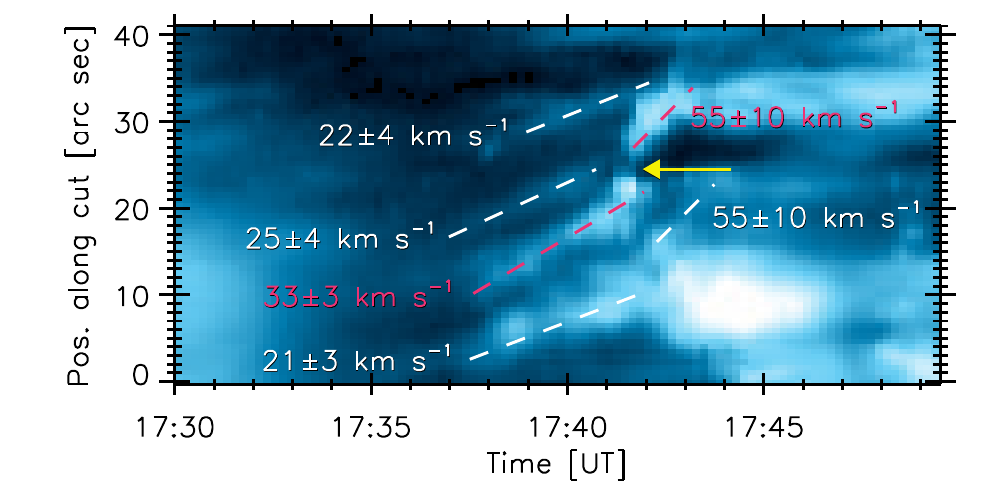} 
\caption{Time--distance diagram produced along the cut plotted in Figure \ref{fig_ribbon_sliploops}(f). The dashed lines plotted in the bottom panel are fits used for estimating the velocity of the apparent slipping motion of flare loops. The yellow arrow points towards an apparent discontinuity between two fits (dashed lines) fitting the motion of the flare loop corresponding to the kernel analyzed later in Section \ref{sec_kernels_spec}. \label{fig_xt_131}}
\end{figure}

AIA 131\,\AA~observations of the slipping flare loops are presented in Figure \ref{fig_ribbon_sliploops}. The first slipping flare loops started to be visible after $\approx$17:37 UT (panel (b)). As can be seen in the animated version of this figure, these flare loops were slipping in the north and north-east direction, indicated by the black arrow shown in panel (f). This arrow marks the cut used for the construction of the time-distance diagram and the analysis of the properties of the slipping motion. The time-distance diagram, presented in the upper panel of Figure \ref{fig_xt_131}, reveals stripes generated by the motion of the slipping flare loops along the cut, which are visible as soon as $\approx$17:35 UT. After $\approx$17:44 UT, the stripes can no longer be distinguished as the emission along the cut started to be dominated by the arcade of flare loops that appeared in the meantime. The velocities of the apparently-slipping flare loops were obtained from linear fits using the dashed lines in the bottom panel of Figure \ref{fig_xt_131} and were found to range between 21 and 55 km\,s$^{-1}$. The slipping velocities of this order of magnitude are commonly found in the literature \citep[e.g.,][]{dudik14, lizhang14, lizhang15, dudik16, lorincik19a}.

\begin{figure*}[!t]
\centering     
\includegraphics[width=4.407cm, clip,   viewport=00 00 169 220]{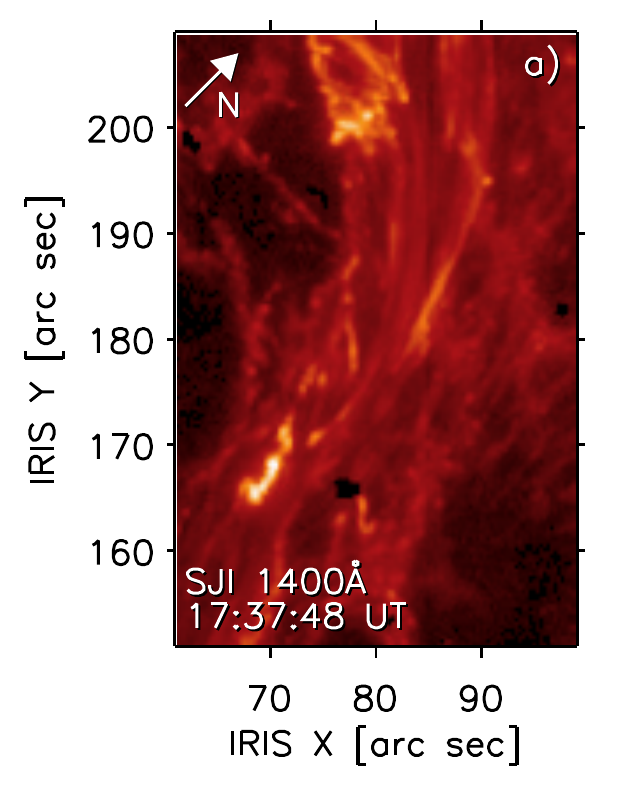} 
\includegraphics[width=3.155cm, clip,   viewport=48 00 169 220]{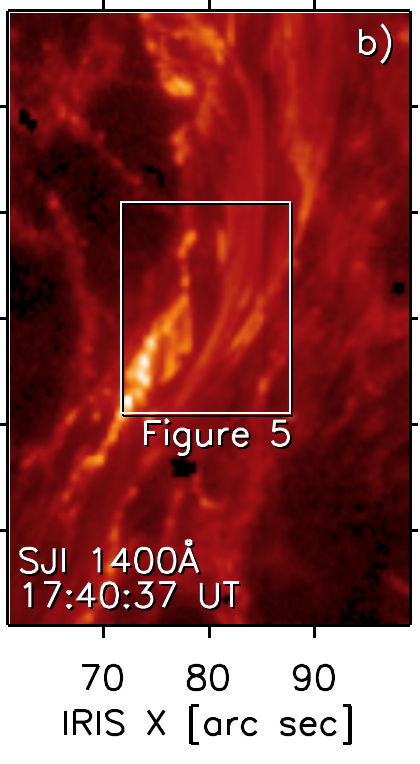}
\includegraphics[width=3.155cm, clip,   viewport=48 00 169 220]{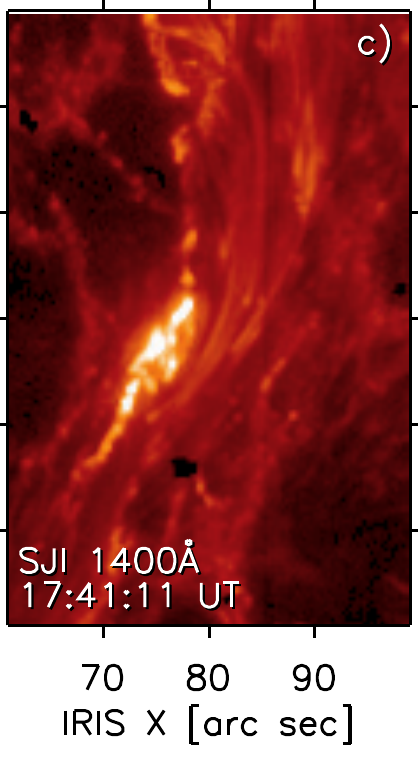}    
\includegraphics[width=3.155cm, clip,   viewport=48 00 169 220]{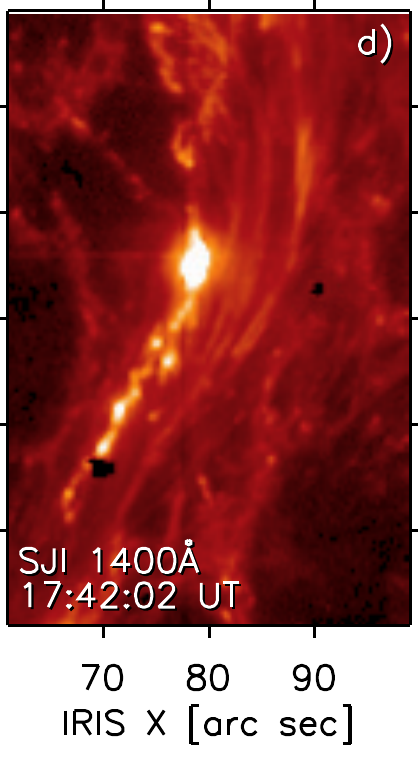}
\includegraphics[width=3.155cm, clip,   viewport=48 00 169 220]{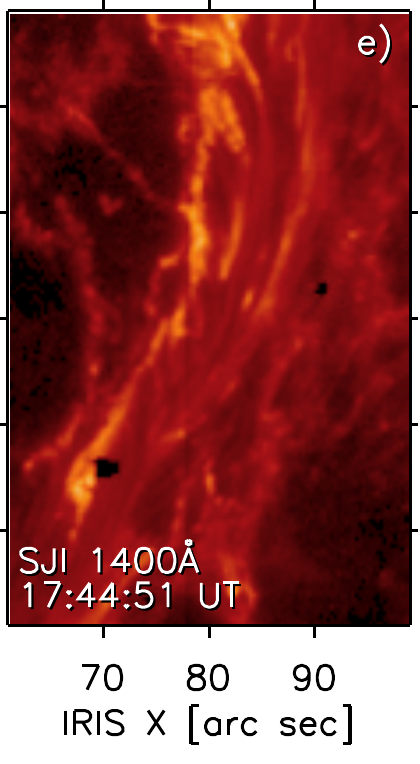}   
\caption{1400\,\AA~SJI observations of the flare kernels {moving} along the elongating ribbon. The arrow shown in panel (a) indicates the direction of the solar north. \\ Animated version of this figure is available online. The video contains observations from the period between 17:30 UT and 17:50 UT and the real-time video duration is 10\,s. \label{fig_ribbon_sji}}
\end{figure*}

IRIS slit-jaw images of the same region, acquired at the same times as discussed above, are shown in Figure \ref{fig_ribbon_sji}. This figure has the same FOV as Figure \ref{fig_ribbon_sliploops} and shows a portion of the north-eastern ribbon during its elongation after the onset of the flare. In panels (a) -- (d), multiple flare kernels {moving} along the elongating ribbon are visible. {These kernels corresponded to the footpoints of the slipping flare loops observed in the same location (see above).} After the kernels appeared in the south (panel (a)), they {moved} towards the north (panels (b), (c)) and north-east direction (panel (d)) until they reached a location where {their motion} ended.


A comparison between Figures \ref{fig_ribbon_sji} and \ref{fig_ribbon_sliploops} reveals that the apparent motion of the footpoint of one of the slipping flare loops observed in AIA 131\,\AA~corresponded to {the motion of} the kernels observed in the SJI images. In Figure \ref{fig_ribbon_sliploops}, this footpoint was marked using the pink arrows in panels (b), (g), (h), and (i). Due to its higher spatial resolution, (compare Figures \ref{fig_ribbon_sji} and \ref{fig_ribbon_sliploops}) IRIS resolves the footpoint of the slipping AIA loop into a series {of moving kernels}. Therefore, had AIA possessed a higher spatial resolution, we would have likely resolved a bundle of slipping loops originating from these kernels instead of a single one. These kernels however still move coherently as one larger brightening, which is consistent with the observations of the AIA 304\,\AA~channel that we discuss later on (Section \ref{sec_ker_detailed}). There, the kernels appear to be a part of a single structure, for which we will refer to them for simplicity as `the kernel'. 

We fitted the slipping velocities of the slipping loop rooted in this kernel using the pink dashed lines in Figure \ref{fig_xt_131}. The slipping velocities resulting from the fits were found to correspond to 33 and 55 km\,s$^{-1}$. The discontinuity between the two linear regressions marked using the yellow arrow is most likely caused by a brief acceleration of the kernel as well as the shape of its trajectory detailed in Sections \ref{sec_ker_detailed} and \ref{sec_discussion_slip}.

Finally, we note that no bright kernels can be distinguished in Figure \ref{fig_ribbon_sji}(e). The time of the observation of this frame is close to that in Figure \ref{fig_ribbon_sliploops}(j), when the flare loop arcade was developed and no slipping flare loops were visible in Figure \ref{fig_xt_131}. 

\section{Spectroscopic analysis of the kernel} \label{sec_kernels_spec}

Next, we analyze spectra formed in the flare kernel {which, as we showed in Section \ref{sec_sliprecco}, corresponds to the footpoint of one of the flare loops slipping along the north-eastern ribbon}. We primarily focus on the characteristics of the \ion{Si}{4} 1402.77\,\AA~line. The line profiles in the ribbon were usually intense, with the peak intensities ranging between $\approx$10$^2$ and 10$^4$\,DN, and easy to distinguish from the FUV continuum. Because of the short 1\,s exposures used in these observations, the peak of this line was usually below the saturation threshold of the instrument ($\approx 1.6 \times 10^4$\,DN). Our inspection of the spectra did not reveal the presence of any identifiable blends to the line, which is in line with other analyzes of this line in flares and active regions \citep[e.g.,][to name a few]{doschek16, polito16a, dudik17_nonmaxw, jeffrey18, young18}. \citet{mulay21} have shown that H$_2$ line emission can sometimes be detected at flare ribbons, with the 1402.64\,\AA~line blending the \ion{Si}{4} 1402.77\,\AA~line. To rule out the presence of this blend, we inspected the \ion{C}{2} spectral window containing the H$_2$ 1333.48\,\AA~and 1333.80\,\AA~lines analyzed by \citet{mulay21} but did not find any traces of these lines. This is likely caused by the relatively-shorter exposure time of our observations (1\,s) compared to the 8\,s exposure time used in the dataset they analyze. Since flare simulations show that the 1402.64\,\AA~line is only by a factor of $\approx2$ stronger than the 1333.80\,\AA~line \citep{jaeggli18}, we conclude that the 1402.64\,\AA~line did not affect our analysis of the \ion{Si}{4} 1402.77\,\AA~line.

Unlike the \ion{Si}{4} 1393.75\,\AA~line routinely observed by IRIS, the 1402.77\,\AA~line does not exhibit signatures of the absorption of chromospheric lines \citep[e.g.,][]{joshi21}. It has been shown that both the intensity and shape of the line can be affected by opacity effects \citep{kerr19}, in some cases leading to self-absorption of the line \citep{yan15}. Since IRIS did not observe the \ion{Si}{4} 1393.75\,\AA~line during this flare, we could not study effects of opacity in the analyzed ribbon. In this work we thus assume that the line is formed under optically thin conditions. Possible implications of the opacity effects on the interpretation of our results are discussed in Section \ref{sec_kernel_rba}.

In the following section, we investigate maps of the integrated line intensity (Section \ref{sec_kernel_raster}), Doppler velocities (Section \ref{sec_kernel_doppler}), and red-blue asymmetries (Section \ref{sec_kernel_rba}). The temporal evolution of this line as well as selected chromospheric lines is discussed in Section \ref{sec_kernel_temporal}. 
\subsection{Spatial characteristics of the \ion{Si}{4} 1402.77\,\AA~line profiles}

In Figure \ref{fig_spec_ribbon} we show the maps of line intensities, Doppler shifts and red-blue asymmetries obtained from the \ion{Si}{4} 1402.77\,\AA~line spectra in five different rasters observed between $\approx$17:40 UT and 17:43 UT. Each of these maps contain all 16 slit positions of each raster plotted between IRIS $Y$ = 171\arcsec and 191\arcsec, containing a portion of the positive-polarity ribbon associated with {the kernel} described in Section \ref{sec_sliprecco}. The times corresponding to the start of the observation of each raster are indicated above the top row of this figure. We limited our analysis of the Doppler velocities as well as red-blue asymmetries to the pixels where the peak intensity of the line exceeded 300\,DN. In such pixels, the entire line profiles including their wings were usually well above the FUV continuum. 

\subsubsection{Maps of the integrated line intensity}
\label{sec_kernel_raster}

\begin{figure*}[t]
\centering
\includegraphics[width=5.60cm, clip,   viewport= 07 32 241 210]{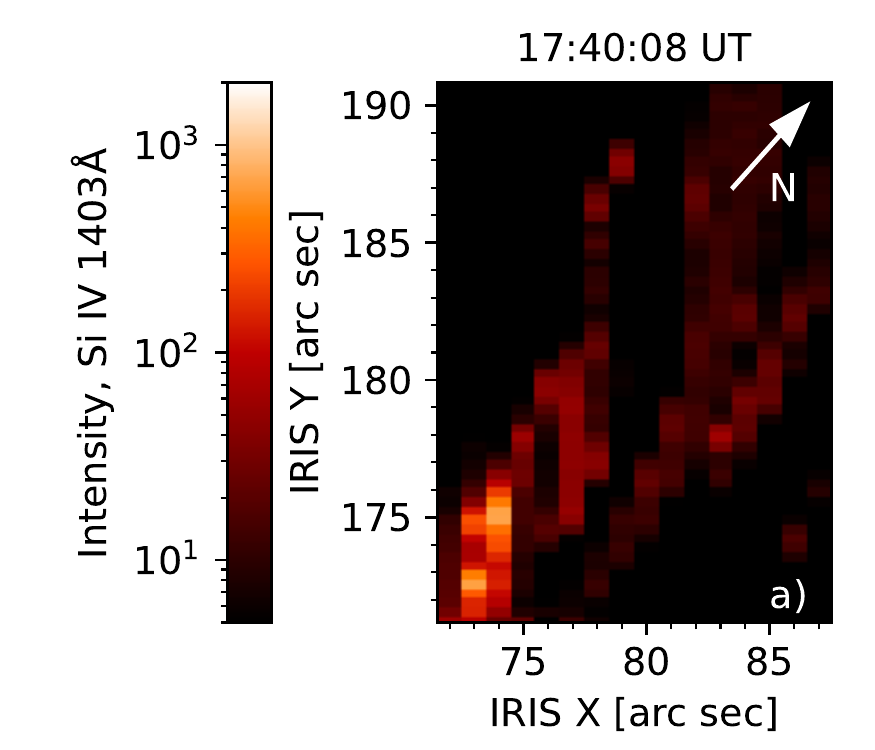}
\includegraphics[width=2.90cm, clip,   viewport= 120 32 241 210]{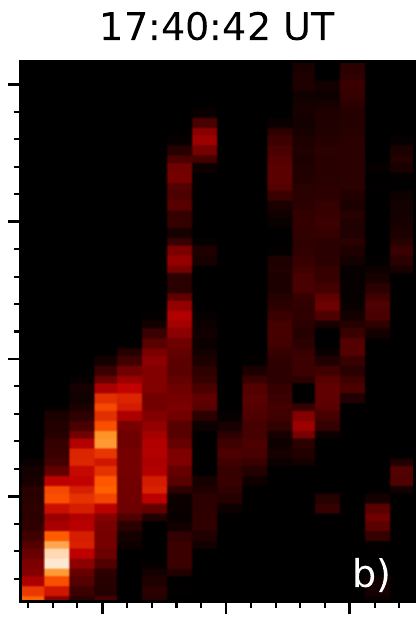}
\includegraphics[width=2.90cm, clip,   viewport= 120 32 241 210]{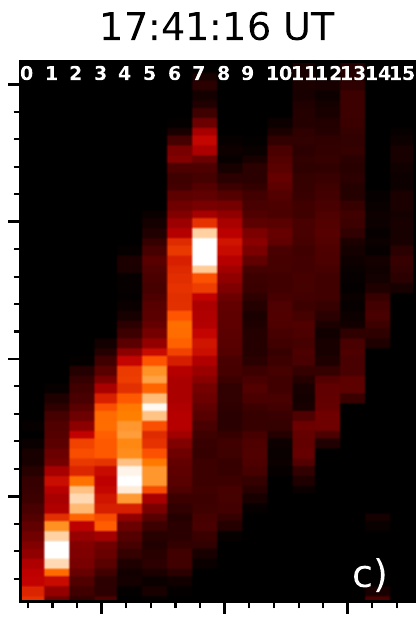}
\includegraphics[width=2.90cm, clip,   viewport= 120 32 241 210]{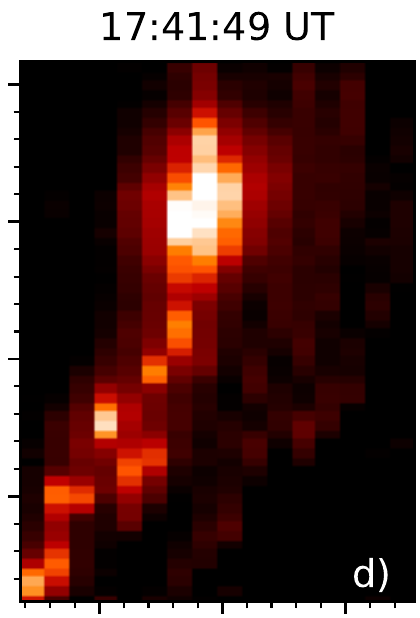}
\includegraphics[width=2.90cm, clip,   viewport= 120 32 241 210]{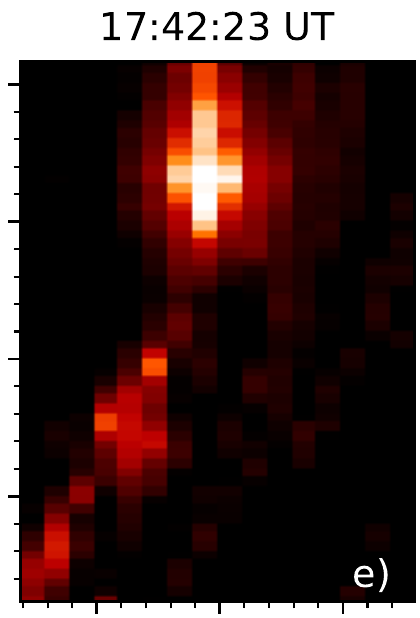}
\\
\includegraphics[width=5.60cm, clip,   viewport= 07 32 241 197]{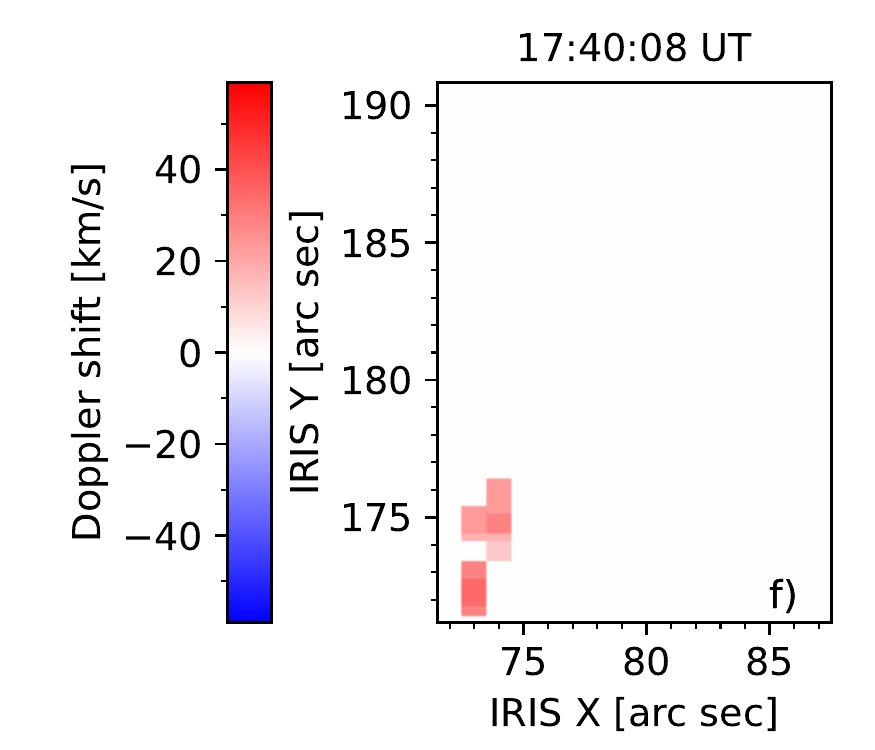}
\includegraphics[width=2.90cm, clip,   viewport= 120 32 241 197]{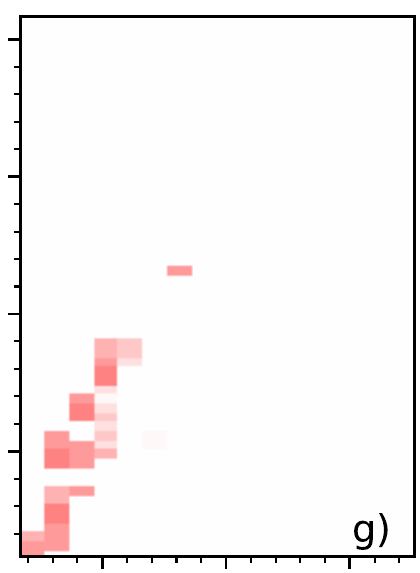}
\includegraphics[width=2.90cm, clip,   viewport= 120 32 241 197]{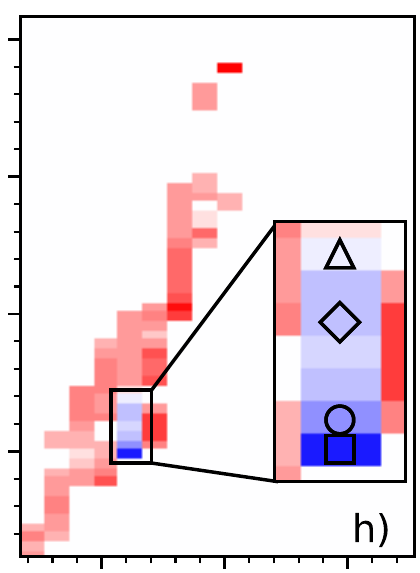}
\includegraphics[width=2.90cm, clip,   viewport= 120 32 241 197]{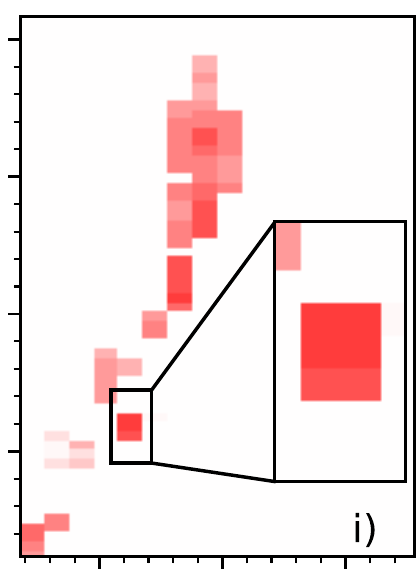}
\includegraphics[width=2.90cm, clip,   viewport= 120 32 241 197]{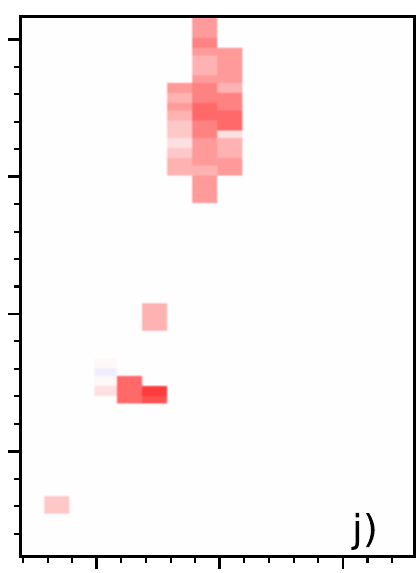}
\\
\includegraphics[width=5.60cm, clip,   viewport= 07 05 241 197]{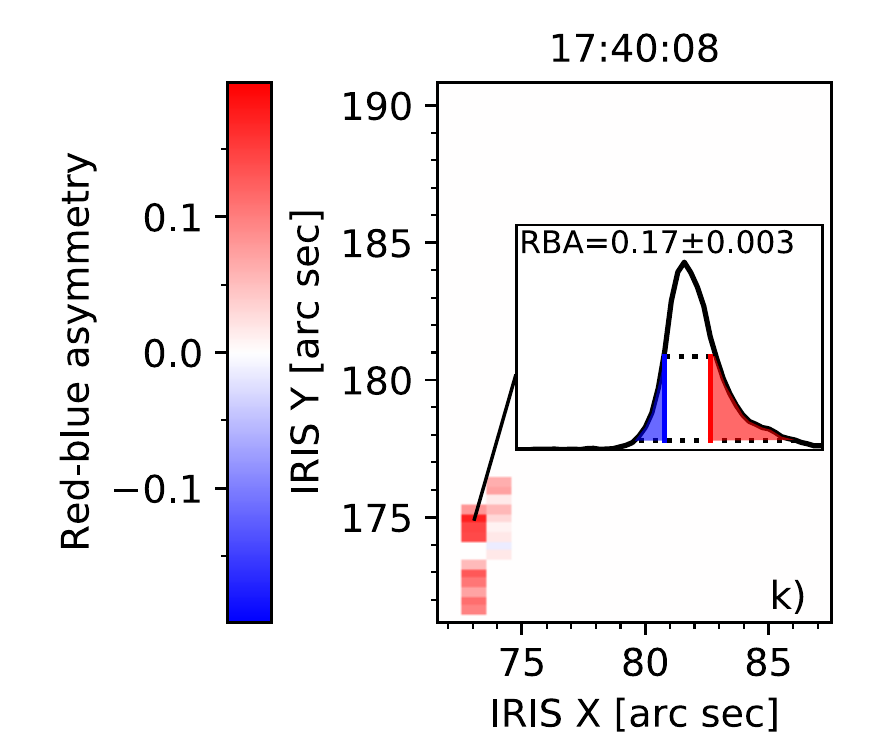}
\includegraphics[width=2.90cm, clip,   viewport= 120 05 241 197]{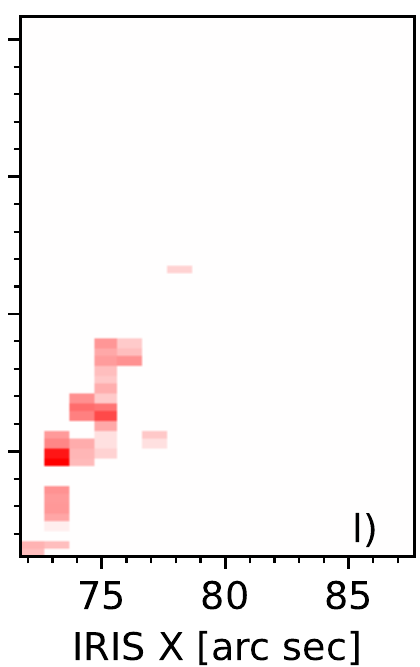}
\includegraphics[width=2.90cm, clip,   viewport= 120 05 241 197]{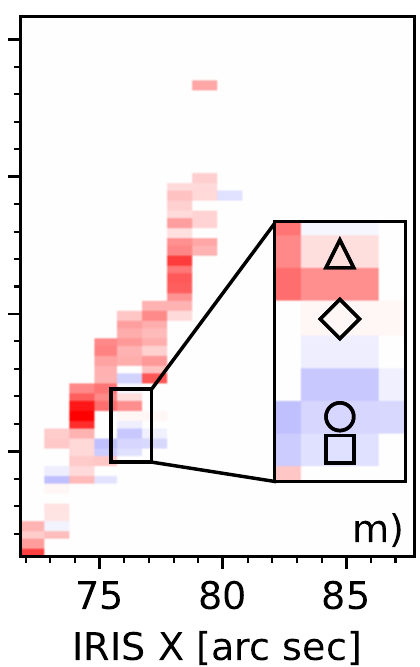}
\includegraphics[width=2.90cm, clip,   viewport= 120 05 241 197]{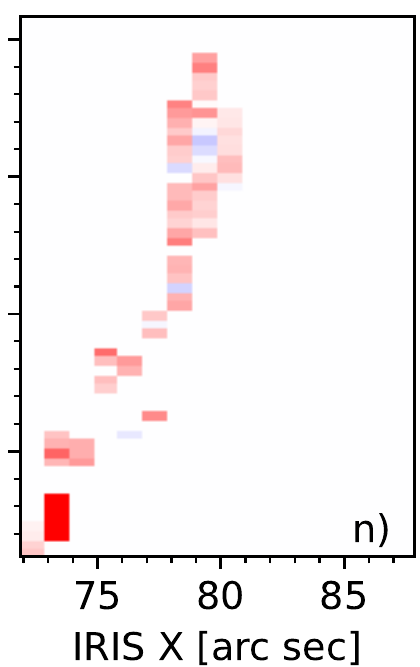}
\includegraphics[width=2.90cm, clip,   viewport= 120 05 241 197]{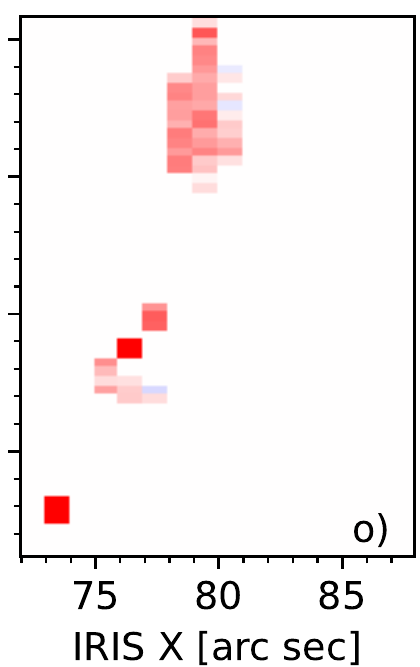}
\caption{Integrated intensity (panels (a)--(e)), Doppler shift of the line centroid (panels (f)--(j)), and red-blue asymmetry maps (panels (k)--(o)) of the Si IV 1402.77\,\AA~line in a flare kernel {moving} along the ribbon. The times listed above the figure indicate the start times of observations of the rasters plotted in each column. The arrow in panel (a) indicates the direction of the solar north. In panel (c), slit positions are indicated at the top of the panel. Inclusions in panels (h), (i), and (m) provide a zoomed view of a small region where the line is blueshifted. The square, circle, diamond, and triangle symbols mark pixels whose spectra are analyzed below. The inclusion in panel (k) shows the line profile in the selected pixel as well as the red and blue areas corresponding to the wings of the line used in the calculation of $RBA$.  \label{fig_spec_ribbon}}
\end{figure*}

The first row of Figure \ref{fig_spec_ribbon} contains maps of the line intensity, which were obtained by integrating the individual intensity bins in the spectral window containing the \ion{Si}{4} 1402.77\,\AA~line in each pixel. This is simpler than performing a multi-Gaussian fitting, and both methods can yield similar results \citep{delzanna19}. The integration was performed over a wavelength interval of $\lambda_0 \pm 1.5$\,\AA~after the subtraction of the FUV continuum averaged outside of this interval. In panel (c) of this figure, positions 0 -- 15 of the slit within the raster are indicated. 

Figure \ref{fig_spec_ribbon} shows that the kernel first appeared in the bottom-left of panel (a). Just as in the imaging observations, the kernel started to move towards the north and north-east directions (panels (b)--(d)) until it stopped (panel (e)). Due to the short 1\,s exposure times, the change of the direction of the kernel's {motion} likely affected its overall appearance in the raster images produced from rasters systematically observed in one direction (towards north-west). In panels (a) and (b) and slit positions 0 -- 5, the kernel moved towards the north, i.e., across different slit positions. Later, in panels (c) -- (e) and slit positions 7 -- 9, the motion occurred primarily towards the north-east, at which the central part of the kernel moved, for the most part, along slit position 8. Panel (c) shows a raster image observed during the change of the direction of the kernel's {motion}. At this time, the kernel accelerated (Section \ref{sec_ker_detailed}), for which it appears to be `smeared' through several slit positions. While most of the brightenings are visible below $\approx$180\arcsec, a small bright part of the kernel is already visible further along the ribbon at \mbox{$\approx$[79\arcsec, 185\arcsec]}.  

\subsubsection{Maps of Doppler velocities}	                   		\label{sec_kernel_doppler}

The second row of Figure \ref{fig_spec_ribbon} presents maps of the Doppler velocities. The Doppler velocities were calculated from line centroids found via analysing the first moment of the \ion{Si}{4} line. The Doppler shifts were calculated using a reference (laboratory) wavelength of 1402.77\,\AA { in the wavelength range limited by the intensities $i_2$ defined in Section \ref{sec_kernel_rba_method}.}

Panels (f)--(j) show that the peaks of the \ion{Si}{4} 1402.77\,\AA~line were usually redshifted. After the kernel appeared in the raster (panel (f)) and {moved} towards the north (panel (g)), the Doppler velocities ($|v_{\text{D}}|$) corresponding to the redshifted peaks were usually below $\approx$20\,km\,s$^{-1}$. In panels (h) and (i), showing the kernel as it {moved} further towards the north-east direction, pixels with relatively higher $|v_{\text{D}}|$ of up to 70 \,km\,s$^{-1}$ can be found. In the northernmost position of the kernel shown in panel (j), the average $|v_{\text{D}}|$ of the line dropped again below \mbox{$\approx$20 km\,s$^{-1}$}.

Interestingly, in a few pixels concentrated in a small region visible in panel (h) at slit position 4 between pixels $y$ = 142 and 148 along the slit, the \ion{Si}{4} 1402.77\,\AA~line exhibited blueshifts. The highest Doppler velocities within this region are found in pixel $y$ = 142 and reach up to $|v_{\text{D}}| \approx $55\,km\,s$^{-1}$. A zoomed view of these pixels is provided in the inserts in panels (h) and (i). These images show that the blueshifts were short-lived, because in the following IRIS raster observed after 17:41:49 UT, the spectra became redshifted again. Note that the transition between the blueshifts to the redshifts is visible in three pixels only, because after the kernel left this slit position, the intensity of the line dropped below the plotting threshold of 300\,DN.

Spectra observed in the pixels marked using the triangle ($y$ = 148), diamond ($y$ = 146), circle ($y$ = 143), and square ($y$ = 142) symbols were selected for a further analysis of the temporal evolution of profiles of selected lines (Section \ref{sec_kernel_temporal}), as well as multi-Gaussian fitting for the \ion{Si}{4} line. As we discuss in Appendix \ref{sec_appendixA}, the Doppler velocities resulting from the Gaussian fitting of the blue wing of the line range roughly between $|v_{\text{D}}| \approx 40$\,km\,s$^{-1}$ and $|v_{\text{D}}| \approx 100$\,km\,s$^{-1}$, with $|v_{\text{D}}| \approx$80\,km\,s$^{-1}$ in the pixel marked using the square symbol. Such large blueshifts are well above those previously reported in the literature (Section \ref{sec_discussion_li19}).

In this work we also investigate the effects of different methods commonly used to localise the line centroid for measurements of the Doppler velocities. These methods include $a)$ the moment analysis, $b)$ single-Gaussian fitting, and $c)$ automatic detection of the wavelength bin corresponding to the peak of the line (pixel with the highest intensity). The largest difference between the line centroids found using these three methods was roughly 0.07\,\AA~in the investigated pixels, translating to a velocity uncertainty of $ \sigma_{v_{\text{D}}}$ $\approx \pm 7.5$\,km\,s$^{-1}$. For the complete analysis we refer the reader to Appendix \ref{sec_appendixB}.
\subsubsection{Red-blue asymmetries: method}       \label{sec_kernel_rba_method}
\begin{figure*}[t]   
\centering
\includegraphics[width=8.50cm, clip,   viewport=20 25 495 315]{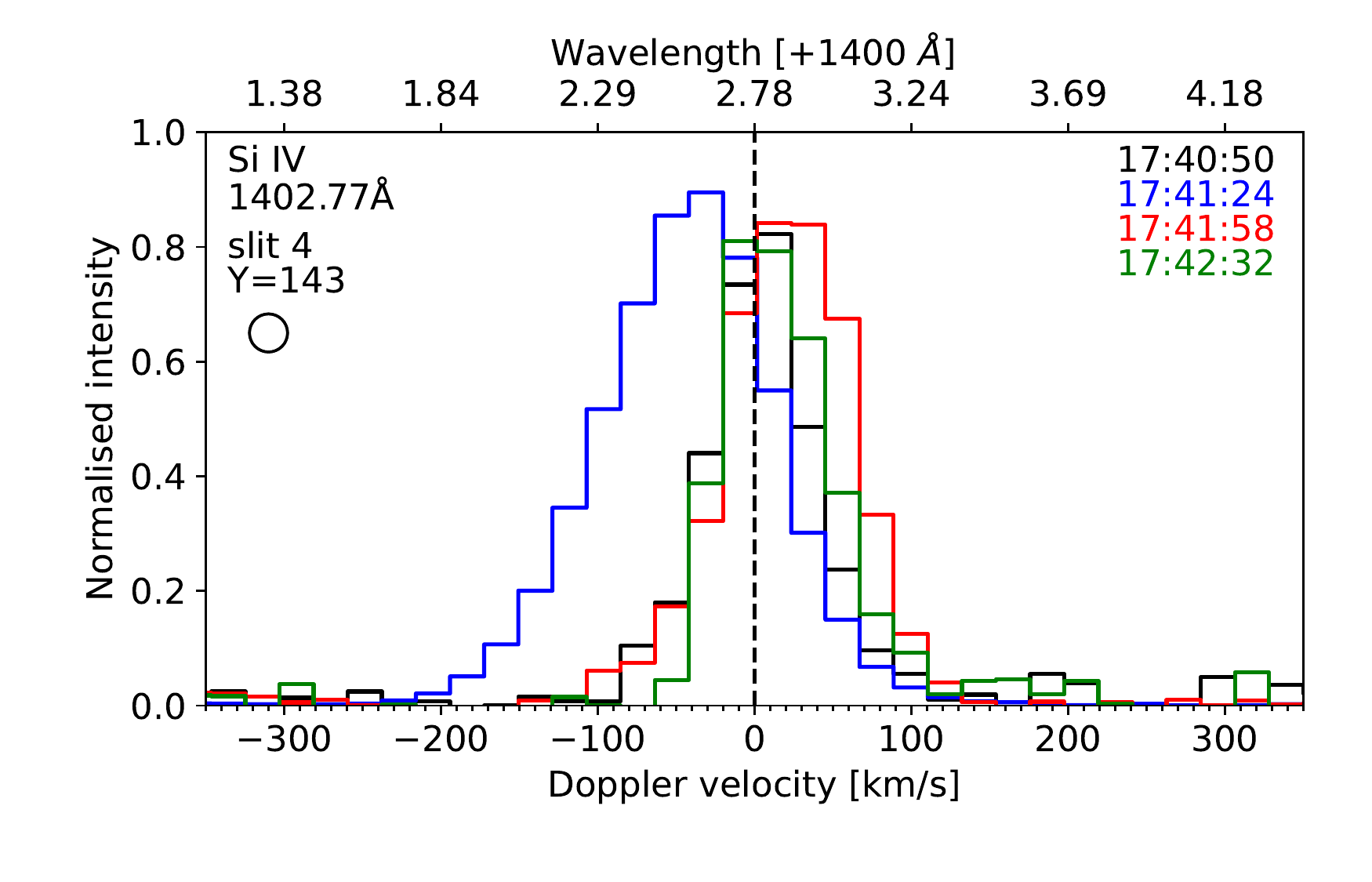} 
\includegraphics[width=8.50cm, clip,   viewport=20 25 495 315]{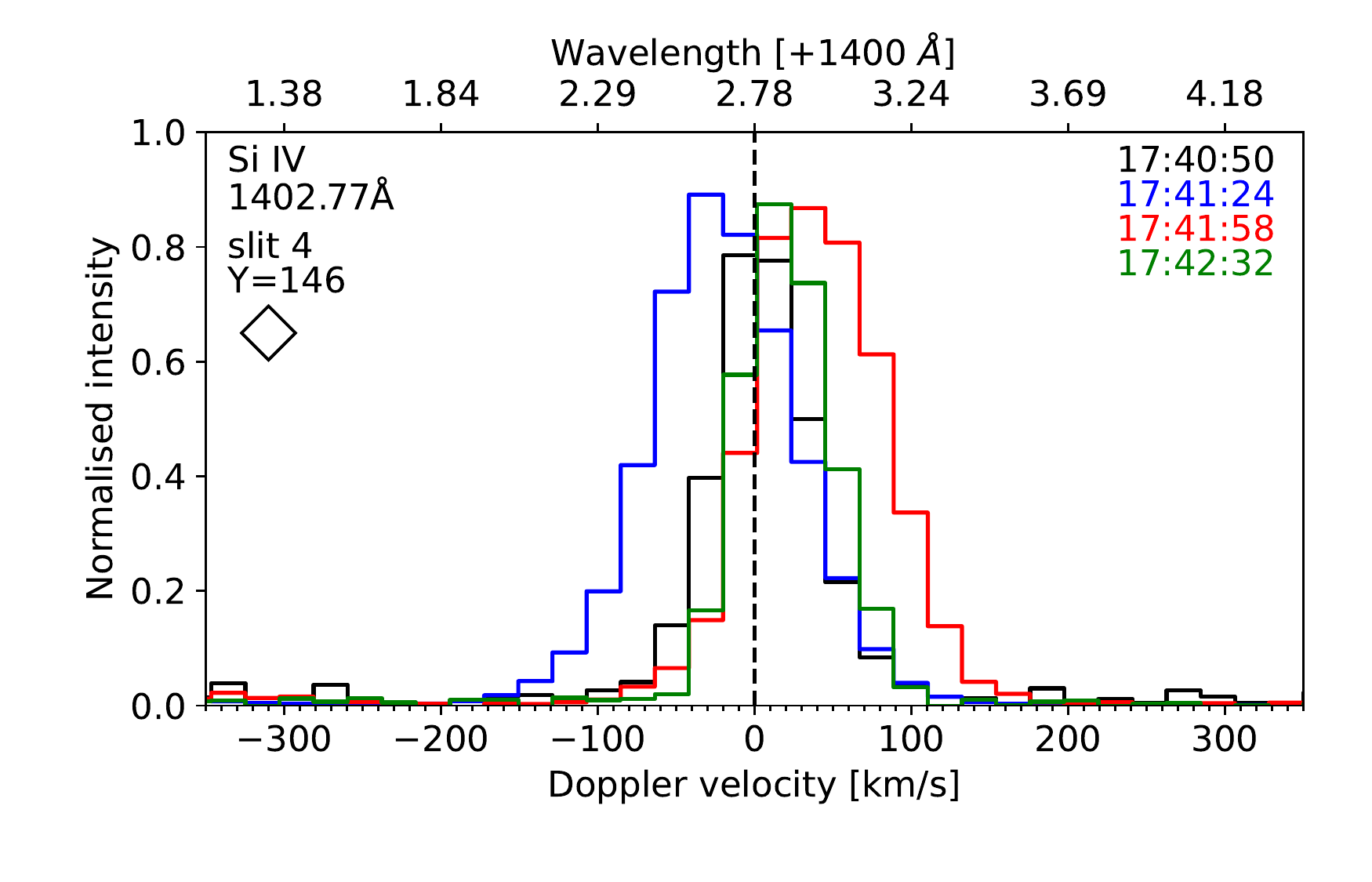} 
\\
\includegraphics[width=8.50cm, clip,   viewport=20 00 495 315]{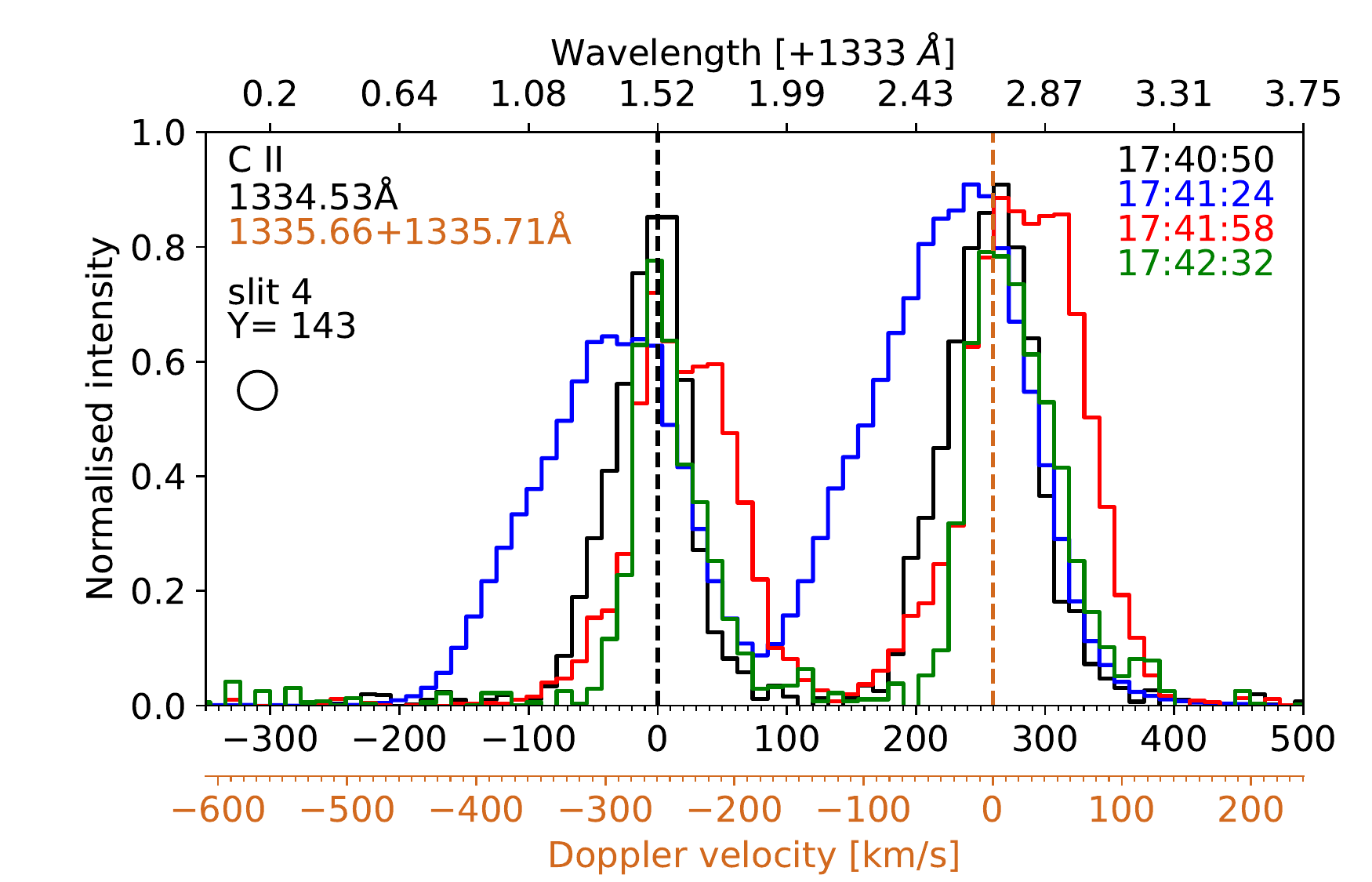}
\includegraphics[width=8.50cm, clip,   viewport=20 00 495 315]{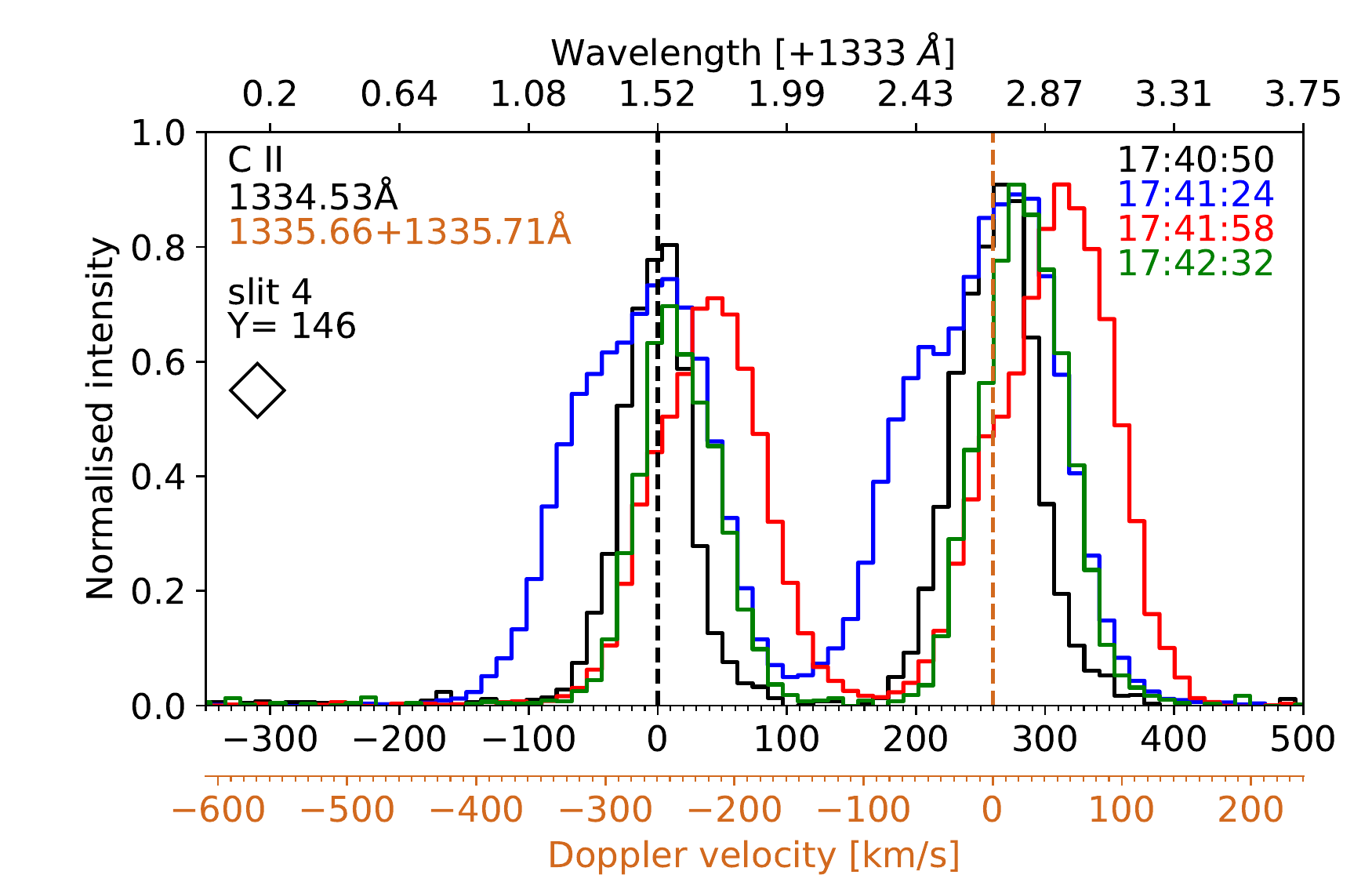}
\\
\includegraphics[width=8.50cm, clip,   viewport=20 25 495 315]{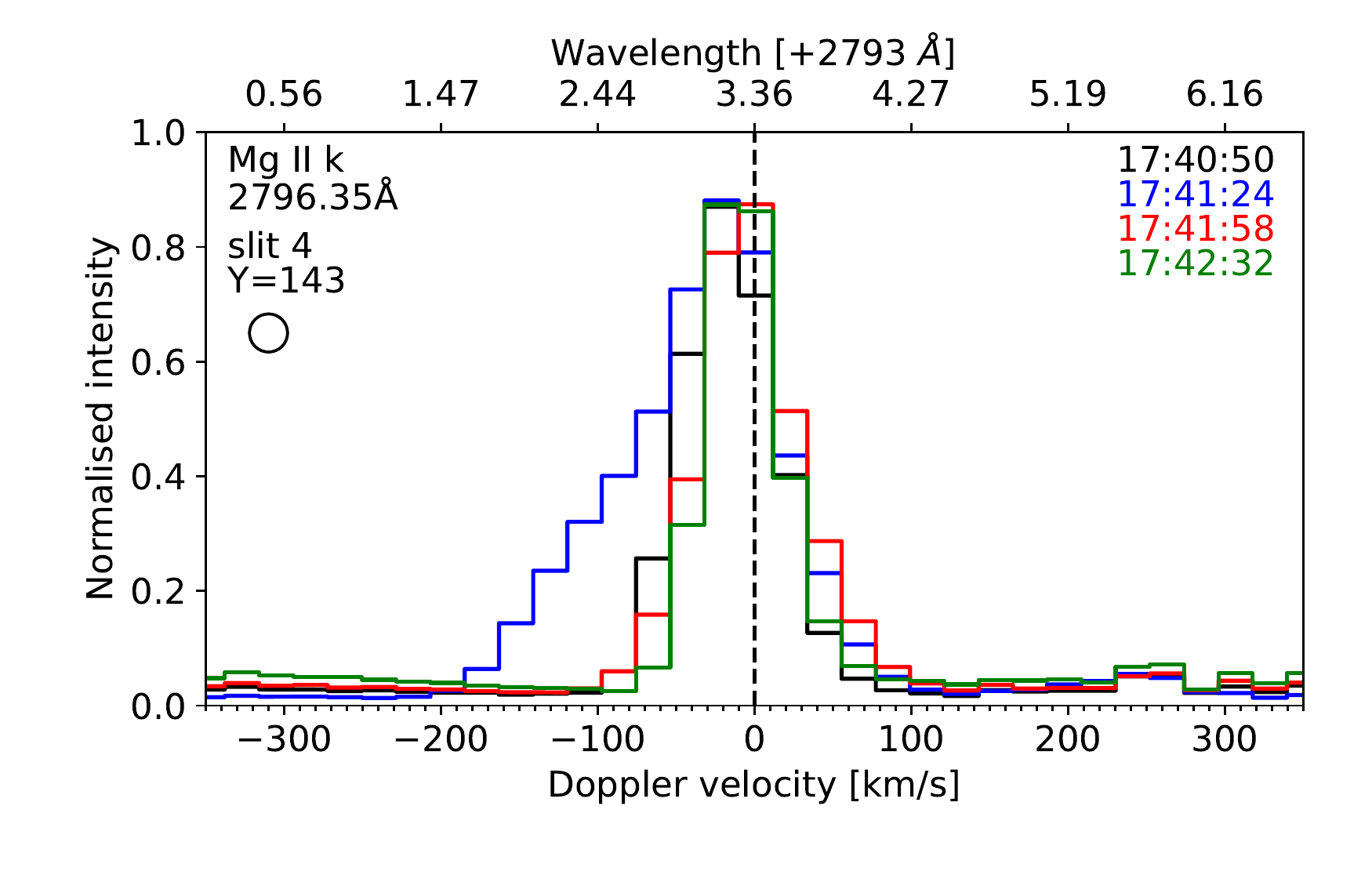}
\includegraphics[width=8.50cm, clip,   viewport=20 25 495 315]{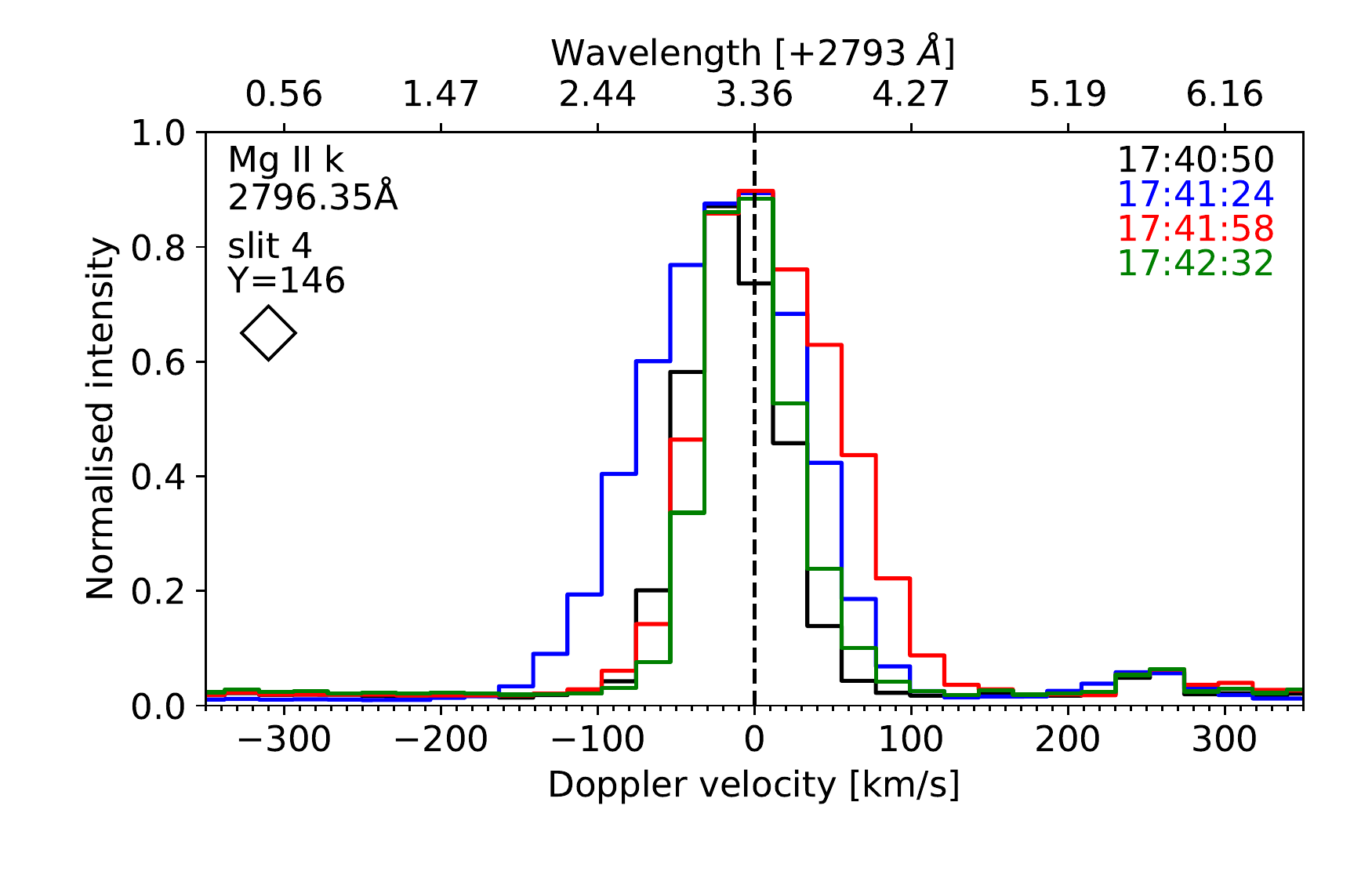}
\\
\includegraphics[width=8.50cm, clip,   viewport=20 25 495 315]{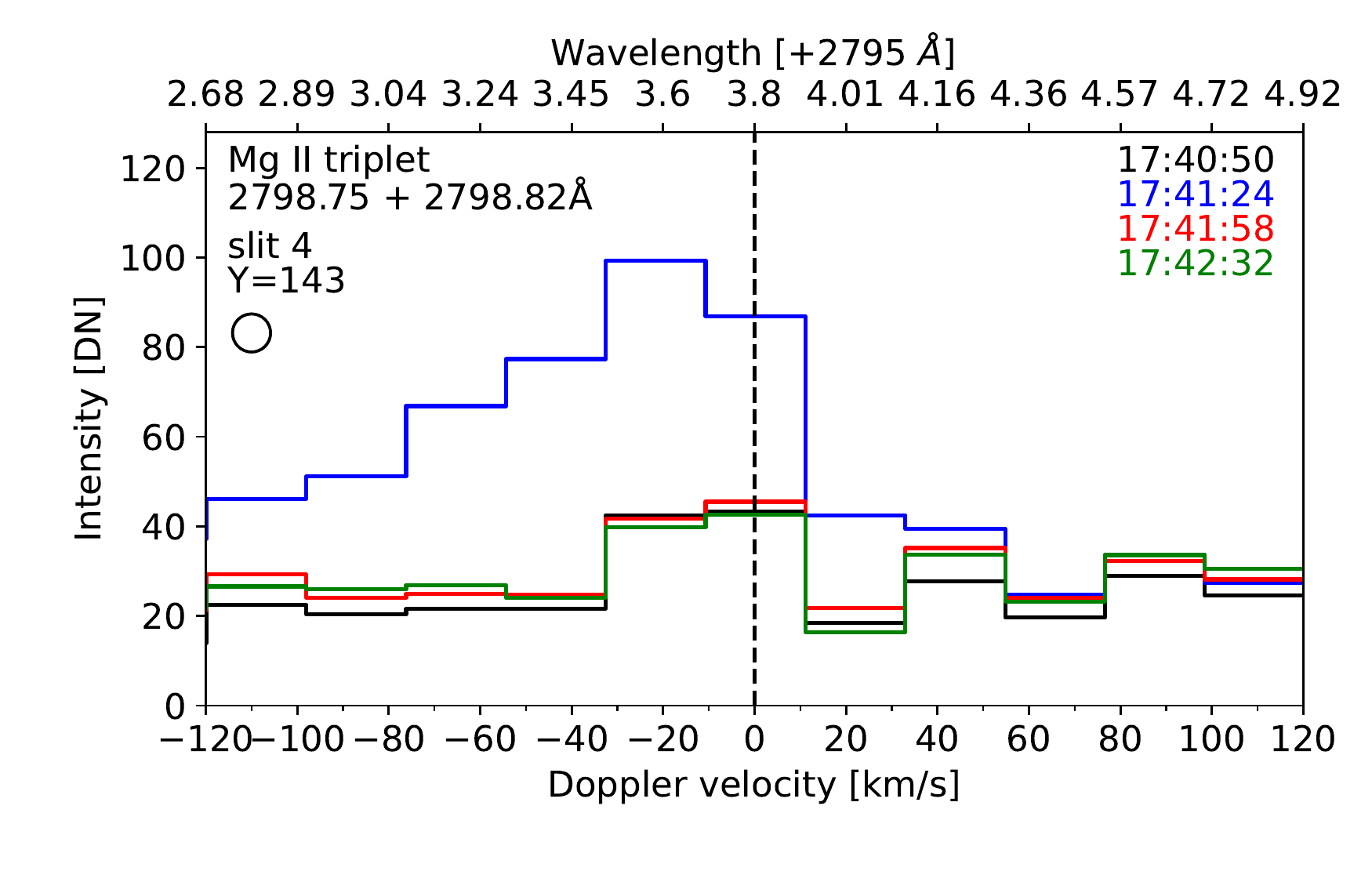}
\includegraphics[width=8.50cm, clip,   viewport=20 25 495 315]{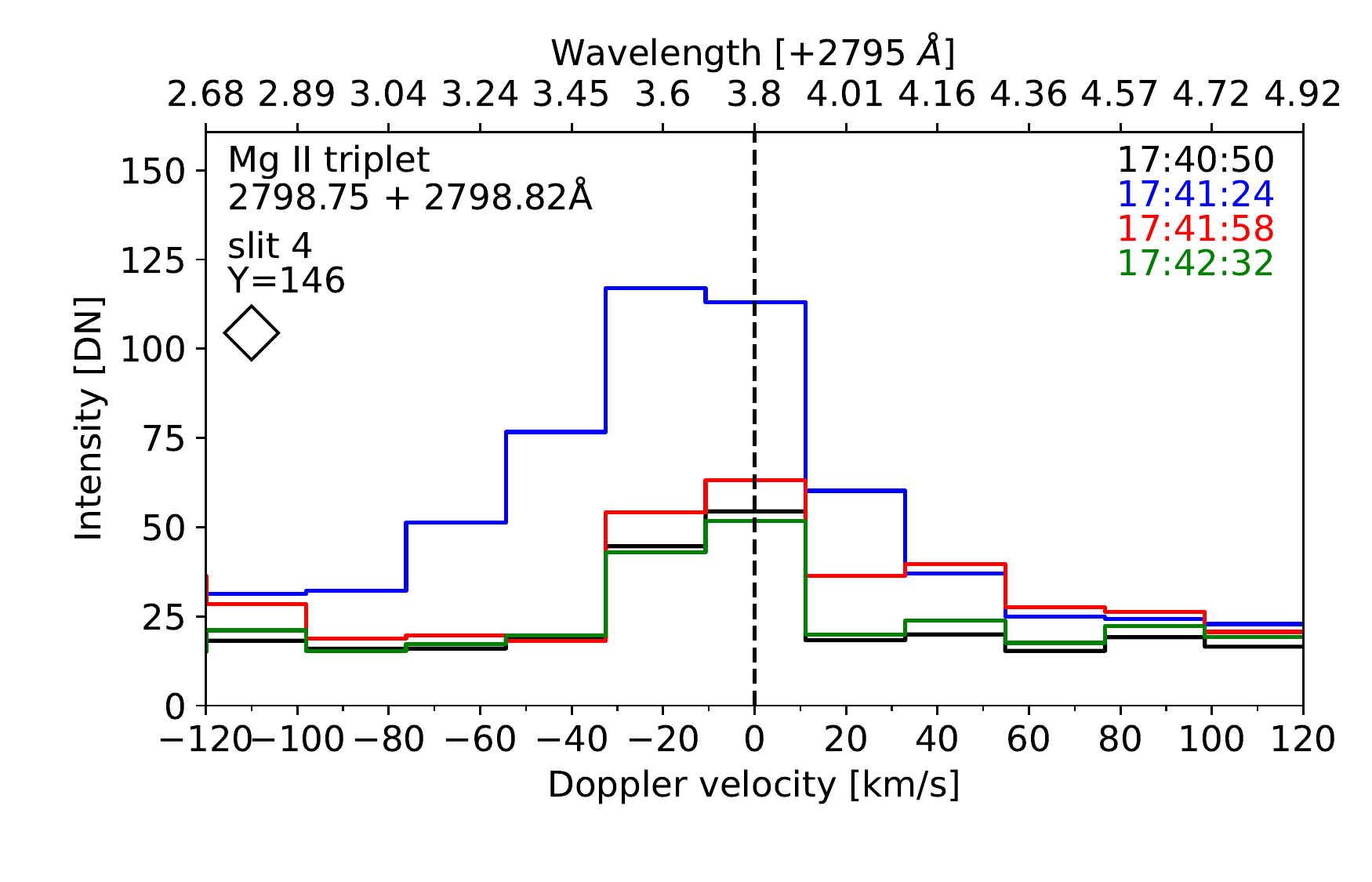}
\caption{Temporal evolution of the \ion{Si}{4} 1402.77\,\AA, \ion{C}{2} 1334.53\,\AA, 1355.66\,\AA, and 1335.71\,\AA, \ion{Mg}{2} k (2796.35\,\AA) line, as well as the 2798.75\,\AA~and 2798.82\,\AA~lines of the \ion{Mg}{2} triplet. The left column shows spectra that were measured in the slit position 4 and the pixel $y$ = 143 marked using the circle symbol. The spectra on the right were observed  along the same slit, in the pixel marked using the diamond symbol ($y$ = 146). The spectra were binned in order to increase the signal-to-noise ratio. The spectra plotted in the first three rows were also normalized to unity. \label{fig_spec_time}}
\end{figure*}

We next investigate the asymmetries of the wings of the \ion{Si}{4} 1402.77\,\AA~line forming as a consequence of unresolved up- or down-flowing components of this line along the line-of-sight. We are particularly interested in the development of the blue and red wing enhancements formed in the kernel as well as the portion of the ribbon along which it {moved}. To do so, we calculate the red-blue asymmetry (hereafter `RBA'), a measure of the relative strength of the line wings normalized to the line intensity \citep[e.g.,][]{depontieu09, tian11, polito19}. In its conventional definition, a negative $RBA$ indicates that the blue wing of the \ion{Si}{4} 1402.77\,\AA~line is stronger than the red wing and vice versa. We define the $RBA$ as
\begin{equation}
RBA=\frac{I_{\text{R}}-I_{\text{B}}}{I_{\text{L}}},
\end{equation}
where $I_{\text{R}}$ and $I_{\text{B}}$ are the intensities of the red and blue wings, respectively, and $I_{\text{L}}$ is the intensity of the line
\begin{equation}
I_{\text{R/B}}=\displaystyle\sum_{\pm i_1}^{\pm i_2} I_{\text{bin}, i},
\end{equation}
\begin{equation}
I_{\text{L}}=\displaystyle\sum_{-i_2}^{+i_2} I_{\text{bin}, i}.
\end{equation}
$I_{\text{bin}, i}$ is the intensity in the $i$-th wavelength bin of the spectral window and $\pm i_1$, $i_2$ are the limits of the summation. They are defined to correspond to 50\% and 5\% of the peak intensity, respectively. The centre of the interval of the summation corresponds to the observed peak wavelength. The $RBA$ are therefore not affected by the Doppler shift of the profile. The propagated uncertainties of the $RBA$ have the form
\begin{equation}
\sigma_{RBA} = RBA \cdot \sqrt{\left(\frac{\sigma_{(I_{\text{R}}-I_{\text{B}})}}{I_{\text{R}}-I_{\text{B}}}\right)^2+\left(\frac{\sigma_{I_{\text{L}}}}{I_{\text{L}}}\right)^2}.
\end{equation}
Our definition of the $RBA$ is similar to the one employed by \citet{polito19}. However, the intensity $I_L$ here stands for the total line intensity obtained by summing the profile between $\pm i_2$, instead of the peak intensity $I_p$ used in the formula of \citet{polito19}. Furthermore, instead of summing the intensities of the line's wings within fixed intervals of wavelengths (velocities), we opted to define the limits of the summation by using fractions of the maximal intensity of the line. This approach makes the $RBA$ independent of the \ion{Si}{4} line width which is important as the line's width varies both in time and space. For example, the top-left panel of Figure \ref{fig_spec_time} indicates that the extent of the blue wing of the line changes from $|v_{\text{D}}| = 100$\,km\,s$^{-1}$ to $|v_{\text{D}}| = 200$\,km\,s$^{-1}$ in the same pixel between two consecutive rasters. Furthermore, the full-width half maximum (FWHM) of the profiles shown in Figure \ref{fig_gaussian_fits} varies by more than 40\% across a spatial scale of $\approx2$\arcsec, with all pixels corresponding to the same flare kernel.

As we only focus on bright pixels in which the peak intensities exceed $300$\,DN, the lower-intensity threshold $\pm i_{\text{2}}$ of the profile is always $> 15$\,DN; i.e., higher than a typical value of the FUV continuum in the spectral window. Nevertheless, to account for the possible occurrence of spectra where the FUV continuum exceeded this value, the average FUV continuum was subtracted from spectra in each spatial pixel prior to the calculation of the $RBA$.  The averaging was performed over the wavelength bins located outside of the wavelength range corresponding to $\lambda_0 \pm 2\,\AA$. An example of the definition of the line wings limited by $i_{\text{1}}$ and $i_{\text{2}}$ in a selected pixel is presented in the inclusion plotted in Figure \ref{fig_spec_ribbon}(k). 
\subsubsection{Maps of red-blue asymmetries}
\label{sec_kernel_rba}
Maps of the red-blue asymmetries calculated using the method described in Section \ref{sec_kernel_rba_method} are plotted in the third row of Figure \ref{fig_spec_ribbon}. In these panels we only plot the value of $RBA$ if $RBA$ $> 3 \sigma_{RBA}$. Panels (k) and (l), respectively, containing the kernel after it appeared in the raster and {moved} towards the north, show that the $RBAs$ were usually positive. This means that the red wings of the line were typically stronger than its blue wings, indicating the presence of an additional downflowing component in profiles which are otherwise consistently redshifted (Figure \ref{fig_spec_ribbon}(f), (g)). An example of one of these pixels is plotted in the inclusion of panel (k). As the kernel {moved} further towards the north-east direction (panel (m)), several pixels with negative $RBAs$, indicative of upflowing components, appeared throughout the raster. The average absolute value of the negative red-blue asymmetries obtained in these pixels was $RBA = -2.7 \times 10^{-2} \pm  1 \times 10^{-3}$, which is lower by a factor $\approx 2.5$ than that obtained in the pixels showing positive RBAs. This indicates that even though the \ion{Si}{4} 1402.77\,\AA~line shown pronounced blue wings (upflows) in some pixels, the blue-wing enhancements were still smaller than their red counterparts (downflows) in the same kernel.

Later, when the kernel {moved} towards the north-east, the number of pixels with negative $RBA$ decreased (panel (n)) until they nearly disappeared (panel (o)). When the kernel {moved} to the final location, the spatial distribution of the $RBA$ became more homogeneous with most of the pixels exhibiting positive $RBA$ (c.f., panels (n) and (o)). The average value of the $RBA$ calculated in the brightest part of the kernel at the location of IRIS $Y > 182\arcsec$ increased by roughly 40\% compared to that shown in panel (n). This means that towards the end of the {kernel's motion}, either the blueshifted/rest components were diminishing, or the redshifted components (and hence the condensation itself) intensified.

Figure \ref{fig_spec_ribbon} also indicates that the number of pixels exhibiting blueshifted profiles (second row) and negative $RBA$ (third row) followed a similar trend. While no blue pixels indicating negative Doppler velocities (panels (f), (g)) and negative $RBA$ (panels (k), (l)) are visible in the first two rasters, they become visible in the subsequent raster plotted in panels (h) and (m). At a later stage of the kernel's {motion} along the ribbon, the number of spectra exhibiting blueshifted profiles and blue-wing asymmetries decreased (panels (i), (j) and (n), (o)). Furthermore, as indicated by the inclusions plotted in panels (h) and (m), the pixels marked using the square and circle symbols exhibited simultaneously blueshifts and the blue-wing enhancements, whereas the line in the triangle pixel was blueshifted with a red-wing enhancement. This likely means that the spectra of the square and circle pixels result from a superposition of a dominant component with a modest upflow and a faint component with a relatively stronger upflow. In the triangle pixel, the core of the line indicated upflows, but the wings of the line already were already showing signatures of downflowing plasma likely induced by the onset of the chromospheric condensation. This behavior evidences that the relative strength of the up- and down-flows varied even among the pixels selected from the same part of the kernel.

We note that peculiar profiles of the \ion{Si}{4} 1402.77\,\AA~line during flares might be also related to optical depth effects rather than a presence of multiple unresolved components. \citet{kerr19} show that both the intensity and the shape of the line vary in RADYN \citep[][]{carlsson92, carlsson95, carlsson97, allred15} flare models with energy fluxes exceeding $5 \times 10^{10}$ ergs/cm$^{2}$/s when the opacity is taken into the account. 

\subsection{Temporal evolution of transition region and chromospheric lines}	\label{sec_kernel_temporal}

As indicated by the maps of spectral characteristics of profiles formed in {the moving kernel} (Section \ref{sec_kernel_doppler} and Figure \ref{fig_spec_ribbon}(f) -- (j)), both blue-shifts and the red-blue asymmetries of the \ion{Si}{4} 1402.77\,\AA~line were relatively short-lived. The temporal evolution of spectra formed in {the kernel} is further {detailed} in Figure \ref{fig_spec_time}, which shows spectra of various lines as a function of the Doppler velocity. Different rows show the profiles of the \ion{Si}{4} 1402.77\,\AA~line, the 1334.53\,\AA, 1335.66\,\AA, 1335.71\,\AA~lines of the \ion{C}{2} ion, the \ion{Mg}{2} k 2796.35\,\AA~line, and finally the 2798.75 \,\AA~and 2798.82\,\AA~lines of the \ion{Mg}{2} triplet. 

The profiles were color-coded to distinguish between the spectra observed before (black), during (blue), and after (red and green) the analyzed kernel entered the slit position 4 where the blueshifts were observed. Spectra observed in two pixels along the slit where the lines were the strongest are shown in this figure. The left column corresponds to the pixel marked using the circle symbol in Figure \ref{fig_spec_ribbon}(h), (m)). The column on the right shows the spectra observed in the pixel marked using the diamond symbol. Note that both pixels are located in the same kernel. Above each panel plotted in Figure \ref{fig_spec_time}, wavelengths (in \AA) corresponding to the Doppler velocities on the horizontal axis below the figure are indicated. The reference wavelengths for the calculation of the Doppler velocities are plotted using the vertical dashed lines and correspond to the laboratory wavelengths of these lines. The panels containing the \ion{C}{2} lines (second row) have two horizontal axes, one with the Doppler velocities for the 1334.53\,\AA~line (black) and the other for the 1335.66\,\AA~and 1335.71\,\AA~lines (brown). For the latter two lines, 1335.71\,\AA~was selected as the reference wavelength for the Doppler velocities, as this line is stronger than its blending line at 1335.66\,\AA~\citep{rathore15a}. For the 2798.75 \,\AA~and 2798.82\,\AA~lines of the \ion{Mg}{2} triplet, the reference wavelength was set to 2798.8\,\AA~\citep{pereira15}. The profiles plotted in the first three rows of Figure \ref{fig_spec_time} were normalized to the peak intensity of each line during the passage of the kernel and binned in wavelength to increase the signal-to-noise ratio. The \ion{Si}{4} and \ion{C}{2} spectral windows, corresponding to the FUV passband of the instrument, were binned by 4 wavelength bins. Binning of 2 was applied to the \ion{Mg}{2} spectral window of the NUV passband. These different bin sizes were used to achieve similar spectral resolutions across the passbands. Since the \ion{Mg}{2} triplet lines (bottom row) were difficult to distinguish from the neighboring NUV continuum, they are shown without the normalisation. 

Initially, as can be seen from the black plots, the lines were either weakly redshifted (the \ion{Si}{4} line as well as the \ion{C}{2} lines), or their peaks corresponded to the rest wavelengths (the \ion{Mg}{2} lines). No enhancements in the line's wings were clear at this stage. 

When the kernel crossed the slit (blue plots), the intensity of these lines increased by factors $\approx$2.5 -- 10, depending on the line. The \ion{Si}{4} line became blueshifted, with a relatively larger blueshift of $|v_{\text{D}}|
= 24$\,km\,s$^{-1}$ found in the circle pixel (left). Both lines of \ion{C}{2}, plotted in the second row of Figure \ref{fig_spec_time}, show pronounced blue wings. Furthermore, in the circle pixel, the peaks of both lines exhibit weak blueshifts with the Doppler velocities estimated as $|v_{\text{D}}|$ = 10 -- 20\,km\,s$^{-1}$. Obtaining the exact Doppler velocity corresponding to the observed peak wavelengths was difficult as the profiles have complicated shapes possibly affected by the opacity effects and the thermal structure of the lower solar atmosphere \citep[see e.g.,][]{rathore15a}. In the diamond pixel (right), the peaks of these lines were redshifted, with $|v_{\text{D}}|$ estimated to range between $\approx$16 and 30\,km\,s$^{-1}$. Unlike those of the \ion{C}{2} lines, the peak of the \ion{Mg}{2} k line was usually found to be at the rest wavelength (third row). Doppler shifts, if any, were within the velocity uncertainty of our measurements ($ \sigma_{v_{\text{D}}} = \pm 7.5$\,km\,s$^{-1}$, see Appendix \ref{sec_appendixB}). This line however still formed pronounced blue wings, stronger in the circle pixel. The enhancement of the blue wing of the two lines of the \ion{Mg}{2} triplet was more clear in the diamond pixel. This was most likely caused by the low counts in this line, leading to the discontinuous shape of the blue wing in the circle pixel. Interestingly, the extent of the blue wings of the \ion{Si}{4}, \ion{C}{2}, and the \ion{Mg}{2} k lines was very similar and depended on the pixel position. While the wings of these lines {extended up to} $|v_{\text{D}}| \approx 200$\,km\,s$^{-1}$ in the circle pixel, the wings were only extending up to $|v_{\text{D}}| \approx 150$\,km\,s$^{-1}$ in the diamond pixel. 

\begin{figure*}[t]
\centering
\includegraphics[width=3.80cm, clip,   viewport= 03 0 155 195]{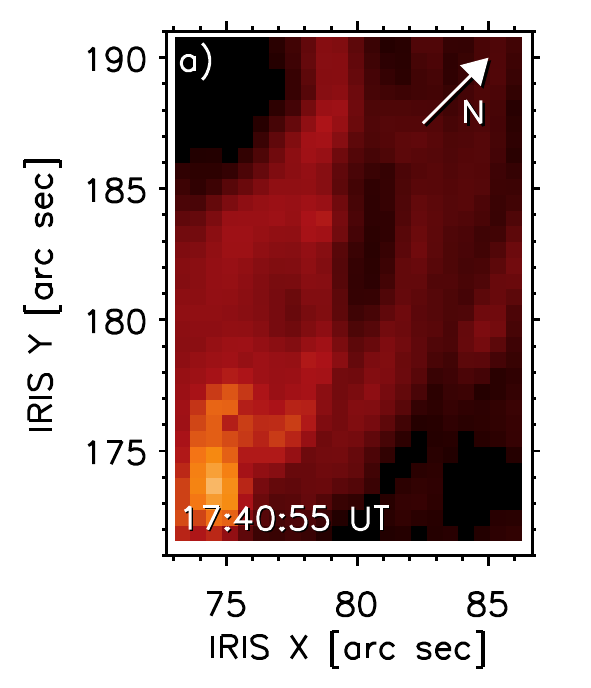}
\includegraphics[width=2.75cm, clip,   viewport= 45 0 155 195]{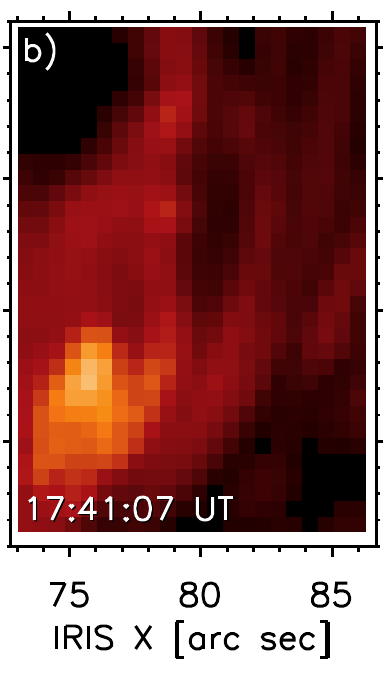}
\includegraphics[width=2.75cm, clip,   viewport= 45 0 155 195]{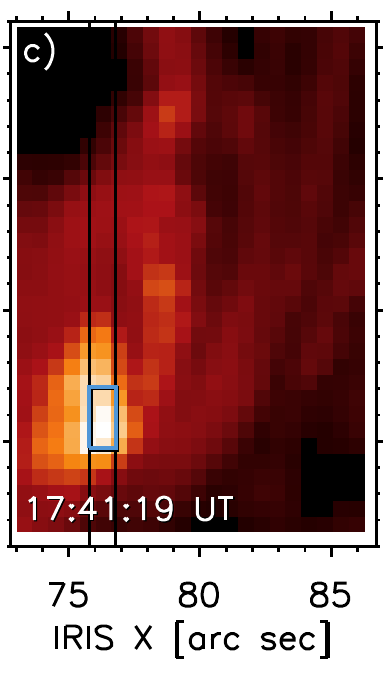}
\includegraphics[width=2.75cm, clip,   viewport= 45 0 155 195]{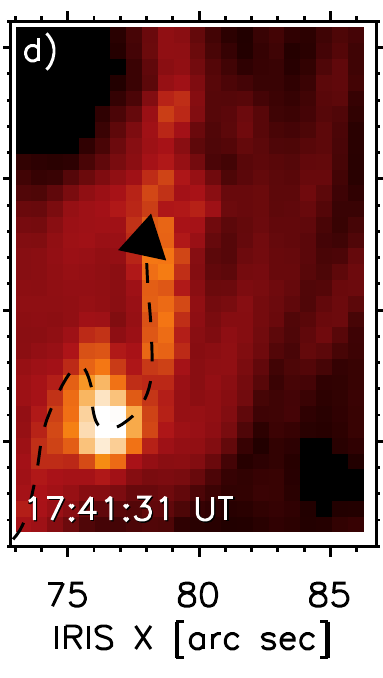}
\includegraphics[width=2.75cm, clip,   viewport= 45 0 155 195]{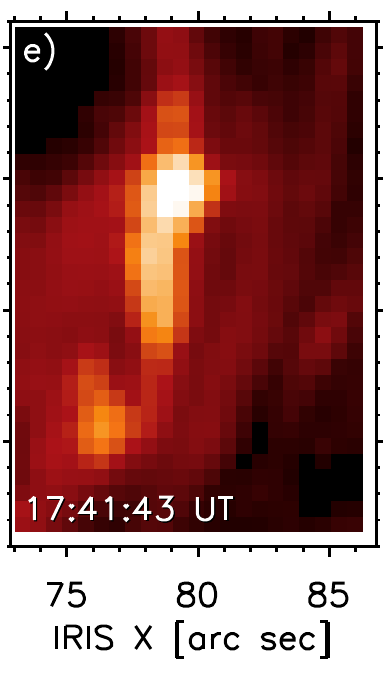}
\includegraphics[width=2.75cm, clip,   viewport= 45 0 155 195]{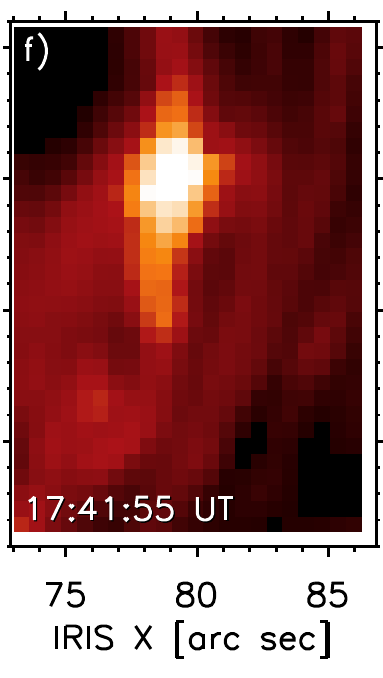}
\\
\includegraphics[width=5cm, clip,   viewport= 05 0 170 198]{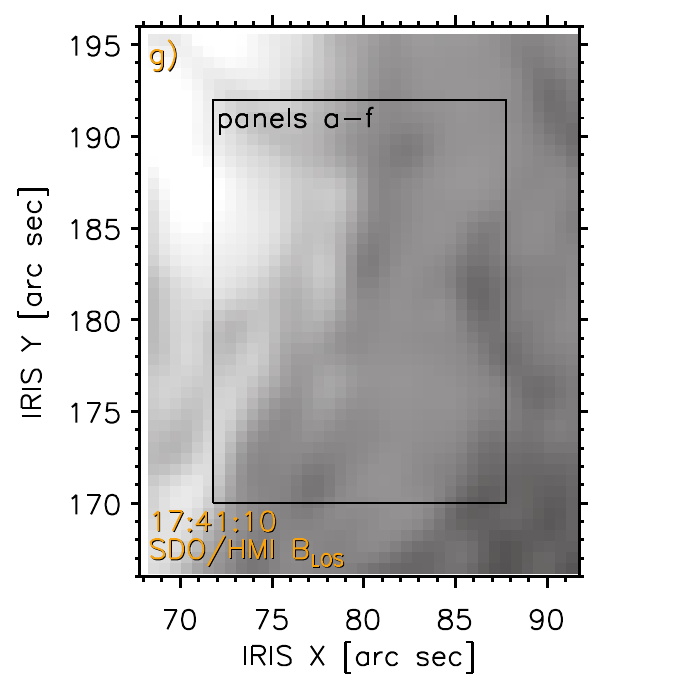}
\includegraphics[width=5cm, clip,   viewport= 05 0 170 198]{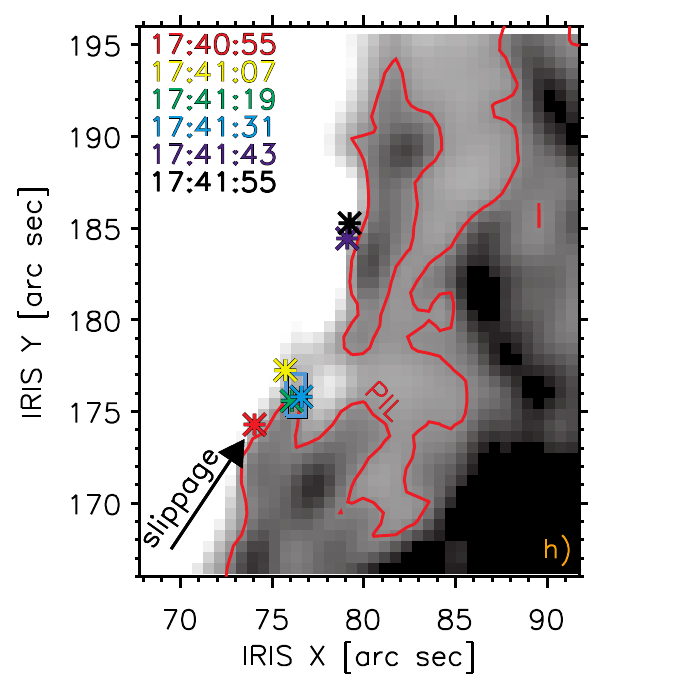}
\caption{The kernel as observed in AIA 304\,\AA. The position of the IRIS slit is indicated using the black lines in panel (c). The blue frame plotted in the same panel marks the location where IRIS observed the blue-shifted spectra. The path of the kernel is indicated in panel (d) using the black dashed arrow. Panels (g) and (h) show the measurements of the $B_{\text{LOS}}$ saturated to $\pm$1500\,G and $\pm$400\,G, respectively. In panel (h), the red contours corresponding to $B_{\text{LOS}} = 0$\,G, the region where the blueshifted profiles were observed, and the direction of the {motion} of the kernel are marked. The colored symbols plotted in this panel mark the positions of the kernel as observed in the 304\,\AA~channel of AIA. Note the direction of the solar north shown by the white arrow on panel (a). \label{fig_ker_detailed}}
\end{figure*}

The profiles observed after the passage of the kernel through the slit are plotted using the red plots. At both positions, the \ion{Si}{4} 1402.77\,\AA~line exhibited clear redshifts corresponding to a few tens km\,s\,$^{-1}$. These are more pronounced in the spectra observed in the diamond pixel plotted in the right column. A similar behavior was exhibited by the lines \ion{C}{2}, again best visible in the diamond pixel. In the same location, while the \ion{Mg}{2} k line exhibited pronounced red wings, the profiles themselves were however not redshifted. Indications of pronounced red wings are also visible in the lines of the \ion{Mg}{2} triplet in the diamond pixel, however, they are hard to distinguish from the NUV continuum due to the low counts in these lines. Finally, two rasters (about 1 minute later; green plots) after the kernel crossed this slit position, the lines nearly entirely returned to the state they were in before the kernel crossed the slit. An exception to this behavior, though weak, was observed in the diamond pixel for the \ion{Si}{4} 1402.77\,\AA~line, which was still redshifted compared to the state before the passage of the kernel (compare the black and the green plots). 

To sum-up the temporal evolution of spectra in the analyzed locations, when the kernel entered the slit, the transition region \ion{Si}{4} 1402.77\,\AA~line exhibited blueshifts of roughly $|v_{\text{D}}| \approx $24\,km\,s$^{-1}$. The chromospheric \ion{C}{2} lines showed weak blueshifts up to $|v_{\text{D}}| = $16\,km\,s$^{-1}$ and blue-wing enhancements, also observed in \ion{Mg}{2} spectra. These signatures of upflows were observed in one raster only, and in the following raster the lines became either redshifted or exhibited red wing enhancements. Roughly one minute later, the profiles started to exhibit the same characteristics they had shown before the passage of the kernel. 

Because of the weak blueshifts of their centroids, the signatures of the upflows visible in the \ion{C}{2} lines were similar to those exhibited by the \ion{Si}{4} line. This can likely be related to the extent of the formation region of the \ion{C}{2} lines, spanning between the upper chromosphere and the lower transition region \citep{rathore15a, rathore15b, bjorgen18}. Note that to our knowledge, this extent of the formation heights of this line has so far been investigated in the quiet Sun and active regions only.

Unlike the \ion{Si}{4} line, the \ion{Mg}{2} triplet and k lines, formed over a range of chromospheric heights \citep[see e.g.,][]{leenaarts13a, leenaarts13b, pereira13, pereira15}, exhibited pronounced blue wings only. This result indicates that the upflows were also present in the chromosphere, which confirms the results of our analysis of the \ion{C}{2} lines. Interpreting the blue asymmetries of the \ion{Mg}{2} lines can however be complicated as they are formed under optically thick conditions \citep[e.g.,][]{kerr15, kuridze15, tei18} and their formation regions vary rapidly during flares \citep{kerr19}.

\subsection{High-cadence observations of the region with blueshifted spectra}
\label{sec_ker_detailed}

The maps of the spatial distribution of the Doppler velocities (Section \ref{sec_kernel_doppler}) indicate that the blueshifts in \ion{Si}{4} 1402.77\,\AA~were located in one small region. Despite their high spatial resolution, the 1400\,\AA~SJI observations of the ribbon (see Figure \ref{fig_ribbon_sji} and its animated version) do not reveal any peculiar behavior of the kernel in the region where the blueshifts occurred. This might be caused either by their cadence (17\,s), or by the timing of their observations, possibly missing the motion of the kernel through the region we investigate. We have therefore inspected {the motion} of the kernel in AIA 304\,\AA~observations carried out at a slightly higher cadence of 12\,s.

The AIA 304\,\AA~images of {the kernel} are presented in Figure \ref{fig_ker_detailed}(a)--(f). The individual panels detail the {motion} of the kernel along the associated portion of the ribbon. The kernel appeared in the bottom-left of panel (a) at which it started {to move} towards the north (panel (b)). Subsequently, instead of continuing further towards the north-east as indicated by the SJI images (Figure \ref{fig_ribbon_sji}), the kernel {moved} slightly to the south-west (panel (c)) to $\approx$[77\arcsec, 176\arcsec]. As is clear from panel (d), the kernel remained almost in the same location 12 seconds later in the consecutive exposure. Later, the kernel was observed roughly 10$\arcsec$ towards the north-east (panel (e)). As seen in panel (f), from this location the kernel {moved} only a short distance, after which it reached its final location. The trajectory of {the kernel}, highlighting the `detour' towards the south-west from its original direction of motion, is indicated using the dashed black arrow in panel (d). 

Based on a manual selection of the pixels corresponding to the brightest part of the kernel, we roughly estimate its velocity $v_{\text{ker}}$ to be of the order of a few tens km\,s$^{-1}$ between the panels (a) and (c). However, the trajectory the kernel {passed} between the panels (d) and (e) is larger, close to 10\arcsec. Given the 12\,s cadence of these observations, the kernel thus had to accelerate to $v_{\text{ker}} \approx 600$\,km\,s$^{-1}$ between these two segments. {We remind the reader than since the kernel corresponds to the footpoint of slipping flare loop (Section \ref{sec_sliprecco}), its motion can be interpreted as the signature of the slipping reconnection. It thus ought to be noted that} the slipping velocity $v_{\text{slip}}$ of this order of magnitude is among the {highest in the} literature \citep[see, e.g.,][]{li18, joshi19, lorincik19a}. Finally, between its positions visible in panels (e) and (f), the kernel {moved} slowly at a velocity even lower than that observed before its acceleration. For a further discussion on the velocity of {the kernel and the associated apparently-slipping flare loops} we refer the reader to Section \ref{sec_discussion_slip}.

The location of the slit position 4, rastered after 17:41:24 UT, is indicated using the black lines in Figure \ref{fig_ker_detailed}(c). In the same panel, the pixels $y$ = 142--148 along the slit, characterized by the presence of the blueshifted spectra, are indicated using the cyan frame. From a comparison between the location of the slit and the direction of the {motion} of the kernel (panels (c) and (d)) we found that the region where the blueshifts occurred corresponds to the `front' of the kernel during its {motion} along the ribbon. Furthermore, by the time the instrument started to observe the subsequent raster at 17:41:49 UT (Figure \ref{fig_spec_ribbon}(d)), the kernel had already {shifted} towards the north-east (Figure \ref{fig_ker_detailed}(e)). This means that some of the slit positions in the raster containing the blueshifted profiles (third column of Figure \ref{fig_spec_ribbon}) were observed during the kernel's acceleration (also see Section \ref{sec_kernel_raster}).

In 3D magnetic reconnection theory, the magnitude of the slipping velocity $v_{\text{slip}}$ and its variations are dictated by the norm $N$ of the magnetic connectivity \citep[][]{janvier13, dudik14}. Magnetic connectivity changes dramatically within the quasi-separatrix layers (QSLs) associated with flares. As pointed out by \citet{lorincik19a}, $v_{\text{slip}}$ can be affected by small-scale variations in the local distribution of the magnetic field.

The maps of $B_{\text{LOS}}$ measured by HMI at a time corresponding to the observations of {the moving kernel}, saturated to $B_{\text{LOS}} = \pm 1500$\,G, are presented in Figure \ref{fig_ker_detailed}(g). These maps were also smoothed using a boxcar of 3 $\times$ 3 pixels to suppress the noise. To the south-east, a large region dominated by the positive-polarity flux concentrations is seen. There, the positive-polarity ribbon and its hook were observed during the impulsive and peak phases of the flare (Section \ref{sec_ribbons}). To the north-west, a region dominated by relatively weaker flux concentrations is located. In order to highlight the variations of the magnetic field there, the map of $B_{\text{LOS}}$ showing the same FOV as panel (g) was saturated to $B_{\text{LOS}} = \pm 400$\,G. The resulting image plotted in panel (h) reveals several blobs of parasitic negative-polarity flux concentrations on the right-hand side of the FOV. To differentiate between the positive- and the negative-polarity flux, the red contours corresponding to $B_{\text{LOS}} = 0$\,G, indicating the locations of the polarity inversion line (PIL), were plotted in the same panel. Locations of {the moving} kernel adopted using the AIA 304\,\AA~observations were also overlaid on this data using asterisks. These asterisks were color-coded to distinguish between different locations of the kernel, from its appearance in the 304\,\AA~images (red) to the location where its motion stopped (black). The black arrow pointing towards the red asterisk indicates the direction of the {kernel's motion} at earlier times as indicated by the SJI observations (Figure \ref{fig_ribbon_sji}(a), (b)). Finally, the blue frame plotted in Figure \ref{fig_ker_detailed}(g) marks the region where IRIS observed the blueshifted profiles. 

{The motion} of the kernel can be summarized as follows: (1) it appeared at $\approx$[70\arcsec, 167\arcsec]; then (2) it started its relatively slow apparent slipping motion along the edge of a large region corresponding to the positive polarity (black arrow), at which it entered the IRIS raster (red and yellow asterisks). (3) When the kernel {moved} slightly towards the south-west direction (green and blue asterisks) it entered a small weak-field region separating the large region of the positive-polarity flux on the left-hand side of the FOV and a small blob of a positive-polarity flux located roughly at [78\arcsec, 176\arcsec]. (4) In the consecutive exposure (violet asterisk), the kernel is again found at the edge of the strong-field region to the north-east. (5) From there, the subsequent motion of the kernel to its final location was slow, indicated by the proximity of the violet and black asterisks. 

This result shows that while the kernel was moving relatively slowly in strong-field regions, it accelerated after approaching a region with relatively lower field. The velocity of {the motion of the kernel} is thus governed by the {underlying} magnetic field, which is consistent with the analysis of kernel dynamics of \citet{lorincik19a}.

The region where the blueshifts and/or the blue-wing enhancements of the analyzed lines briefly appeared corresponds to the green and blue asterisks. Since these mark the weak-field region in which the kernel accelerated, the unique characteristics of spectra we report on in this manuscript were likely affected by the properties of the motion governed by the local magnetic field. On the other hand, spectra forming in the kernel during its relatively slower motion through high-$B$ regions were usually redshifted, as is common for transition region and chromospheric lines observed during large flares.

\section{Discussion}
\label{sec_discussion}

We now discuss our observations and their interpretation in detail. Specifically, our observations are discussed in Section \ref{sec_discussion_slip}. In Section \ref{sec_discussion_li19} we discuss the existing literature on blueshifts of the \ion{Si}{4} line, and their relative paucity. Finally, Section \ref{sec_discussion_models} discusses our observations from the viewpoint of numerical models.

\subsection{Slipping reconnection and characteristics of spectra} \label{sec_discussion_slip}

According to the analysis of {the velocity of its apparent slipping motion (Section \ref{sec_ker_detailed}),} the kernel was initially moving through a strong-field region at speeds $v_\mathrm{ker}$ of few tens km\,s$^{-1}$. Later, as it entered the weak-field region, {its velocity} increased by an order of magnitude. Then, the kernel reached another strong-field region where it slowed down and eventually stopped.

The velocities of the kernel can be related to the velocities of the apparently-slipping flare loop rooted in this kernel, whose signatures were studied using the time-distance diagram in Figure \ref{fig_xt_131} (see also Section \ref{sec_sliprecco}). The counterpart of {the moving kernel} during the slower stage of its motion was the flare loop whose motion was fitted using the pink dash line indicating the velocity $v_{\text{slip}}$ = 33\,km\,s$^{-1}$. The second pink dashed line in Figure \ref{fig_xt_131} fits the motion of a slipping loop located at the coordinate $\approx$30\arcsec of the cut and results in the speed $v_{\text{slip}}$\,=\,55\,km\,s$^{-1}$. This motion corresponds to the flare loop when the kernel was {moving} along a very short trajectory close to its final location, marked using the pink arrow in Figure \ref{fig_ribbon_sliploops}(i). The discontinuity between the two pink fitting lines, marked using the yellow arrow in Figure \ref{fig_xt_131}, is most likely related to the brief acceleration of the kernel reported in Section \ref{sec_ker_detailed}. This small region is roughly located at IRIS $Y$ = 180\arcsec~(Figure \ref{fig_ribbon_sliploops}(f)) through which the kernel {moved} rapidly after its acceleration (Figure \ref{fig_ker_detailed}(e)). 

A possible explanation for the discontinuity in the {apparent slipping motion of the} flare loops, as well as their velocities before and after the acceleration of the kernel, concerns the amounts of energy deposited by the slipping reconnection. If the apparent slipping velocity of the kernel is too fast, there are no flare loops observed \citep[see also][]{lorincik19a}. This is likely due to the decrease of the amount of energy deposited per unit area per time along the slipping magnetic field lines associated with the increase in the apparent slipping velocity. Therefore, fast slippage translates into low energy deposition into the chromosphere, which may be insufficient to drive chromospheric evaporation of a dense enough plasma for the slipping flare loop to become observable against the background corona. The increase of the footpoint motion velocity could also be interpreted in terms of an increase of the reconnection rate conventionally defined as a product of the velocity of field line footpoints away from the PIL and the normal component of the photospheric magnetic field \citep{forbes2000}. The apparent slipping motion of flare loops and kernels however occurs inherently along the PIL, which does not conform to the classical definition of the reconnection rate \citep[see Section 1. of][]{lorincik19a}.

It ought to be noted that the cadence of the current instrumentation might not be sufficiently high for precise addressing of manifestations of the slipping reconnection \citep[see e.g., Section 4.2. in][]{dudik16}. The seemingly continuous motion of the kernel in between its initial and terminal locations, indicated using the red and black asterisks in Figure \ref{fig_ker_detailed}, could in principle result from motions of two separate kernels, one moving towards the weak-field region and the other away from this region.
\subsection{Comparison with previously-reported observations of the \ion{Si}{4} line blueshifts}    
\label{sec_discussion_li19}

As mentioned in Section \ref{sec_introduction}, during large solar flares, lines formed in the solar transition region such as the \ion{Si}{4} 1402.77\,\AA~line we analyze here are usually redshifted presumably as a consequence of the downflow-inducing chromospheric condensation. Observations of \ion{Si}{4} blueshifts are very rare, examples of studies reporting on them include e.g., \citet{jeffrey18} and \citet{li19} for the same B1.6-class flare, or \citet{li15} for an X1.0-class flare.

The blueshifts reported by \citet{li15} occurred in one location and were rather weak, roughly of $|v_{\text{D}}| = 10$km\,s$^{-1}$, which is by a factor of 5 lower than what we found in one of the pixels at the front of the kernel. The blueshifts found by \citet{jeffrey18} and \citet{li19} were slightly stronger, reaching $|v_{\text{D}}| \approx 20$km\,s$^{-1}$, which is however still lower than the velocities we report on here. On the other hand, some signatures of the upflows observed during the B-class flare were similar to those we report on. For example, according to the bottom-right panel of Figure 4 in \citet{li19}, the blueshifts of the \ion{Si}{4} line were cotemporal with those of the \ion{Mg}{2} k line. This corresponds to our observations, as the chromospheric lines were exhibiting weak blueshifts and/or pronounced blue wings at the same time and in the same location as the \ion{Si}{4} line (Section \ref{sec_kernel_temporal}). Further, the observations of \citet{li19} indicate that the \ion{Si}{4} blueshifts were short-lived, persisting for less than $\approx$20\,s. This timescale is within our upper estimate on the lifetimes of the blueshifts (34\,s) adopted from the raster cadence.

The blueshifts of \citet{li15} and \citet{li19} were interpreted as signatures of the so-called `gentle evaporation' \citep[see e.g.,][and Section \ref{sec_discussion_models}]{fischer85b, fischer85c}. We suggest that this interpretation is not applicable to our observations because the upflow velocities we report on here are larger than those observed by \citet{li15} and \citet{li19} or predicted in the past. Furthermore, observations of upflows interpreted in terms of the gentle evaporation in the era of Coronal Diagnostic Spectrometer (CDS) indicated that blueshifts of chromospheric and transition region lines can persist over periods spanning a few minutes, in some cases eventually reappearing later on \citep{brosius04, brosius09}. Blueshift timescales of this order of magnitude do not correspond to the \ion{Si}{4} blueshifts observed by IRIS. It also ought to be noted that \citet{jeffrey18} suggested that the \ion{Si}{4} blueshifts observed during the same event as analyzed by \citet{li19} were not caused by the evaporation because they had occurred before the intensity of the line peaked.

We finally note that due to the short lifetimes of the blueshifts ($< 34$\,s), IRIS would have possibly missed their development had its slit been pointing to a different position within the raster. We suggest that the rarity of the observations of the blueshifted \ion{Si}{4} line profiles is most likely related to the short lifetime of the upflows at the corresponding temperatures. Their detection is likely contingent on the mutual position of the slit of IRIS and the portion of the ribbon where the blueshifts develop at the very instant of the exposure. Likely, short-lived features such as the blueshifted profiles of cool lines might disappear in datasets observed at longer exposure times. By analogy, similar conclusions can be postulated for the occurrence of blueshifts in different observational modes of IRIS. As we have shown in our data observed in a 16$\arcsec$-wide raster, the blueshifts might develop in a region only a few arc-sec large while the spectra in the rest of the raster might still exhibit redshifts.

\subsection{Numerical models used to interpret blueshifts in chromospheric and transition region lines}

\label{sec_discussion_models}

The behavior of Doppler shifts in transition-region and chromospheric lines observed by IRIS has also been studied from the viewpoint of numerical modeling of the flaring atmosphere. It has been shown that the magnitude and behavior of flows at different temperatures crucially depends on the heights at which the accelerated electrons deposit their energy. These in turn depend on the parameters of the electron distributions (in particular the low-energy cut-off and flux) as well as the initial conditions of the atmosphere \citep{reep15, polito18, testa20}.

The blueshifts in \ion{Si}{4}  have been successfully reproduced in the numerical simulations of coronal nanoflare heating of \citet{testa14} performed using the RADYN code. The authors found that their observations can be reproduced when the model atmosphere is heated by nonthermal electrons, depositing their energy in the lower atmosphere. According to Figure 4 in \citet{testa14}, such blueshifts are short-lived, persisting for few tens of seconds. On the other hand, blueshifts of this line were not reproduced in models including atmospheres heated via thermal conduction. RADYN models with a larger parameter space were analyzed in \citet{polito18}, who found that nonthermal electrons whose high-energy cutoff exceeds 10 keV {with a total energy larger than} $10^{24}$\,erg can reproduce blueshifts in \ion{Si}{4} reaching up to 30\,km\,s$^{-1}$. In addition, Figure 15 of \citet{polito18} shows that the threshold in the low-energy cut-off between up- and down-flows for \ion{Si}{4} depends on the total energy (or energy flux) of the simulation, and increases for larger energies. The agreement between our observations of the blueshifted \ion{Si}{4} line and the prediction of RADYN models of \citet{testa14,polito18} with relatively weak energy fluxes ($\approx~10^{9}$ ergs/cm$^{2}$/s) might suggest that we are observing a site of more modest energy deposition as compared to what is usually observed in flare ribbons. As a comparison, note that larger M--X class flares are usually modelled using fluxes of about $\approx~10^{11}$--$10^{12}$ ergs/cm$^{2}$/s, \citet[e.g.,][]{polito16a}. On the other hand, our observed chromospheric spectra, and in particular the \ion{Mg}{2} blue wing enhancements, do not seem to match those in RADYN models of smaller heating events. 

As mentioned in Section \ref{sec_introduction}, several studies focused on blueshifts of the chromospheric \ion{Mg}{2} line can be found in the literature. \citet{tei18} reported on blue-wing enhancements of the \ion{Mg}{2} h (2803.52\,\AA) line, roughly corresponding to 10\,km\,s$^{-1}$, observed at the leading edge of a flare kernel in a C-class flare. The blueshifts {existed} for periods spanning between 9 to 48\,s and were followed by redshifts with corresponding velocities reaching 50\,km\,s$^{-1}$. These blueshifts were interpreted by a cloud model, in which accelerated electrons deposited their energy into the chromosphere. The target region then expanded and triggered upflows of the overlying cool chromospheric material. In addition, persisting blueshifts, however not followed by the development of redshifts, were observed during the decaying phase of the kernel. According to the authors, this occurs when the kinetic energy of the {precipitating} particles is not large enough to produce a downflowing component in the emitting plasma. Note that similar interpretations for blue wing enhancements of the \ion{Mg}{2} lines observed by IRIS were obtained in the numerical works of \citet{zhu19} and \citet{hong20}. 

Doppler shifts of the \ion{Mg}{2} k line during the peak phase of the same flare as the one analyzed in this manuscript were also studied by \citet{huang19}. They found strong blue wing enhancements at the leading edge of the propagating ribbon close to the location where we observed the blueshifts in \ion{Si}{4}, \ion{C}{2}, and \ion{Mg}{2}, however, during the peak of the flare roughly 15 minutes after the period we focused on. The emission of the \ion{Mg}{2} k line was subsequently reproduced in a RADYN model with a total flux of 10$^{12}$ ergs/cm$^{2}$/s, selected after analysis of RHESSI HXR spectra. The authors suggest that the blueshifts might be caused by the decrease of temperature and increase of the electron density induced by the precipitating particles. 

As results of the studies discussed in this section indicate, blueshifts of lines formed in the transition region or the chromosphere can be reproduced in certain models. Whether blueshifts of the chromospheric and transition region lines formed at different heights can be reproduced simultaneously, and under what conditions, remains to be resolved in the future. 

At the moment, no models of the emission of flare ribbons and flare loops forming as a consequence of the slipping reconnection exist. To some extent, the heating by accelerated particles precipitating into the lower atmosphere along the apparently-slipping field lines can be approximated using a series of 1D models, each depicting the evolution of plasma in a single pixel along heated ribbons. The observed manifestations of the field line slippage could then be used to constraint the duration and the periods of the heating. We plan to carry out such analysis in the future. 

\section{Summary} \label{sec_summary}

In this work we studied in detail spectra of flare ribbons which formed during the 2015 June 22 M6.5-class flare observed at a high resolution by IRIS and SDO/AIA. We focused on the characteristics of transition region and chromospheric lines, in particular the \ion{Si}{4} 1402.77\,\AA~line in a flare kernel which was observed to {move} along flare ribbon as a consequence of the magnetic slipping reconnection. Our main results can be summarized in the following points:

\begin{enumerate}

\item{The flare led to the formation of a pair of $J$-shaped (hooked) flare ribbons. IRIS observed most of the hooked ribbon to the north-east, spatially coincident with the positive-polarity flux. This ribbon hook was highly structured and exhibited dynamical evolution, as the hook completely relocated within a short period of $\approx$15 minutes during the impulsive and peak phases of the flare. In accordance with the predictions imposed by the standard flare model in 3D, this phenomenon was accompanied by drifting of the hot channel's footpoints along the drifting hook observed in the 131\,\AA~channel of AIA.}

\item{The elongation of the ribbon was accompanied by the apparent slipping motion of flare loops, best visible in AIA 131\,\AA. The episode of the slippage we focused on occurred along the straight part of the ribbon forming after the onset of the flare. The velocities of the slippage measured using a time-distance diagram ranged between 21 and 55 km\,s$^{-1}$ and correspond to those typically found in the literature. {In the SJI 1400 images and AIA 304\,\AA~channel, flare kernels moving along the same ribbon were observed}. {The motion of some of these kernels was coherent, seemingly forming one larger kernel in AIA observations.} The motion of this kernel corresponded to the motion of the footpoint of one of the slipping flare loops observed in AIA 131\,\AA.}

\item{The \ion{Si}{4} 1402.77\,\AA~line formed in this kernel was usually redshifted with corresponding Doppler velocities $|v_{\text{D}}| < 70$\,km\,s$^{-1}$. In a small region corresponding to 7 pixels along one slit position, the profile broadened to $\approx$ 1.5\,\AA~and its peak was blueshifted with corresponding Doppler velocities $|v_{\text{D}}|$ of roughly $\approx 50$\,km\,s$^{-1}$. The blueshifts were short-lived as in the following raster, acquired 34\,s later, they were no longer observed.}

\item{In {this kernel}, the same line usually exhibited pronounced red wings, indicative of a presence of additional downflowing components existing irrespective of the Doppler shifts of the line. In the raster in which the blueshifts were observed, a few pixels exhibiting pronounced blue wings, indicating additional upflowing components, appeared. The number of the pixels exhibiting the upflowing components decreased towards the end of the kernel's {motion}.}

\item{In the region characterized by the presence of the blueshifted \ion{Si}{4} 1402.77\,\AA~line profiles, the \ion{C}{2} and \ion{Mg}{2} lines exhibited weak blueshifts up to $|v_{\text{D}}| < 20$\,km\,s$^{-1}$ and blue wing enhancements, respectively. These characteristics were again observed in one raster only. In the subsequent raster, the lines were observed to exhibit redshifts and/or pronounced red wings.}

\item{Using high (12\,s) cadence observations from the AIA 304\,\AA~channel we found that the {motion} of the kernel was governed by the variations of $B_{\text{LOS}}$ measured by HMI, in the same manner as reported by \citet{lorincik19a}. Upon entering a small weak-field region, the kernel accelerated and {the speed of its apparent slipping motion} increased by an order of magnitude. The small region where IRIS observed the blueshifted lines corresponded to the front of the kernel during its acceleration. }

\item{Our observations, combined with the results reported by \citet{lorincik19a}, evidence that the spatial variations in the distribution of the local magnetic field affect not only the properties of the slipping motion, but also the characteristics of spectra observed in flare ribbons.}
\end{enumerate}

Observations of blueshifted lines of ions such as \ion{Si}{4}, \ion{C}{2}, and \ion{Mg}{2} are significant, as they evidence a rather exceptional response of the lower solar atmosphere to the flare heating.

From the observational point of view, the question how common the blueshifts of cool lines really are remains to be answered. While several pixels exhibit the presence of blueshifted spectra, the spectra in a vast majority of the pixels we investigated here are still redshifted. The detection of the blueshifts is complicated by their short lifetimes, which in our observations as well as those reported on in the literature are of the order of a few tens of seconds. Indeed, as pointed out in Section \ref{sec_discussion}, the analyzed raster would have probably missed the short-lived blueshifted spectra if the slit of IRIS had not been imaging the front of the kernel where signatures of the upflows in cool lines were detected. Based on our results we suggest that observing the blueshifted spectra of cool lines in sit-and-stare observations limited to one slit position \citep[as in][]{jeffrey18, li19} likely seems to be a matter of coincidence. This highlights the importance of observations of rasters at a cadence as high as possible, or even measuring the spectra from different slits simultaneously, an option offered by the forthcoming Multi-slit Solar Explorer \citep[MUSE;][]{depontieu20, cheung21} mission.

In the future, we plan to extend our analysis to entire datasets, containing flare kernels moving inherently along flare ribbons as a consequence of the magnetic slipping reconnection. Such analysis would greatly benefit from the involvement of machine learning methods for clustering of profiles of different lines, searching for common characteristics of their spectra, and relating the occurrence of blueshifted profiles of cool lines to the properties of the apparent slipping motion of both flare kernels and the hot flare loops rooted in them. 
\vspace{1 cm}
\\
J.D. acknowledges the project 20-07908S of the Grant Agency of Czech Republic as well as insitutional support RVO: 67985815 from the Czech Academy of Sciences. J.L. was supported by NASA under contract NNG09FA40C ({\it IRIS}). V.P. acknowledges support from NASA grants 80NSSC20K0716 and NNG09FA40C ({\it IRIS}). IRIS is a NASA small explorer mission developed and operated by LMSAL with mission operations executed at NASA Ames Research center and major contributions to downlink communications funded by ESA and the Norwegian Space Centre. AIA and data are provided courtesy of NASA/SDO and the AIA science team. 

\appendix 

\section{Gaussian fitting of the \ion{Si}{4} 1402.77\,\AA~line}
\label{sec_appendixA}

The \ion{Si}{4} 1402.77\,\AA~line often shows asymmetric profiles related to various physical mechanisms occurring in the emitting plasma. In flare ribbons, the profiles usually exhibit red-wing enhancements caused by the downflows of plasma along the line of sight \citep[e.g.,][]{tian15}. With the exception of a few pixels where the line exhibited blue-wing enhancements, this was also the case of the profiles we observed in the {moving} flare kernel (Section \ref{sec_kernel_rba}).

Approximations of asymmetric profiles can be obtained by fitting with multiple Gaussian components. We fitted the observed spectra using the \texttt{xcfit.pro} routine included in the SolarSoft package within IDL. Note that in active regions, the \ion{Si}{4} 1402.77\,\AA~profiles exhibit strong wings but are symmetric \citep{dudik17_nonmaxw}, i.e., their kurtosis is larger than 3 (corresponding to the Gaussian) and their skewness (or the red-blue asymmetry; $RBA$) is close to zero, and they are best approximated by the $\kappa$ distribution. Such fits were not applicable in our analysis, as profiles were usually far from symmetric. Therefore, we considered only multiple Gaussian components. To evaluate the goodness of the fits, the reduced $\chi^2_\mathrm{r}$ statistic was evaluated in all of the fitted pixels.

We performed fitting of the profiles which exhibited partial blueshifts and the presence of pronounced blue wings (Sections \ref{sec_kernel_doppler} and \ref{sec_kernel_temporal}). These profiles were found in a small region of 7 pixels along one slit position, highlighted in the zoomed inclusions in Figure \ref{fig_spec_ribbon}(h), (m). We started our analysis by fitting the \ion{Si}{4} 1402.77\,\AA~profiles using single-Gaussian fits whose parameters (amplitudes, wavelengths, and widths) were not constrained. Single-Gaussian fits were found to be very poor approximations of the spectra, with $\chi^2_\mathrm{r}$ up to 11 in some cases. The fits were significantly improved by adding the second Gaussian, resulting in $\chi^2_\mathrm{r} = 1.5 - 5$ depending on the pixel. Obtaining fits with $\chi^2_\mathrm{r} \rightarrow 1$ in all pixels in this region was only possible after adding an additional, third Gaussian to the fit of the \ion{Si}{4} line. Furthermore, two additional Gaussians were necessary to fit the relatively weak \ion{O}{4} 1401.16\,\AA~as well as the \ion{O}{4} 1404.81\,\AA~and \ion{S}{4} 1404.81\,\AA~lines found in the spectral window of the \ion{Si}{4} line. The strength of these lines to the spectra varied among the fitted pixels. They were observed to be the weakest in the pixel $y$ = 142 marked using the square symbol where the overall fits to the \ion{Si}{4} spectral window were reliable even without fitting these lines. On the other hand, accounting for their contribution in the pixels $y = $145 and 146 (diamond symbol) lowers $\chi^2_\mathrm{r}$ from $\approx 5$ to $\approx 1$.

Since in this study we focused on the asymmetries of the line related to the presence of up- and down- flowing components along the line-of-sight, each of the three Gaussians included in the fit of the \ion{Si}{4} line was restricted to either one of the wings or the central part of the line. The range limiting the centroid $\lambda_\mathrm{c}$ of the Gaussian fitting the blue (left) wing of the line was selected as $\lambda_\mathrm{c, b} \in \langle 1401.5\,\AA, 1402.77\,\AA \rangle$. The Gaussian fitting the central part of the line close to the rest wavelength $\lambda_0$ was restricted to lie within $\lambda_\mathrm{c, 0} \in \langle 1402\,\AA, 1403.5\,\AA \rangle$. Finally, the range for the Gaussian fitting the red (right) wing of the line was selected as $\lambda_\mathrm{c, r} \in \langle 1402.77\,\AA, 1404\,\AA \rangle$. The span of these wavelength ranges was adjusted to account for the possible Doppler shifts of the entire profile. The minimal FWHM of the fitting Gaussians was set to 0.2\,\AA. This value was adopted from the FWHM of quiet-Sun profiles measured in a large region observed in the FOV outside of the flare.  

\begin{figure*}[!h]
\centering   
\includegraphics[width=18.00cm, clip,   viewport=00 05 1150 350]{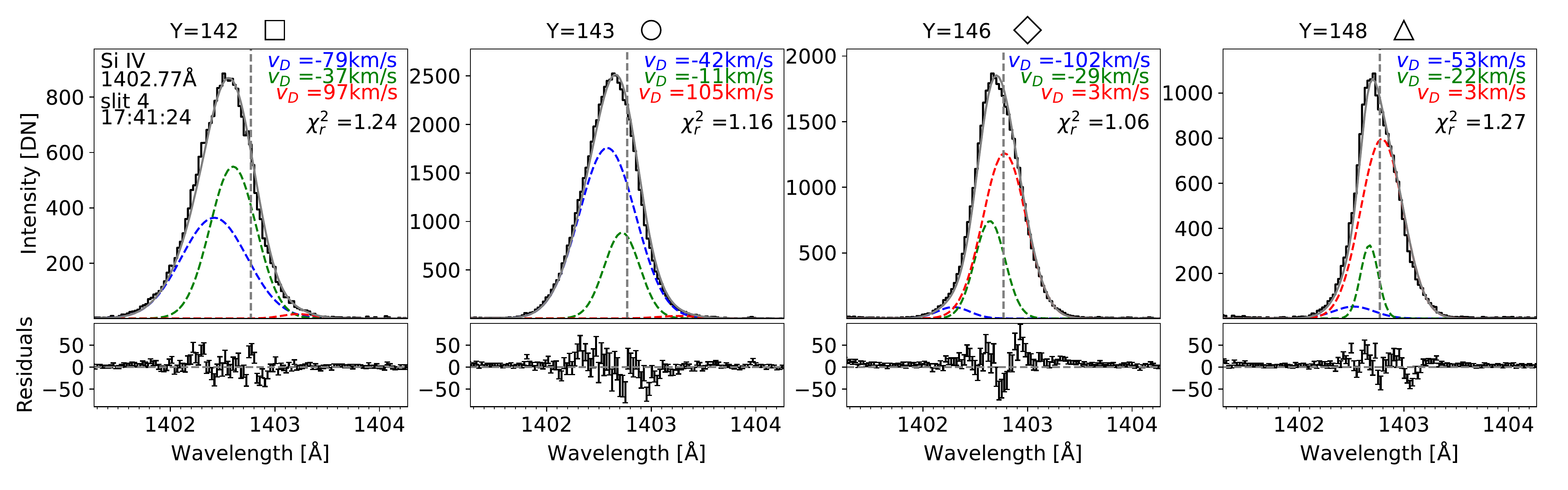}
\caption{Three-Gaussian fits of the Si IV 1402.77\,\AA~line and their residuals at four different locations marked using the symbols in Figure \ref{fig_spec_ribbon}(h) and (m). The Doppler velocities corresponding to the centroids of the fitting Gaussians are indicated at the top right of each panel together with the $\chi^2_\mathrm{r}$ statistics of the fit. \label{fig_gaussian_fits}}
\end{figure*}

Four example fits of spectra observed in the pixels marked using the triangle ($y$ = 148), diamond ($y$ = 146), circle ($y$ = 143), and square ($y$ = 142) symbols in Figure \ref{fig_spec_ribbon}(h), (m) are presented in Figure \ref{fig_gaussian_fits}. Each panel contains the observed spectrum (black) overplotted by the blue, green, and red Gaussians fitting the blue wing, the central part, and the red wing of the line. The final fit corresponding to the sum of the Gaussians is indicated using the grey color.

In the first two pixels marked using the square and the circle symbols, the Gaussians fitting the blue wing and the central part of the line were dominating. The blueshifts of the blue and green Gaussians correspond to $|v_{\text{D}}| = 79$\,km\,s$^{-1}$, $|v_{\text{D}}| = 37$\,km\,s$^{-1}$ and $|v_{\text{D}}| = 37$\,km\,s$^{-1}$, $|v_{\text{D}}| = 11$\,km\,s$^{-1}$ in the square and circle pixels, respectively. This result confirms that the strongest blueshifts within this small region are found in the square pixel (Section \ref{sec_kernel_doppler}). The red Gaussians, fitting the red wings of the line, indicate rather large Doppler velocities of roughly $|v_{\text{D}}| = 100$\,km\,s$^{-1}$. Their amplitudes are however very low,  of $\approx30$\,DN and $\approx40$\,DN, in square and circle pixels, respectively. As such, they were below the 5\% intensity threshold for the detection of the line's wings in our calculation of the $RBA$ (Section \ref{sec_kernel_rba}). Therefore, in the map of the $RBA$ plotted in Figure \ref{fig_spec_ribbon}(m), these two pixels appear as blue. 

Even though the lines observed in the triangle and diamond pixels have blueshifted peaks, they are clearly asymmetric due to the strong red components. As can be seen from the relative strength of the Gaussians fitting these profiles, this indeed is the case there. In the fits of the spectra in these two pixels, the weakly-redshifted red Gaussians shifted to $|v_{\text{D}}| = 3$\,km\,s$^{-1}$ are dominating. The amplitude of the blue Gaussians, corresponding to $|v_{\text{D}}| = 102$\,km\,s$^{-1}$ and $|v_{\text{D}}| = 53$\,km\,s$^{-1}$ in the diamond and triangle pixels has lowered to $\approx190$\,DN and $\approx100$\,DN, respectively. Even though these Gaussians are weak, they still affect the result of the calculation of the $RBA$ of these profiles. The $RBA$ obtained in the diamond pixel is close to zero (Figure \ref{fig_spec_ribbon}(m)) despite the strong red component. 

\section{Doppler velocities from line centroids obtained using different methods}
\label{sec_appendixB}

In the analyzes of the Doppler velocities of lines observed by IRIS, the centroids (or peak wavelengths) of the line are often derived from either single-Gaussian fitting or analysis of line moments. Both methods are used widely % throughout the literature
\citep[e.g.,][]{li15, lizhang15, warren16, brosius18, li19}, with some studies comparing their performance on the same datasets. While some studies conclude that the two methods yield comparable results \citep[e.g.,][]{jeffrey18}, in some cases issues with the {accuracy} of the single-Gaussian fits to spectra are debated \citep[see e.g.,][]{li15, warren16}. Here, we quantify the differences between the Doppler velocities calculated from the moment analysis and the single-Gaussian fitting. Furthermore, we extend this analysis to a third method based on automatic detection of the peak wavelength corresponding to the highest intensity measured in the \ion{Si}{4} line.

\begin{figure}[!h]
\centering   
\includegraphics[width=8.8cm, clip,   viewport=10 00 420 250]{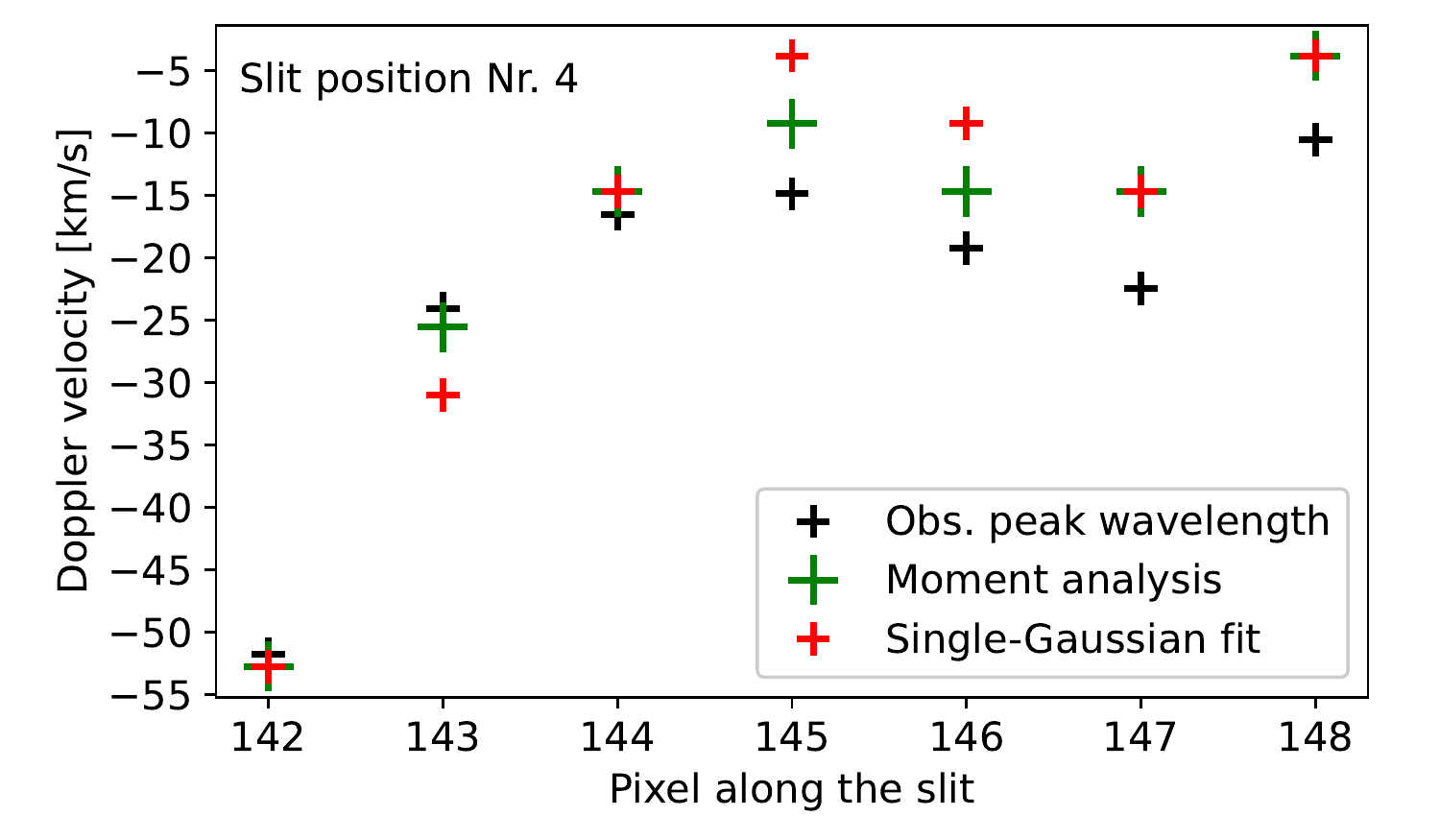}
\caption{The Doppler velocities measured in the individual pixels in the zoomed inclusion plotted in Figure \ref{fig_spec_ribbon}(c) and (h). The different colors indicate which method was used to measure the peak wavelength of the line for the calculation of the Doppler velocity. \label{fig_method_comparison}}
\end{figure}

The Doppler velocities within the small region as a function of the pixel position along the slit are plotted using the plus signs in Figure \ref{fig_method_comparison}. The green plus signs indicate the velocities resulting from the moment analysis, the black plus signs mark the velocities resulting from the observed peak wavelengths, and the  red plus signs stand for those obtained using the single-Gaussian fits.

According to this figure, the resulting velocities are generally close one to another, with the largest difference of $\approx 11$\,km\,s$^{-1}$ found in the pixel denoted by the diamond symbol in Figure \ref{fig_spec_ribbon}. On the other hand, the smallest difference between the velocities is found in the square pixel, which also shows the largest blueshifts roughly corresponding to $|v_{\text{D}}| = 50 - 55$\,km\,s$^{-1}$. In four of the pixels plotted in this figure, the velocities obtained from the moment analysis and the single-Gaussian fits are the same, as indicated by the intersecting red and green plus signs. In the remaining pixels, the velocities obtained from observed peak wavelengths were closer to those resulting from the moment analysis. Apart from the circle pixel, the Doppler velocities resulting from the single-Gaussian fitting were consistently shifted towards the red part of the spectrum. This is most likely caused by the red-wing enhancements the fitting method attempts to approximate in order to reduce the $\chi^2_\mathrm{r}$ statistic. As a result, the centroids of the fitting Gaussians are shifted towards higher wavelengths. Similar results were obtained in all of the pixels we randomly selected for the investigation. The largest difference in the peak wavelengths obtained using these three methods we found was $\approx0.07$\,\AA, which roughly translates to the velocity uncertainty $ \sigma_{v_{\text{D}}}$ of $\pm 7.5$\,km\,s$^{-1}$. The velocity uncertainties of this order of magnitude are common in IRIS analyzes of the Doppler shifts \citep[e.g.,][]{li15, jeffrey18, li19}. 

\clearpage

\bibliographystyle{aasjournal}    %% Author (year)
\bibliography{bibliography}

\begin{thebibliography}{}
\expandafter\ifx\csname natexlab\endcsname\relax\def\natexlab#1{#1}\fi
\providecommand{\url}[1]{\href{#1}{#1}}
\providecommand{\dodoi}[1]{doi:~\href{http://doi.org/#1}{\nolinkurl{#1}}}
\providecommand{\doeprint}[1]{\href{http://ascl.net/#1}{\nolinkurl{http://ascl.net/#1}}}
\providecommand{\doarXiv}[1]{\href{https://arxiv.org/abs/#1}{\nolinkurl{https://arxiv.org/abs/#1}}}

\bibitem[{{Acton} {et~al.}(1982){Acton}, {Leibacher}, {Canfield}, {Gunkler},
  {Hudson}, \& {Kiplinger}}]{acton82}
{Acton}, L.~W., {Leibacher}, J.~W., {Canfield}, R.~C., {et~al.} 1982, \apj,
  263, 409, \dodoi{10.1086/160513}

\bibitem[{{Allred} {et~al.}(2015){Allred}, {Kowalski}, \&
  {Carlsson}}]{allred15}
{Allred}, J.~C., {Kowalski}, A.~F., \& {Carlsson}, M. 2015, \apj, 809, 104,
  \dodoi{10.1088/0004-637X/809/1/104}

\bibitem[{{Antonucci} \& {Dennis}(1983)}]{antonucci83}
{Antonucci}, E., \& {Dennis}, B.~R. 1983, \solphys, 86, 67,
  \dodoi{10.1007/BF00157175}

\bibitem[{{Aulanier} \& {Dud{\'\i}k}(2019)}]{aulanier19}
{Aulanier}, G., \& {Dud{\'\i}k}, J. 2019, Astronomy and Astrophysics, 621, A72,
  \dodoi{10.1051/0004-6361/201834221}

\bibitem[{{Aulanier} {et~al.}(2012){Aulanier}, {Janvier}, \&
  {Schmieder}}]{aulanier12}
{Aulanier}, G., {Janvier}, M., \& {Schmieder}, B. 2012, Astronomy and
  Astrophysics, 543, A110, \dodoi{10.1051/0004-6361/201219311}

\bibitem[{{Aulanier} {et~al.}(2006){Aulanier}, {Pariat}, {D{\'e}moulin}, \&
  {DeVore}}]{aulanier06}
{Aulanier}, G., {Pariat}, E., {D{\'e}moulin}, P., \& {DeVore}, C.~R. 2006,
  \solphys, 238, 347, \dodoi{10.1007/s11207-006-0230-2}

\bibitem[{{Berlicki} {et~al.}(2004){Berlicki}, {Schmieder}, {Vilmer},
  {Aulanier}, \& {Del Zanna}}]{berlicki04}
{Berlicki}, A., {Schmieder}, B., {Vilmer}, N., {Aulanier}, G., \& {Del Zanna},
  G. 2004, \aap, 423, 1119, \dodoi{10.1051/0004-6361:20040259}

\bibitem[{{Bj{\o}rgen} {et~al.}(2018){Bj{\o}rgen}, {Sukhorukov}, {Leenaarts},
  {Carlsson}, {de la Cruz Rodr{\'\i}guez}, {Scharmer}, \&
  {Hansteen}}]{bjorgen18}
{Bj{\o}rgen}, J.~P., {Sukhorukov}, A.~V., {Leenaarts}, J., {et~al.} 2018, \aap,
  611, A62, \dodoi{10.1051/0004-6361/201731926}

\bibitem[{{Boerner} {et~al.}(2012){Boerner}, {Edwards}, {Lemen}, {Rausch},
  {Schrijver}, {Shine}, {Shing}, {Stern}, {Tarbell}, {Title}, {Wolfson},
  {Soufli}, {Spiller}, {Gullikson}, {McKenzie}, {Windt}, {Golub}, {Podgorski},
  {Testa}, \& {Weber}}]{boerner12}
{Boerner}, P., {Edwards}, C., {Lemen}, J., {et~al.} 2012, Solar Physics, 275,
  41, \dodoi{10.1007/s11207-011-9804-8}

\bibitem[{{Brosius}(2003)}]{brosius03}
{Brosius}, J.~W. 2003, \apj, 586, 1417, \dodoi{10.1086/367958}

\bibitem[{{Brosius}(2009)}]{brosius09}
---. 2009, \apj, 701, 1209, \dodoi{10.1088/0004-637X/701/2/1209}

\bibitem[{{Brosius}(2013)}]{brosius13}
---. 2013, \apj, 762, 133, \dodoi{10.1088/0004-637X/762/2/133}

\bibitem[{{Brosius} \& {Daw}(2015)}]{brosius15}
{Brosius}, J.~W., \& {Daw}, A.~N. 2015, \apj, 810, 45,
  \dodoi{10.1088/0004-637X/810/1/45}

\bibitem[{{Brosius} \& {Inglis}(2018)}]{brosius18}
{Brosius}, J.~W., \& {Inglis}, A.~R. 2018, The Astrophysical Journal, 867, 85,
  \dodoi{10.3847/1538-4357/aae5f5}

\bibitem[{{Brosius} \& {Phillips}(2004)}]{brosius04}
{Brosius}, J.~W., \& {Phillips}, K. J.~H. 2004, \apj, 613, 580,
  \dodoi{10.1086/422873}

\bibitem[{{Carlsson} \& {Stein}(1992)}]{carlsson92}
{Carlsson}, M., \& {Stein}, R.~F. 1992, \apjl, 397, L59, \dodoi{10.1086/186544}

\bibitem[{{Carlsson} \& {Stein}(1995)}]{carlsson95}
---. 1995, \apjl, 440, L29, \dodoi{10.1086/187753}

\bibitem[{{Carlsson} \& {Stein}(1997)}]{carlsson97}
---. 1997, \apj, 481, 500, \dodoi{10.1086/304043}

\bibitem[{{Carmichael}(1964)}]{carmichael64}
{Carmichael}, H. 1964, NASA Special Publication, 50, 451

\bibitem[{{Cheng} \& {Ding}(2016)}]{cheng16}
{Cheng}, X., \& {Ding}, M.~D. 2016, The Astrophysical Journal Letters, 823, L4,
  \dodoi{10.3847/2041-8205/823/1/L4}

\bibitem[{{Cheung} {et~al.}(2021){Cheung}, {Mart{\'\i}nez-Sykora}, {Testa}, {De
  Pontieu}, {Chintzoglou}, {Rempel}, {Polito}, {Kerr}, {Reeves}, {Fletcher},
  {Jin}, {N{\'o}brega-Siverio}, {Danilovic}, {Antolin}, {Allred}, {Hansteen},
  {Ugarte-Urra}, {DeLuca}, {Longcope}, {Takasao}, {DeRosa}, {Boerner},
  {Jaeggli}, {Nitta}, {Daw}, {Carlsson}, {Golub}, \& {the MUSE
  team}}]{cheung21}
{Cheung}, M. C.~M., {Mart{\'\i}nez-Sykora}, J., {Testa}, P., {et~al.} 2021,
  arXiv e-prints, arXiv:2106.15591.
\newblock \doarXiv{2106.15591}

\bibitem[{{Czaykowska} {et~al.}(1999){Czaykowska}, {De Pontieu}, {Alexander},
  \& {Rank}}]{czaykowska99}
{Czaykowska}, A., {De Pontieu}, B., {Alexander}, D., \& {Rank}, G. 1999, \apjl,
  521, L75, \dodoi{10.1086/312176}

\bibitem[{{De Pontieu} {et~al.}(2020){De Pontieu}, {Mart{\'\i}nez-Sykora},
  {Testa}, {Winebarger}, {Daw}, {Hansteen}, {Cheung}, \&
  {Antolin}}]{depontieu20}
{De Pontieu}, B., {Mart{\'\i}nez-Sykora}, J., {Testa}, P., {et~al.} 2020, \apj,
  888, 3, \dodoi{10.3847/1538-4357/ab5b03}

\bibitem[{{De Pontieu} {et~al.}(2009){De Pontieu}, {McIntosh}, {Hansteen}, \&
  {Schrijver}}]{depontieu09}
{De Pontieu}, B., {McIntosh}, S.~W., {Hansteen}, V.~H., \& {Schrijver}, C.~J.
  2009, The Astrophysical Journal Letters, 701, L1,
  \dodoi{10.1088/0004-637X/701/1/L1}

\bibitem[{{De Pontieu} {et~al.}(2014){De Pontieu}, {Title}, {Lemen}, {Kushner},
  {Akin}, {Allard}, {Berger}, {Boerner}, {Cheung}, {Chou}, {Drake}, {Duncan},
  {Freeland}, {Heyman}, {Hoffman}, {Hurlburt}, {Lindgren}, {Mathur}, {Rehse},
  {Sabolish}, {Seguin}, {Schrijver}, {Tarbell}, {W{\"u}lser}, {Wolfson},
  {Yanari}, {Mudge}, {Nguyen-Phuc}, {Timmons}, {van Bezooijen}, {Weingrod},
  {Brookner}, {Butcher}, {Dougherty}, {Eder}, {Knagenhjelm}, {Larsen},
  {Mansir}, {Phan}, {Boyle}, {Cheimets}, {DeLuca}, {Golub}, {Gates}, {Hertz},
  {McKillop}, {Park}, {Perry}, {Podgorski}, {Reeves}, {Saar}, {Testa}, {Tian},
  {Weber}, {Dunn}, {Eccles}, {Jaeggli}, {Kankelborg}, {Mashburn}, {Pust},
  {Springer}, {Carvalho}, {Kleint}, {Marmie}, {Mazmanian}, {Pereira}, {Sawyer},
  {Strong}, {Worden}, {Carlsson}, {Hansteen}, {Leenaarts}, {Wiesmann},
  {Aloise}, {Chu}, {Bush}, {Scherrer}, {Brekke}, {Martinez-Sykora}, {Lites},
  {McIntosh}, {Uitenbroek}, {Okamoto}, {Gummin}, {Auker}, {Jerram}, {Pool}, \&
  {Waltham}}]{depontieu14}
{De Pontieu}, B., {Title}, A.~M., {Lemen}, J.~R., {et~al.} 2014, Solar Physics,
  289, 2733, \dodoi{10.1007/s11207-014-0485-y}

\bibitem[{{De Pontieu} {et~al.}(2021){De Pontieu}, {Polito}, {Hansteen},
  {Testa}, {Reeves}, {Antolin}, {Nobrega-Siverio}, {Kowalski},
  {Martinez-Sykora}, {Carlsson}, {McIntosh}, {Liu}, {Daw}, \&
  {Kankelborg}}]{depontieu21}
{De Pontieu}, B., {Polito}, V., {Hansteen}, V., {et~al.} 2021, arXiv e-prints,
  arXiv:2103.16109.
\newblock \doarXiv{2103.16109}

\bibitem[{{del Zanna} {et~al.}(2006){del Zanna}, {Berlicki}, {Schmieder}, \&
  {Mason}}]{delzanna06}
{del Zanna}, G., {Berlicki}, A., {Schmieder}, B., \& {Mason}, H.~E. 2006,
  \solphys, 234, 95, \dodoi{10.1007/s11207-006-0016-6}

\bibitem[{{Del Zanna} {et~al.}(2019){Del Zanna}, {Gupta}, \&
  {Mason}}]{delzanna19}
{Del Zanna}, G., {Gupta}, G.~R., \& {Mason}, H.~E. 2019, Astronomy and
  Astrophysics, 631, A163, \dodoi{10.1051/0004-6361/201834625}

\bibitem[{{Del Zanna} {et~al.}(2002){Del Zanna}, {Landini}, \&
  {Mason}}]{delzanna02}
{Del Zanna}, G., {Landini}, M., \& {Mason}, H.~E. 2002, Astronomy and
  Astrophysics, 385, 968, \dodoi{10.1051/0004-6361:20020164}

\bibitem[{{Doschek} {et~al.}(2016){Doschek}, {Warren}, \& {Young}}]{doschek16}
{Doschek}, G.~A., {Warren}, H.~P., \& {Young}, P.~R. 2016, \apj, 832, 77,
  \dodoi{10.3847/0004-637X/832/1/77}

\bibitem[{{Dud{\'{\i}}k} {et~al.}(2014){Dud{\'{\i}}k}, {Janvier}, {Aulanier},
  {Del Zanna}, {Karlick{\'y}}, {Mason}, \& {Schmieder}}]{dudik14}
{Dud{\'{\i}}k}, J., {Janvier}, M., {Aulanier}, G., {et~al.} 2014, The
  Astrophysical Journal, 784, 144, \dodoi{10.1088/0004-637X/784/2/144}

\bibitem[{{Dud{\'\i}k} {et~al.}(2017){Dud{\'\i}k}, {Polito},
  {Dzif{\v{c}}{\'a}kov{\'a}}, {Del Zanna}, \& {Testa}}]{dudik17_nonmaxw}
{Dud{\'\i}k}, J., {Polito}, V., {Dzif{\v{c}}{\'a}kov{\'a}}, E., {Del Zanna},
  G., \& {Testa}, P. 2017, The Astrophysical Journal, 842, 19,
  \dodoi{10.3847/1538-4357/aa71a8}

\bibitem[{{Dud{\'{\i}}k} {et~al.}(2016){Dud{\'{\i}}k}, {Polito}, {Janvier},
  {Mulay}, {Karlick{\'y}}, {Aulanier}, {Del Zanna}, {Dzif{\v c}{\'a}kov{\'a}},
  {Mason}, \& {Schmieder}}]{dudik16}
{Dud{\'{\i}}k}, J., {Polito}, V., {Janvier}, M., {et~al.} 2016, The
  Astrophysical Journal, 823, 41, \dodoi{10.3847/0004-637X/823/1/41}

\bibitem[{{Fisher} {et~al.}(1985{\natexlab{a}}){Fisher}, {Canfield}, \&
  {McClymont}}]{fischer85a}
{Fisher}, G.~H., {Canfield}, R.~C., \& {McClymont}, A.~N. 1985{\natexlab{a}},
  \apj, 289, 414, \dodoi{10.1086/162901}

\bibitem[{{Fisher} {et~al.}(1985{\natexlab{b}}){Fisher}, {Canfield}, \&
  {McClymont}}]{fischer85b}
---. 1985{\natexlab{b}}, \apj, 289, 425, \dodoi{10.1086/162902}

\bibitem[{{Fisher} {et~al.}(1985{\natexlab{c}}){Fisher}, {Canfield}, \&
  {McClymont}}]{fischer85c}
---. 1985{\natexlab{c}}, \apj, 289, 434, \dodoi{10.1086/162903}

\bibitem[{{Fletcher} {et~al.}(2004){Fletcher}, {Pollock}, \&
  {Potts}}]{fletcher04}
{Fletcher}, L., {Pollock}, J.~A., \& {Potts}, H.~E. 2004, Solar Physics, 222,
  279, \dodoi{10.1023/B:SOLA.0000043580.89730.4d}

\bibitem[{{Fletcher} {et~al.}(2011){Fletcher}, {Dennis}, {Hudson}, {Krucker},
  {Phillips}, {Veronig}, {Battaglia}, {Bone}, {Caspi}, {Chen}, {Gallagher},
  {Grigis}, {Ji}, {Liu}, {Milligan}, \& {Temmer}}]{fletcher11}
{Fletcher}, L., {Dennis}, B.~R., {Hudson}, H.~S., {et~al.} 2011, Space Science
  Reviews, 159, 19, \dodoi{10.1007/s11214-010-9701-8}

\bibitem[{{Forbes} \& {Lin}(2000)}]{forbes2000}
{Forbes}, T.~G., \& {Lin}, J. 2000, Journal of Atmospheric and
  Solar-Terrestrial Physics, 62, 1499, \dodoi{10.1016/S1364-6826(00)00083-3}

\bibitem[{{Graham} \& {Cauzzi}(2015)}]{grahamcauzzi15}
{Graham}, D.~R., \& {Cauzzi}, G. 2015, The Astrophysical Journal Letters, 807,
  L22, \dodoi{10.1088/2041-8205/807/2/L22}

\bibitem[{{Graham} {et~al.}(2020){Graham}, {Cauzzi}, {Zangrilli}, {Kowalski},
  {Sim{\~o}es}, \& {Allred}}]{graham20}
{Graham}, D.~R., {Cauzzi}, G., {Zangrilli}, L., {et~al.} 2020, The
  Astrophysical Journal, 895, 6, \dodoi{10.3847/1538-4357/ab88ad}

\bibitem[{{Heyvaerts} {et~al.}(1977){Heyvaerts}, {Priest}, \&
  {Rust}}]{heyvaerts77}
{Heyvaerts}, J., {Priest}, E.~R., \& {Rust}, D.~M. 1977, \apj, 216, 123,
  \dodoi{10.1086/155453}

\bibitem[{{Hirayama}(1974)}]{hirayama74}
{Hirayama}, T. 1974, Solar Physics, 34, 323, \dodoi{10.1007/BF00153671}

\bibitem[{{Hong} {et~al.}(2020){Hong}, {Li}, {Ding}, \& {Zhou}}]{hong20}
{Hong}, J., {Li}, Y., {Ding}, M.~D., \& {Zhou}, Y.-H. 2020, \apj, 890, 115,
  \dodoi{10.3847/1538-4357/ab6d05}

\bibitem[{{Huang} {et~al.}(2019){Huang}, {Xu}, {Sadykov}, {Jing}, \&
  {Wang}}]{huang19}
{Huang}, N., {Xu}, Y., {Sadykov}, V.~M., {Jing}, J., \& {Wang}, H. 2019, \apjl,
  878, L15, \dodoi{10.3847/2041-8213/ab2330}

\bibitem[{{Jaeggli} {et~al.}(2018){Jaeggli}, {Judge}, \& {Daw}}]{jaeggli18}
{Jaeggli}, S.~A., {Judge}, P.~G., \& {Daw}, A.~N. 2018, \apj, 855, 134,
  \dodoi{10.3847/1538-4357/aaafd5}

\bibitem[{{Janvier}(2017)}]{janvier17}
{Janvier}, M. 2017, Journal of Plasma Physics, 83, 535830101,
  \dodoi{10.1017/S0022377817000034}

\bibitem[{{Janvier} {et~al.}(2013){Janvier}, {Aulanier}, {Pariat}, \&
  {D{\'e}moulin}}]{janvier13}
{Janvier}, M., {Aulanier}, G., {Pariat}, E., \& {D{\'e}moulin}, P. 2013,
  Astronomy and Astrophysics, 555, A77, \dodoi{10.1051/0004-6361/201321164}

\bibitem[{{Jeffrey} {et~al.}(2018){Jeffrey}, {Fletcher}, {Labrosse}, \&
  {Sim{\~o}es}}]{jeffrey18}
{Jeffrey}, N.~L.~S., {Fletcher}, L., {Labrosse}, N., \& {Sim{\~o}es}, P.~J.~A.
  2018, Science Advances, 4, 2794, \dodoi{10.1126/sciadv.aav2794}

\bibitem[{{Joshi} {et~al.}(2019){Joshi}, {Zhu}, {Schmieder}, {Aulanier},
  {Janvier}, {Joshi}, {Magara}, {Chandra}, \& {Inoue}}]{joshi19}
{Joshi}, N.~C., {Zhu}, X., {Schmieder}, B., {et~al.} 2019, \apj, 871, 165,
  \dodoi{10.3847/1538-4357/aaf3b5}

\bibitem[{{Joshi} {et~al.}(2021){Joshi}, {Schmieder}, {Tei}, {Aulanier},
  {L{\"o}rin{\v{c}}{\'\i}k}, {Chandra}, \& {Heinzel}}]{joshi21}
{Joshi}, R., {Schmieder}, B., {Tei}, A., {et~al.} 2021, Astronomy and
  Astrophysics, 645, A80, \dodoi{10.1051/0004-6361/202039229}

\bibitem[{{Kang} {et~al.}(2019){Kang}, {Inoue}, {Kusano}, {Park}, \&
  {Moon}}]{kang19}
{Kang}, J., {Inoue}, S., {Kusano}, K., {Park}, S.-H., \& {Moon}, Y.-J. 2019,
  \apj, 887, 263, \dodoi{10.3847/1538-4357/ab5582}

\bibitem[{{Kerr} {et~al.}(2019){Kerr}, {Carlsson}, {Allred}, {Young}, \&
  {Daw}}]{kerr19}
{Kerr}, G.~S., {Carlsson}, M., {Allred}, J.~C., {Young}, P.~R., \& {Daw}, A.~N.
  2019, The Astrophysical Journal, 871, 23, \dodoi{10.3847/1538-4357/aaf46e}

\bibitem[{{Kerr} {et~al.}(2015){Kerr}, {Sim{\~o}es}, {Qiu}, \&
  {Fletcher}}]{kerr15}
{Kerr}, G.~S., {Sim{\~o}es}, P.~J.~A., {Qiu}, J., \& {Fletcher}, L. 2015,
  Astronomy and Astrophysics, 582, A50, \dodoi{10.1051/0004-6361/201526128}

\bibitem[{{Kopp} \& {Pneuman}(1976)}]{kopp76}
{Kopp}, R.~A., \& {Pneuman}, G.~W. 1976, Solar Physics, 50, 85,
  \dodoi{10.1007/BF00206193}

\bibitem[{{Kuridze} {et~al.}(2015){Kuridze}, {Mathioudakis}, {Sim{\~o}es},
  {Rouppe van der Voort}, {Carlsson}, {Jafarzadeh}, {Allred}, {Kowalski},
  {Kennedy}, {Fletcher}, {Graham}, \& {Keenan}}]{kuridze15}
{Kuridze}, D., {Mathioudakis}, M., {Sim{\~o}es}, P.~J.~A., {et~al.} 2015, \apj,
  813, 125, \dodoi{10.1088/0004-637X/813/2/125}

\bibitem[{{Kuroda} {et~al.}(2018){Kuroda}, {Gary}, {Wang}, {Fleishman}, {Nita},
  \& {Jing}}]{kuroda18}
{Kuroda}, N., {Gary}, D.~E., {Wang}, H., {et~al.} 2018, The Astrophysical
  Journal, 852, 32, \dodoi{10.3847/1538-4357/aa9d98}

\bibitem[{{Leenaarts} {et~al.}(2013{\natexlab{a}}){Leenaarts}, {Pereira},
  {Carlsson}, {Uitenbroek}, \& {De Pontieu}}]{leenaarts13a}
{Leenaarts}, J., {Pereira}, T.~M.~D., {Carlsson}, M., {Uitenbroek}, H., \& {De
  Pontieu}, B. 2013{\natexlab{a}}, \apj, 772, 89,
  \dodoi{10.1088/0004-637X/772/2/89}

\bibitem[{{Leenaarts} {et~al.}(2013{\natexlab{b}}){Leenaarts}, {Pereira},
  {Carlsson}, {Uitenbroek}, \& {De Pontieu}}]{leenaarts13b}
---. 2013{\natexlab{b}}, \apj, 772, 90, \dodoi{10.1088/0004-637X/772/2/90}

\bibitem[{{Lemen} {et~al.}(2012){Lemen}, {Title}, {Akin}, {Boerner}, {Chou},
  {Drake}, {Duncan}, {Edwards}, {Friedlaender}, {Heyman}, {Hurlburt}, {Katz},
  {Kushner}, {Levay}, {Lindgren}, {Mathur}, {McFeaters}, {Mitchell}, {Rehse},
  {Schrijver}, {Springer}, {Stern}, {Tarbell}, {Wuelser}, {Wolfson}, {Yanari},
  {Bookbinder}, {Cheimets}, {Caldwell}, {Deluca}, {Gates}, {Golub}, {Park},
  {Podgorski}, {Bush}, {Scherrer}, {Gummin}, {Smith}, {Auker}, {Jerram},
  {Pool}, {Soufli}, {Windt}, {Beardsley}, {Clapp}, {Lang}, \&
  {Waltham}}]{lemen12}
{Lemen}, J.~R., {Title}, A.~M., {Akin}, D.~J., {et~al.} 2012, Solar Physics,
  275, 17, \dodoi{10.1007/s11207-011-9776-8}

\bibitem[{{Li} {et~al.}(2015{\natexlab{a}}){Li}, {Ning}, \& {Zhang}}]{lining15}
{Li}, D., {Ning}, Z.~J., \& {Zhang}, Q.~M. 2015{\natexlab{a}}, \apj, 813, 59,
  \dodoi{10.1088/0004-637X/813/1/59}

\bibitem[{{Li} {et~al.}(2018{\natexlab{a}}){Li}, {Hou}, {Yang}, \&
  {Zhang}}]{li18b}
{Li}, T., {Hou}, Y., {Yang}, S., \& {Zhang}, J. 2018{\natexlab{a}}, \apj, 869,
  172, \dodoi{10.3847/1538-4357/aaefee}

\bibitem[{{Li} {et~al.}(2016){Li}, {Yang}, {Hou}, \& {Zhang}}]{li16}
{Li}, T., {Yang}, K., {Hou}, Y., \& {Zhang}, J. 2016, \apj, 830, 152,
  \dodoi{10.3847/0004-637X/830/2/152}

\bibitem[{{Li} {et~al.}(2018{\natexlab{b}}){Li}, {Yang}, {Zhang}, {Hou}, \&
  {Zhang}}]{li18}
{Li}, T., {Yang}, S., {Zhang}, Q., {Hou}, Y., \& {Zhang}, J.
  2018{\natexlab{b}}, \apj, 859, 122, \dodoi{10.3847/1538-4357/aabe84}

\bibitem[{{Li} \& {Zhang}(2014)}]{lizhang14}
{Li}, T., \& {Zhang}, J. 2014, The Astrophysical Journal Letters, 791, L13,
  \dodoi{10.1088/2041-8205/791/1/L13}

\bibitem[{{Li} \& {Zhang}(2015)}]{lizhang15}
---. 2015, The Astrophysical Journal Letters, 804, L8,
  \dodoi{10.1088/2041-8205/804/1/L8}

\bibitem[{{Li} {et~al.}(2019){Li}, {Ding}, {Hong}, {Li}, \& {Gan}}]{li19}
{Li}, Y., {Ding}, M.~D., {Hong}, J., {Li}, H., \& {Gan}, W.~Q. 2019, The
  Astrophysical Journal, 879, 30, \dodoi{10.3847/1538-4357/ab245a}

\bibitem[{{Li} {et~al.}(2015{\natexlab{b}}){Li}, {Ding}, {Qiu}, \&
  {Cheng}}]{li15}
{Li}, Y., {Ding}, M.~D., {Qiu}, J., \& {Cheng}, J.~X. 2015{\natexlab{b}}, \apj,
  811, 7, \dodoi{10.1088/0004-637X/811/1/7}

\bibitem[{{Liu} {et~al.}(2018){Liu}, {Cao}, {Chae}, {Ahn}, {Prasad Choudhary},
  {Lee}, {Liu}, {Deng}, {Wang}, \& {Wang}}]{liu18}
{Liu}, C., {Cao}, W., {Chae}, J., {et~al.} 2018, The Astrophysical Journal,
  869, 21, \dodoi{10.3847/1538-4357/aaecd0}

\bibitem[{{L{\"o}rin{\v{c}}{\'\i}k}
  {et~al.}(2019{\natexlab{a}}){L{\"o}rin{\v{c}}{\'\i}k}, {Aulanier},
  {Dud{\'\i}k}, {Zemanov{\'a}}, \& {Dzif{\v{c}}{\'a}kov{\'a}}}]{lorincik19a}
{L{\"o}rin{\v{c}}{\'\i}k}, J., {Aulanier}, G., {Dud{\'\i}k}, J.,
  {Zemanov{\'a}}, A., \& {Dzif{\v{c}}{\'a}kov{\'a}}, E. 2019{\natexlab{a}}, The
  Astrophysical Journal, 881, 68, \dodoi{10.3847/1538-4357/ab298f}

\bibitem[{{L{\"o}rin{\v{c}}{\'\i}k}
  {et~al.}(2019{\natexlab{b}}){L{\"o}rin{\v{c}}{\'\i}k}, {Dud{\'\i}k}, \&
  {Aulanier}}]{lorincik19b}
{L{\"o}rin{\v{c}}{\'\i}k}, J., {Dud{\'\i}k}, J., \& {Aulanier}, G.
  2019{\natexlab{b}}, The Astrophysical Journal, 885, 83,
  \dodoi{10.3847/1538-4357/ab4519}

\bibitem[{{Mason} {et~al.}(1986){Mason}, {Shine}, {Gurman}, \&
  {Harrison}}]{mason86}
{Mason}, H.~E., {Shine}, R.~A., {Gurman}, J.~B., \& {Harrison}, R.~A. 1986,
  \apj, 309, 435, \dodoi{10.1086/164615}

\bibitem[{{Miku{\l}a} {et~al.}(2017){Miku{\l}a}, {Heinzel}, {Liu}, \&
  {Berlicki}}]{mikula17}
{Miku{\l}a}, K., {Heinzel}, P., {Liu}, W., \& {Berlicki}, A. 2017, The
  Astrophysical Journal, 845, 30, \dodoi{10.3847/1538-4357/aa7d4e}

\bibitem[{{Milligan}(2015)}]{milligan15}
{Milligan}, R.~O. 2015, \solphys, 290, 3399, \dodoi{10.1007/s11207-015-0748-2}

\bibitem[{{Milligan} {et~al.}(2006{\natexlab{a}}){Milligan}, {Gallagher},
  {Mathioudakis}, {Bloomfield}, {Keenan}, \& {Schwartz}}]{milligan06a}
{Milligan}, R.~O., {Gallagher}, P.~T., {Mathioudakis}, M., {et~al.}
  2006{\natexlab{a}}, \apjl, 638, L117, \dodoi{10.1086/500555}

\bibitem[{{Milligan} {et~al.}(2006{\natexlab{b}}){Milligan}, {Gallagher},
  {Mathioudakis}, \& {Keenan}}]{milligan06b}
{Milligan}, R.~O., {Gallagher}, P.~T., {Mathioudakis}, M., \& {Keenan}, F.~P.
  2006{\natexlab{b}}, \apjl, 642, L169, \dodoi{10.1086/504592}

\bibitem[{{Mulay} \& {Fletcher}(2021)}]{mulay21}
{Mulay}, S.~M., \& {Fletcher}, L. 2021, \mnras, 504, 2842,
  \dodoi{10.1093/mnras/stab367}

\bibitem[{{Panos} {et~al.}(2018){Panos}, {Kleint}, {Huwyler}, {Krucker},
  {Melchior}, {Ullmann}, \& {Voloshynovskiy}}]{panos18}
{Panos}, B., {Kleint}, L., {Huwyler}, C., {et~al.} 2018, The Astrophysical
  Journal, 861, 62, \dodoi{10.3847/1538-4357/aac779}

\bibitem[{{Pereira} {et~al.}(2015){Pereira}, {Carlsson}, {De Pontieu}, \&
  {Hansteen}}]{pereira15}
{Pereira}, T. M.~D., {Carlsson}, M., {De Pontieu}, B., \& {Hansteen}, V. 2015,
  The Astrophysical Journal, 806, 14, \dodoi{10.1088/0004-637X/806/1/14}

\bibitem[{{Pereira} {et~al.}(2013){Pereira}, {Leenaarts}, {De Pontieu},
  {Carlsson}, \& {Uitenbroek}}]{pereira13}
{Pereira}, T.~M.~D., {Leenaarts}, J., {De Pontieu}, B., {Carlsson}, M., \&
  {Uitenbroek}, H. 2013, \apj, 778, 143, \dodoi{10.1088/0004-637X/778/2/143}

\bibitem[{{Pesnell} {et~al.}(2012){Pesnell}, {Thompson}, \&
  {Chamberlin}}]{pesnell12}
{Pesnell}, W.~D., {Thompson}, B.~J., \& {Chamberlin}, P.~C. 2012, Solar
  Physics, 275, 3, \dodoi{10.1007/s11207-011-9841-3}

\bibitem[{{Poduval} {et~al.}(2013){Poduval}, {DeForest}, {Schmelz}, \&
  {Pathak}}]{poduval13}
{Poduval}, B., {DeForest}, C.~E., {Schmelz}, J.~T., \& {Pathak}, S. 2013, \apj,
  765, 144, \dodoi{10.1088/0004-637X/765/2/144}

\bibitem[{{Polito} {et~al.}(2016{\natexlab{a}}){Polito}, {Del Zanna},
  {Dud{\'\i}k}, {Mason}, {Giunta}, \& {Reeves}}]{polito16b}
{Polito}, V., {Del Zanna}, G., {Dud{\'\i}k}, J., {et~al.} 2016{\natexlab{a}},
  Astronomy and Astrophysics, 594, A64, \dodoi{10.1051/0004-6361/201628965}

\bibitem[{{Polito} {et~al.}(2016{\natexlab{b}}){Polito}, {Reep}, {Reeves},
  {Sim{\~o}es}, {Dud{\'\i}k}, {Del Zanna}, {Mason}, \& {Golub}}]{polito16a}
{Polito}, V., {Reep}, J.~W., {Reeves}, K.~K., {et~al.} 2016{\natexlab{b}}, The
  Astrophysical Journal, 816, 89, \dodoi{10.3847/0004-637X/816/2/89}

\bibitem[{{Polito} {et~al.}(2015){Polito}, {Reeves}, {Del Zanna}, {Golub}, \&
  {Mason}}]{polito15}
{Polito}, V., {Reeves}, K.~K., {Del Zanna}, G., {Golub}, L., \& {Mason}, H.~E.
  2015, \apj, 803, 84, \dodoi{10.1088/0004-637X/803/2/84}

\bibitem[{{Polito} {et~al.}(2018){Polito}, {Testa}, {Allred}, {De Pontieu},
  {Carlsson}, {Pereira}, {Go{\v{s}}i{\'c}}, \& {Reale}}]{polito18}
{Polito}, V., {Testa}, P., {Allred}, J., {et~al.} 2018, \apj, 856, 178,
  \dodoi{10.3847/1538-4357/aab49e}

\bibitem[{{Polito} {et~al.}(2019){Polito}, {Testa}, \& {De Pontieu}}]{polito19}
{Polito}, V., {Testa}, P., \& {De Pontieu}, B. 2019, The Astrophysical Journal
  Letters, 879, L17, \dodoi{10.3847/2041-8213/ab290b}

\bibitem[{{Qiu} {et~al.}(2002){Qiu}, {Lee}, {Gary}, \& {Wang}}]{qiu02}
{Qiu}, J., {Lee}, J., {Gary}, D.~E., \& {Wang}, H. 2002, \apj, 565, 1335,
  \dodoi{10.1086/324706}

\bibitem[{{Rathore} \& {Carlsson}(2015)}]{rathore15a}
{Rathore}, B., \& {Carlsson}, M. 2015, \apj, 811, 80,
  \dodoi{10.1088/0004-637X/811/2/80}

\bibitem[{{Rathore} {et~al.}(2015){Rathore}, {Pereira}, {Carlsson}, \& {De
  Pontieu}}]{rathore15b}
{Rathore}, B., {Pereira}, T. M.~D., {Carlsson}, M., \& {De Pontieu}, B. 2015,
  \apj, 814, 70, \dodoi{10.1088/0004-637X/814/1/70}

\bibitem[{{Reep} {et~al.}(2015){Reep}, {Bradshaw}, \& {Alexander}}]{reep15}
{Reep}, J.~W., {Bradshaw}, S.~J., \& {Alexander}, D. 2015, \apj, 808, 177,
  \dodoi{10.1088/0004-637X/808/2/177}

\bibitem[{{Sahu} {et~al.}(2020){Sahu}, {Joshi}, {Mitra}, {Veronig}, \&
  {Yurchyshyn}}]{sahu20}
{Sahu}, S., {Joshi}, B., {Mitra}, P.~K., {Veronig}, A.~M., \& {Yurchyshyn}, V.
  2020, The Astrophysical Journal, 897, 157, \dodoi{10.3847/1538-4357/ab962b}

\bibitem[{{Scherrer} {et~al.}(2012){Scherrer}, {Schou}, {Bush}, {Kosovichev},
  {Bogart}, {Hoeksema}, {Liu}, {Duvall}, {Zhao}, {Title}, {Schrijver},
  {Tarbell}, \& {Tomczyk}}]{scherrer12}
{Scherrer}, P.~H., {Schou}, J., {Bush}, R.~I., {et~al.} 2012, Solar Physics,
  275, 207, \dodoi{10.1007/s11207-011-9834-2}

\bibitem[{{Sturrock}(1966)}]{sturrock66}
{Sturrock}, P.~A. 1966, Nature, 211, 695, \dodoi{10.1038/211695a0}

\bibitem[{{SunPy Community} {et~al.}(2020){SunPy Community}, {Barnes}, {Bobra},
  {Christe}, {Freij}, {Hayes}, {Ireland}, {Mumford}, {Perez-Suarez}, {Ryan},
  {Shih}, {Chanda}, {Glogowski}, {Hewett}, {Hughitt}, {Hill}, {Hiware},
  {Inglis}, {Kirk}, {Konge}, {Mason}, {Maloney}, {Murray}, {Panda}, {Park},
  {Pereira}, {Reardon}, {Savage}, {Sip{\H{o}}cz}, {Stansby}, {Jain}, {Taylor},
  {Yadav}, {Rajul}, \& {Dang}}]{sunpy20}
{SunPy Community}, {Barnes}, W.~T., {Bobra}, M.~G., {et~al.} 2020, The
  Astrophysical Journal, 890, 68, \dodoi{10.3847/1538-4357/ab4f7a}

\bibitem[{{Tei} {et~al.}(2018){Tei}, {Sakaue}, {Okamoto}, {Kawate}, {Heinzel},
  {UeNo}, {Asai}, {Ichimoto}, \& {Shibata}}]{tei18}
{Tei}, A., {Sakaue}, T., {Okamoto}, T.~J., {et~al.} 2018, Publications of the
  Astronomical Society of Japan, 70, 100, \dodoi{10.1093/pasj/psy047}

\bibitem[{{Testa} {et~al.}(2020){Testa}, {Polito}, \& {De Pontieu}}]{testa20}
{Testa}, P., {Polito}, V., \& {De Pontieu}, B. 2020, The Astrophysical Journal,
  889, 124, \dodoi{10.3847/1538-4357/ab63cf}

\bibitem[{{Testa} {et~al.}(2014){Testa}, {De Pontieu}, {Allred}, {Carlsson},
  {Reale}, {Daw}, {Hansteen}, {Martinez-Sykora}, {Liu}, {DeLuca}, {Golub},
  {McKillop}, {Reeves}, {Saar}, {Tian}, {Lemen}, {Title}, {Boerner},
  {Hurlburt}, {Tarbell}, {Wuelser}, {Kleint}, {Kankelborg}, \&
  {Jaeggli}}]{testa14}
{Testa}, P., {De Pontieu}, B., {Allred}, J., {et~al.} 2014, Science, 346,
  1255724, \dodoi{10.1126/science.1255724}

\bibitem[{{Tian} \& {Chen}(2018)}]{tian18}
{Tian}, H., \& {Chen}, N.~H. 2018, The Astrophysical Journal, 856, 34,
  \dodoi{10.3847/1538-4357/aab15a}

\bibitem[{{Tian} {et~al.}(2011){Tian}, {McIntosh}, {De Pontieu},
  {Mart{\'\i}nez-Sykora}, {Sechler}, \& {Wang}}]{tian11}
{Tian}, H., {McIntosh}, S.~W., {De Pontieu}, B., {et~al.} 2011, The
  Astrophysical Journal, 738, 18, \dodoi{10.1088/0004-637X/738/1/18}

\bibitem[{{Tian} {et~al.}(2015){Tian}, {Young}, {Reeves}, {Chen}, {Liu}, \&
  {McKillop}}]{tian15}
{Tian}, H., {Young}, P.~R., {Reeves}, K.~K., {et~al.} 2015, The Astrophysical
  Journal, 811, 139, \dodoi{10.1088/0004-637X/811/2/139}

\bibitem[{{Wang} {et~al.}(2018){Wang}, {Liu}, {Deng}, \& {Wang}}]{wang18}
{Wang}, J., {Liu}, C., {Deng}, N., \& {Wang}, H. 2018, \apj, 853, 143,
  \dodoi{10.3847/1538-4357/aaa712}

\bibitem[{{Warren} {et~al.}(2016){Warren}, {Reep}, {Crump}, \&
  {Sim{\~o}es}}]{warren16}
{Warren}, H.~P., {Reep}, J.~W., {Crump}, N.~A., \& {Sim{\~o}es}, P. J.~A. 2016,
  The Astrophysical Journal, 829, 35, \dodoi{10.3847/0004-637X/829/1/35}

\bibitem[{{Watanabe} {et~al.}(2010){Watanabe}, {Hara}, {Sterling}, \&
  {Harra}}]{watanabe10}
{Watanabe}, T., {Hara}, H., {Sterling}, A.~C., \& {Harra}, L.~K. 2010, The
  Astrophysical Journal, 719, 213, \dodoi{10.1088/0004-637X/719/1/213}

\bibitem[{{Wuelser} {et~al.}(1994){Wuelser}, {Canfield}, {Acton}, {Culhane},
  {Phillips}, {Fludra}, {Sakao}, {Masuda}, {Kosugi}, \& {Tsuneta}}]{wuelser94}
{Wuelser}, J.-P., {Canfield}, R.~C., {Acton}, L.~W., {et~al.} 1994, The
  Astrophysical Journal, 424, 459, \dodoi{10.1086/173903}

\bibitem[{{Yan} {et~al.}(2015){Yan}, {Peter}, {He}, {Tian}, {Xia}, {Wang},
  {Tu}, {Zhang}, {Chen}, \& {Barczynski}}]{yan15}
{Yan}, L., {Peter}, H., {He}, J., {et~al.} 2015, \apj, 811, 48,
  \dodoi{10.1088/0004-637X/811/1/48}

\bibitem[{{Young} {et~al.}(2013){Young}, {Doschek}, {Warren}, \&
  {Hara}}]{young13}
{Young}, P.~R., {Doschek}, G.~A., {Warren}, H.~P., \& {Hara}, H. 2013, \apj,
  766, 127, \dodoi{10.1088/0004-637X/766/2/127}

\bibitem[{{Young} {et~al.}(2018){Young}, {Tian}, {Peter}, {Rutten}, {Nelson},
  {Huang}, {Schmieder}, {Vissers}, {Toriumi}, {Rouppe van der Voort},
  {Madjarska}, {Danilovic}, {Berlicki}, {Chitta}, {Cheung}, {Madsen},
  {Reardon}, {Katsukawa}, \& {Heinzel}}]{young18}
{Young}, P.~R., {Tian}, H., {Peter}, H., {et~al.} 2018, Space Science Reviews,
  214, 120, \dodoi{10.1007/s11214-018-0551-0}

\bibitem[{{Zarro} {et~al.}(1988){Zarro}, {Slater}, \& {Freeland}}]{zarro88}
{Zarro}, D.~M., {Slater}, G.~L., \& {Freeland}, S.~L. 1988, The Astrophysical
  Journal Letters, 333, L99, \dodoi{10.1086/185296}

\bibitem[{{Zemanov{\'a}} {et~al.}(2019){Zemanov{\'a}}, {Dud{\'\i}k},
  {Aulanier}, {Thalmann}, \& {G{\"o}m{\"o}ry}}]{zemanova19}
{Zemanov{\'a}}, A., {Dud{\'\i}k}, J., {Aulanier}, G., {Thalmann}, J.~K., \&
  {G{\"o}m{\"o}ry}, P. 2019, The Astrophysical Journal, 883, 96,
  \dodoi{10.3847/1538-4357/ab3926}

\bibitem[{{Zhang} {et~al.}(2012){Zhang}, {Cheng}, \& {Ding}}]{zhang12}
{Zhang}, J., {Cheng}, X., \& {Ding}, M.-D. 2012, Nature Communications, 3, 747,
  \dodoi{10.1038/ncomms1753}

\bibitem[{{Zhu} {et~al.}(2019){Zhu}, {Kowalski}, {Tian}, {Uitenbroek},
  {Carlsson}, \& {Allred}}]{zhu19}
{Zhu}, Y., {Kowalski}, A.~F., {Tian}, H., {et~al.} 2019, \apj, 879, 19,
  \dodoi{10.3847/1538-4357/ab2238}

\end{thebibliography}

\end{document}